\documentclass[12pt,a4paper,twocolumn,aps,fleqn,nofootinbib,superscriptaddress,longbibliography]{revtex4}
\usepackage[utf8]{inputenc}
\usepackage[english]{babel}
\usepackage{epsf}
\usepackage{latexsym,amssymb,euscript}
\usepackage[dvips]{graphicx}
\usepackage[fleqn]{amsmath}
\usepackage{nicefrac}
\usepackage{slashed}
\usepackage{booktabs}
\usepackage{hyperref}
\usepackage{braket}
\usepackage{chngcntr}
\usepackage{bbold}
\usepackage{graphics}
\usepackage{graphicx}
\usepackage{mciteplus}
\usepackage{pdfpages}
\usepackage{setspace}
\usepackage[titletoc]{appendix}
\graphicspath{{./figures/}}
\hypersetup{
 linktocpage = true,
 urlcolor = citegreen,
 colorlinks = true,
 linkcolor = urlblue,
 anchorcolor = citegreen,
 citecolor = citegreen,
 pdfstartview = {XYZ null null 1.25} 
           }
\usepackage[left=1.5cm, right=1.5cm, top=3.0cm, bottom=2.5cm]{geometry}
\usepackage{pstricks}
\usepackage{color}
\usepackage{xcolor}
\definecolor{urlblue}{rgb}{0.2,0.4,0.7}
\definecolor{citegreen}{rgb}{0,0.4,0.2}
\definecolor{linkred}{rgb}{0.9,0.2,0.1}
\usepackage{float}
\usepackage{academicons}
\definecolor{orcidlogocol}{HTML}{A6CE39}
\usepackage{fancyhdr}
\newcommand{\drv}{{\rm d}}

\newcommand{\scalarp}{\hspace{-0.05cm}\mathbin{\vcenter{\hbox{\scalebox{.3}{$\bullet$}}}}\hspace{-0.05cm}}

\newcommand{\LQCD}{\Lambda_{\rm QCD}}
\newcommand{\MSb}{\overline{\rm MS}}

\newcommand{\DY}{\Delta Y}

\newcommand{\tcite}[1]{~\cite{#1}}
\newcommand{\tref}[1]{~\ref{#1}}
\newcommand{\eref}[1]{~\eqref{#1}}

\newcommand{\tarr}{
\begin{array}}
\newcommand{\earr}{\end{array}}

\newcommand{\orcidFGC}{\href{https://orcid.org/0000-0003-3299-2203}{\includegraphics[scale=0.1]{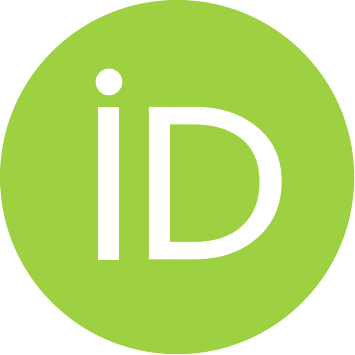}}}

\newcommand{\orcidMF}{\href{https://orcid.org/0000-0002-2408-2210}{\includegraphics[scale=0.1]{figures/logo-orcid.pdf}}}

\newcommand{\orcidMMAM}{\href{http://orcid.org/0000-0002-6449-2257}{\includegraphics[scale=0.1]{figures/logo-orcid.pdf}}}

\newcommand{\orcidAP}{\href{https://orcid.org/0000-0001-8984-3036}{\includegraphics[scale=0.1]{figures/logo-orcid.pdf}}}

\begin{document}


\title{
  {\Large \bf Ultraforward production of a charmed hadron \\ plus a Higgs boson in unpolarized proton collisions}
}

\author{Francesco Giovanni Celiberto \orcidFGC}
\email{fceliberto@ectstar.eu}
\affiliation{\setstretch{1.0}European Centre for Theoretical Studies in Nuclear Physics and Related Areas (ECT*), I-38123 Villazzano, Trento, Italy}
\affiliation{\setstretch{1.0}Fondazione Bruno Kessler (FBK), I-38123 Povo, Trento, Italy}
\affiliation{\setstretch{1.0}INFN-TIFPA Trento Institute of Fundamental Physics and Applications, I-38123 Povo, Trento, Italy}

\author{Michael Fucilla \orcidMF}
\email{michael.fucilla@unical.it}
\affiliation{\setstretch{1.0}Dipartimento di Fisica, Universit\`a della Calabria, I-87036 Arcavacata di Rende, Cosenza, Italy}
\affiliation{\setstretch{1.0}Istituto Nazionale di Fisica Nucleare, Gruppo collegato di Cosenza I-87036 Arcavacata di Rende, Cosenza, Italy}
\affiliation{\setstretch{1.0}Université Paris-Saclay, CNRS, IJCLab, 91405 Orsay, France}


\author{Mohammed M.A. Mohammed \orcidMMAM}
\email{mohammed.maher@unical.it}
\affiliation{\setstretch{1.0}Dipartimento di Fisica, Universit\`a della Calabria, I-87036 Arcavacata di Rende, Cosenza, Italy}
\affiliation{\setstretch{1.0}Istituto Nazionale di Fisica Nucleare, Gruppo collegato di Cosenza I-87036 Arcavacata di Rende, Cosenza, Italy}

\author{Alessandro Papa \orcidAP \;}
\email{alessandro.papa@fis.unical.it}
\affiliation{\setstretch{1.0}Dipartimento di Fisica, Universit\`a della Calabria, I-87036 Arcavacata di Rende, Cosenza, Italy}
\affiliation{\setstretch{1.0}Istituto Nazionale di Fisica Nucleare, Gruppo collegato di Cosenza I-87036 Arcavacata di Rende, Cosenza, Italy}

\begin{abstract}
\vspace{0.75cm}
\begin{center}
 {\bf Abstract}
\end{center}
\vspace{0.75cm}
\hrule \vspace{0.50cm}
\footnotesize \setstretch{1.0}

We investigate the inclusive emission in unpolarized proton collisions of a charm-flavored hadron in association with a Higgs boson, featuring large transverse momenta and produced with a large rapidity distance.
Taking advantage of a narrow timing coincidence between the ATLAS detector and the future FPF ones, we study the behavior of cross sections and azimuthal correlations for ultraforward rapidities of the detected hadron.
We provide evi\-den\-ce that the hybrid high-energy and collinear factorization, encoding the BFKL resummation of large energy logarithms and supplemented by collinear densities and fragmentation functions, offers a fair description of this process and comes out as an important tool to deepen our understanding of strong interactions in ultraforward production regimes.

\vspace{0.50cm} \hrule
\vspace{0.75cm}
{
 \setlength{\parindent}{0pt}
}
\vspace{1.50cm}
\end{abstract}


\maketitle

\newpage

\begingroup
 \hypersetup{linktoc = page, 
             }
 \phantom{.}\\\phantom{.}\\\phantom{.}
 { 
 \tableofcontents
 }
\endgroup

\vspace{0.75cm}
\hrule \vspace{0.50cm}


\section{Introduction}
\label{sec:intro}

The validity of Quantum Chromodynamics (QCD) as the standard theory of strong interactions comes out from a long history of experimental evidences and theoretical achievements. Although being founded upon simple concepts and elegant ideas, QCD exhibits a complicated dynamics within its high-energy limits. This calls for an enhancement of the theoretical description that goes beyond the pure perturbative expansion in the strong coupling constant, $\alpha_s$.

On one side, the applicability of perturbative QCD (pQCD) techniques takes place in kinematic regimes where the momentum transfers accessed in the hard-scattering process are larger than the characteristic hadronic scale, $\LQCD$, at which non-perturbative effects become dominant. 
On the other side, even though pQCD fixed-order calculations within the collinear-factorization approach have been successful in describing experimental data, there are kinematic regions, accessed at high energies, were the pure collinear approach needs to be improved.

One of these kinematic sectors is the so-called \emph{semi-hard} regime\tcite{Gribov:1983ivg}, which is characterized by the hierarchy of energy scales $s\gg \{Q\}^2 \gg \LQCD^2$, with $s$ the center-of-mass energy squared and $\{Q\}$ one or a set of hard scales given by the considered final states. 
In this regime the convergence of the perturbative expansion, truncated at a certain order in the strong coupling $\alpha_s$, is spoiled. This is due to the appearance of large logarithms of the form $\ln(s/Q^2)$, that enter the perturbative series with a power increasing with the order of $\alpha_{s}$. The most adequate and powerful mechanism to perform the all-order resummation of these energy logarithms is provided by the Balitsky--Fadin--Kuraev--Lipatov (BFKL) framework~\cite{Fadin:1975cb,Kuraev:1976ge,Kuraev:1977fs,Balitsky:1978ic}. It allows for the systematic inclusion of all terms proportional to $(\alpha_s\ln(s))^n$, the
so-called leading logarithmic approximation or LLA, and of the ones proportional to $\alpha_s(\alpha_s\ln(s))^n$, the next-to-leading logarithmic approximation or NLA.

BFKL cross sections take the form of an elegant convolution between a process-independent Green's function, which embodies the resummation of energy logarithms, and two impact factors, depicting the fragmentation of each incoming particle to an identified outgoing one. The BFKL Green's function evolves according to an integral equation. Its kernel is known up to the next-to-leading order (NLO) in the perturbative expansion for any fixed, not growing with $s$, momentum transfer $t$ and for any possible two-gluon color $t$-channel exchange\tcite{Fadin:1998py,Ciafaloni:1998gs,Fadin:1998jv,Fadin:2000kx,Fadin:2000hu,Fadin:2004zq,Fadin:2005zj}.
Impact factors represent the process-dependent part of the cross section. In the context of hadron-initiated collisions with two particles emitted with a large rapidity separation (the setup matter of investigation in this article), impact factors are in turn written as a convolution between collinear parton density functions (PDFs)\footnote{See Ref.\tcite{Amoroso:2022eow} for a recent review of progresses and challenges in the determination of distribution functions in the proton at the precision frontier.} and collinear inputs describing the production mechanisms of identified final-state hadrons, such as fragmentation functions (FFs). Therefore, in this particular case we refer to our formalism as a hybrid high-energy and collinear factorization.\footnote{The first application of the hybrid factorization for forward-backward final states was in the context of Mueller--Navelet jets\tcite{Colferai:2010wu}. Another approach, close in spirit with our hybrid factorization for single forward emissions, was build in Ref.\tcite{Deak:2009xt} (see also Refs.\tcite{vanHameren:2015uia,Deak:2018obv,VanHaevermaet:2020rro,Blanco:2020akb,vanHameren:2020rqt,Guiot:2021vnp}). There, a general framework for building NLO impact factors was recently set up\tcite{vanHameren:2022mtk}.}
So far, the number of impact factors calculated within NLO accuracy is quite small.

Over the last years, semi-hard inclusive processes at high energies have been considered as a promising running track for the search of BFKL signals at current and future colliders (see Refs.\tcite{Celiberto:2017ius,Hentschinski:2022xnd,AlexanderAryshev:2022pkx} for an overview of recent applications). The larger is the rapidity interval between the tagged objects, the greater is our chance to unveil the inner aspects of high-energy dynamics. 
An incomplete list of these channels consists in: the inclusive emissions of two light-flavored jets at large transverse momenta and being widely separated in rapidity (Mueller--Navelet reactions\tcite{Mueller:1986ey}), for which several phenomenological studies have been conducted by different Collaborations~(see, \emph{e.g.},~Refs.\tcite{Colferai:2010wu,Caporale:2012ih,Ducloue:2013hia,Ducloue:2013bva,Caporale:2013uva,Caporale:2014gpa,Colferai:2015zfa,Caporale:2015uva,Ducloue:2015jba,Celiberto:2015yba,Celiberto:2015mpa,Celiberto:2016ygs,Celiberto:2016vva,Caporale:2018qnm}), the inclusive production of a light di-hadron system\tcite{Celiberto:2016hae,Celiberto:2016zgb,Celiberto:2017ptm,Celiberto:2017uae,Celiberto:2017ydk}, three and four light-jet tags\tcite{Caporale:2015vya,Caporale:2015int,Caporale:2016soq,Caporale:2016vxt,Caporale:2016xku,Celiberto:2016vhn,Caporale:2016djm,Chachamis:2016lyi,Chachamis:2016qct,Caporale:2016pqe,Caporale:2016lnh,Caporale:2016zkc,Caporale:2017jqj}, hadron-jet\tcite{Bolognino:2018oth,Bolognino:2019cac,Bolognino:2019yqj,Celiberto:2020wpk,Celiberto:2020rxb} and Higgs-jet\tcite{Celiberto:2020tmb,Mohammed:2022gbk,Celiberto:2021fjf,Celiberto:2021tky,Celiberto:2021txb,Celiberto:2021xpm} high-energy correlations, heavy-light di-jet systems\tcite{Bolognino:2021mrc,Bolognino:2021hxx}, Drell--Yan~\cite{Golec-Biernat:2018kem} and heavy-hadron studies\tcite{Boussarie:2017oae,Celiberto:2017nyx,Bolognino:2019yls,Bolognino:2019ccd,Celiberto:2021dzy,Celiberto:2021fdp,Celiberto:2022rfj,Bolognino:2022wgl,Celiberto:2022dyf,Celiberto:2022keu}.
Notably, the BFKL resummation offers us a unique chance to unravel the proton structure at small-$x$.
More in particular, exclusive as well as inclusive single forward emissions in lepton-proton or proton-proton collisions are golden channels for the study of the
\emph{unintegrated gluon distribution} (UGD) in the proton, suitably defined
as a $p_T$-convolution of the BFKL Green's function, which determines the small-$x$ evolution of the struck gluon, and a non-perturbative proton impact factor.
The UGD was recently investigated \emph{via} polarized amplitudes and cross sections for the diffractive exclusive leptoproduction of light vector mesons at HERA\tcite{Anikin:2011sa,Besse:2013muy,Bolognino:2018rhb,Bolognino:2018mlw,Bolognino:2019bko,Bolognino:2019pba,Celiberto:2019slj} and the EIC\tcite{Bolognino:2021niq,Bolognino:2021gjm,Bolognino:2022uty,Celiberto:2022fam}, and through the helicity structure functions for the inclusive forward Drell--Yan di-lepton production at LHCb\tcite{Motyka:2014lya,Motyka:2016lta,Brzeminski:2016lwh,Celiberto:2018muu}.
Determinations of NLO and next-to-NLO (NNLO) collinear PDFs improved with the inclusion of NLA small-$x$ effects were obtained in Refs.\tcite{Ball:2017otu,Abdolmaleki:2018jln,Bonvini:2019wxf}. Model calculations for polarized \emph{transverse-momentum-dependent} (TMD) gluon distributions\tcite{Mulders:2000sh,Meissner:2007rx,Lorce:2013pza,Boer:2016xqr} effectively embodying BFKL inputs were achieved in Refs.\tcite{Bacchetta:2020vty,Celiberto:2021zww,Bacchetta:2021oht,Bacchetta:2021lvw,Bacchetta:2021twk,Bacchetta:2022esb,Arbuzov:2020cqg,AbdulKhalek:2021gbh,Khalek:2022bzd,Bacchetta:2022crh}.

It is widely recognized that the theoretical description of semi-hard inclusive observables suffers from instabilities emerging in the BFKL series. This happens because NLA corrections both to the BFKL Green’s function and impact factors turn out to be of the same size and with opposite sign of pure LLA contributions. This issue brings to large uncertainties arising from the renormalization and factorization scale choice.
In particular, instabilities affecting azimuthal correlations for light jet and hadron emissions\tcite{Ducloue:2013bva,Caporale:2014gpa,Celiberto:2016hae,Celiberto:2017ptm} are too strong to prevent any possibility of setting these scales at the natural values provided by kinematics, namely the observed transverse momenta.
As a possible solution, it was proposed the adoption of some optimization procedures for scale fixing. 
The Brodsky--Lepage--Mackenzie (BLM) method~\cite{Brodsky:1996sg,Brodsky:1997sd,Brodsky:1998kn,Brodsky:2002ka} effectively helped to partially suppress these instabilities. As a side effect, however, BLM scales are definitely larger than the natural ones\tcite{Celiberto:2020wpk}. This translates in a loss of one or more orders of magnitude for the corresponding cross sections, and any attempt at reaching the precision level fails.

Recently, the study of semi-hard reactions featuring the emission of heavy-flavored hadrons was proposed as a new direction toward restoring the stability of the high-energy series under higher-order corrections and scale variations.
The first evidence that the peculiar behavior of heavy-flavor collinear FFs\tcite{Kniehl:2008zza,Kramer:2018vde,Kniehl:2020szu}, determined in a variable-flavor number-scheme (VFNS) framework\tcite{Mele:1990cw,Cacciari:1993mq}, leads to a remarkable stabilization of the NLA series came out in the context of single-charmed\tcite{Celiberto:2021dzy,Celiberto:2022rfj} and single-bottomed\tcite{Celiberto:2021fdp,Celiberto:2022rfj} hadron studies at the LHC.
This stabilizing effect emerges when a non-decreasing, smooth-behaved in energy gluon FF is convoluted with proton PDFs.

Remarkably, a strong stabilization fairly arises also when a forward quarkonium state ($J/\psi$ or $\Upsilon$) is inclusively emitted in association with a backward jet\tcite{Celiberto:2022dyf}.
In that work the quarkonium production at large transverse momentum was described at the hand of a DGLAP-evolving heavy-quark FF, modeled on the basis of the non-relativistic QCD (NRQCD) framework\tcite{Caswell:1985ui,Thacker:1990bm,Bodwin:1994jh,Braaten:1993mp,Zheng:2019dfk}.

\begin{figure*}[!t]
\centering

\includegraphics[width=0.60\textwidth]{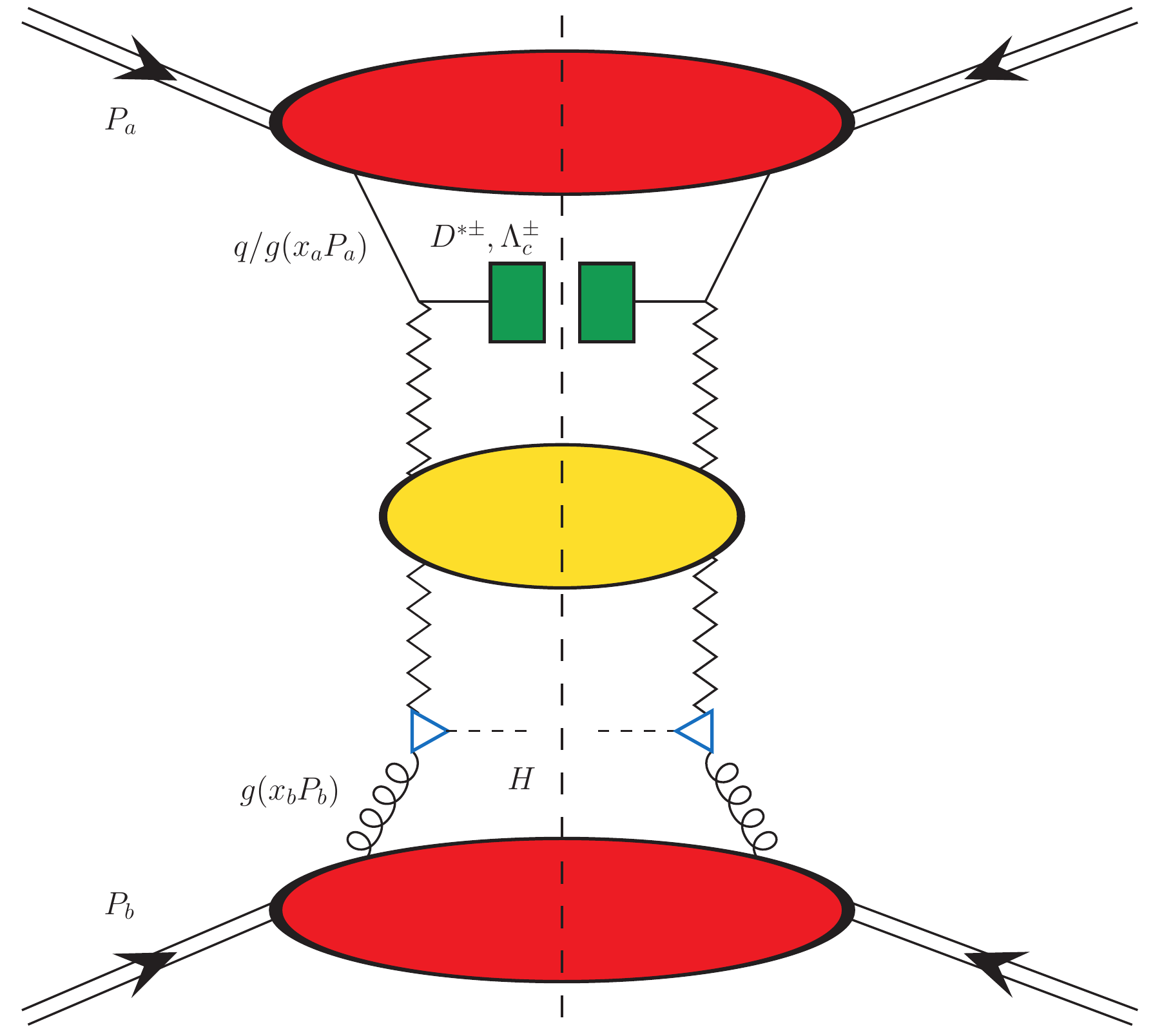}

\caption{Hybrid high-energy and collinear factorization in force. Diagrammatic representation of the ultraforward inclusive hadron plus Higgs production. Red blobs depict proton collinear PDFs, green rectangles stand for charmed-hadron collinear FFs, and blue triangles denote top-quark fermion loops. The BFKL Green's function, portrayed by the yellow blob, is connected to off-shell impact factors \emph{via} Reggeon waggle lines. The diagram was obtained through the {\tt JaxoDraw 2.0} package~\cite{Binosi:2008ig}.}
\label{fig:process}
\end{figure*}

The emergence of such a fair stability through two independent FF descriptions, namely both \emph{via} standard determinations of single-charmed hadron FFs from global fits in the collinear approach\tcite{Kniehl:2008zza,Kramer:2018vde,Kniehl:2020szu} and \emph{via} heavy-quarkonium NRQCD fragmentation models\tcite{Braaten:1993mp,Zheng:2019dfk}, strongly corroborates our statement that intrinsic features of heavy-flavored emissions offer us a way to perform reliable studies of semi-hard observables at the NLA level and at natural energy scales.

Another evidence of similar stabilizing effects came recently out from another sub-class of semi-hard reactions, where the large transverse masses of detected objects lead to a reduction of values assumed by the running coupling and thus to a systematic suppression of instabilities as\-so\-cia\-ted to the high-energy series.
These effects were quantified in a study on resummed distributions in rapidity and transverse momentum for the inclusive production of a Higgs and a light-flavored jet\tcite{Celiberto:2020tmb}, the large boson transverse mass acting as a high-energy stabilizer.

More in general, the analysis proposed in Ref.\tcite{Celiberto:2020tmb} is one of the most recent advances in the study of Higgs differential distributions at high energies.
Pioneering analyses of Higgs emissions in multi-jet events were done in Refs.\tcite{DelDuca:1993ga,DelDuca:2003ba}.
The effect of hard-rescattering corrections from gluon ladders in the central exclusive Higgs hadroproduction was assessed in Ref.\tcite{Bartels:2006ea}.
Combined BFKL and Sudakov effects in the almost back-to-back production of a Higgs-jet system in proton collisions were recently investigated\tcite{Xiao:2018esv}.
Ref.\tcite{Pasechnik:2006du} contains the LO calculation for the doubly off-shell impact factor portraying the production of a Higgs boson from gluon-gluon fusion in the small-$x$ limit.
The small-$x$ resummation based on the Altarelli--Ball--Forte (ABF) approach\tcite{Ball:1995vc,Ball:1997vf,Altarelli:2001ji,Altarelli:2003hk,Altarelli:2005ni,Altarelli:2008aj,White:2006yh} was successfully employed first in the description of central Higgs production\tcite{Marzani:2008az,Caola:2011wq}, then in rapidity\tcite{Caola:2010kv} and momentum\tcite{Forte:2015gve} distributions.\footnote{The ABF formalism relies on high-energy factorization schemes\tcite{Catani:1990xk,Catani:1990eg,Collins:1991ty,Catani:1993ww,Catani:1993rn,Catani:1994sq,Ball:2007ra,Caola:2010kv} and has been numerically implemented in the {\tt HELL} work environment\tcite{Bonvini:2016wki,Bonvini:2017ogt,Bonvini:2018iwt}.}
First doubly resummed predictions for central-Higgs production accounting for both the small-$x$ and the threshold\tcite{Bonciani:2003nt,deFlorian:2005fzc,Muselli:2017bad,Forte:2021wxe} resummations were presented in Ref.\tcite{Bonvini:2018ixe} (see also Ref.\tcite{Ball:2013bra}).
Small-$x$ double-logarithmic contributions to Higgs production in the large top-mass limit were considered in Ref.\tcite{Hautmann:2002tu}, whereas high-energy azimuthal signatures in Higgs angular distri\-bu\-tions were investigated in Ref.\tcite{Cipriano:2013ooa}.

In the present work we combine in our hybrid high-energy and collinear factorization the two main ingredients that have brought to the emergence of the \emph{natural stabilization} of resummed semi-hard distributions.
We consider the ultraforward inclusive emission of a single-charmed hadron, a $\Lambda_c^\pm$ or a $D^{*\pm}$ meson, accompanied by a Higgs boson produced in more central regions of rapidity (see Fig.\tref{fig:process}).
We expect that distinctive signals of the mentioned natural sta\-bi\-li\-ty will fairly emerge from the analysis of the rapidity and azimuthal-angle spectrum of the tagged heavy-flavor~$+$~Higgs-boson final state, thus making this reaction one of the most po\-wer\-ful channels to probe the high-energy QCD regime.

We add to our analysis a third, novel ingredient coming from very recent developments in the context of the Forward Physics Community.
Taking advantage of forthcoming studies doable at future detectors of the planned Forward Physics Facility (FPF)\tcite{Anchordoqui:2021ghd,Feng:2022inv}, we allow the charmed particle to be detected in so far unexplored ultraforward windows of rapidity, which are definitely beyond the acceptances of current LHC ones.
More in particular, we benefit from the opportunity of making FPF detectors to work in coincidence with ATLAS. 
According to current plans, FPF will detect decays products (mainly neutrinos) of light- or heavy-flavored hadrons, thus permitting us to reconstruct the production signal of our charmed hadron. However, the FPF will not be able to see Higgs bosons.
Therefore, a very precise timing procedure will make feasible the simultaneous FPF ultraforward hadron detection and the ATLAS tag of the Higgs. Techni\-cal details on the FPF~$+$~ATLAS narrow timing coincidence are reported in Section~VI~E of Ref.\tcite{Anchordoqui:2021ghd}.

The intersection corner between the charm-flavor physics and the Higgs sector opens up fascinating windows toward frontier research lines. 
A novel mechanism for the exclusive diffractive Higgs-boson production from intrinsic heavy-flavor components in the proton was proposed in Refs.\tcite{Brodsky:2006wb,Brodsky:2007yz}.
Shedding light on the intrinsic charm contribution\tcite{Brodsky:1980pb,Brodsky:2015fna,Kniehl:2009ar,Bednyakov:2013zta,Ball:2015dpa}, it plays a crucial role in obtaining precise determinations of proton PDFs\tcite{Dulat:2013hea,Ball:2015tna}, which in turn are needed to improve the description of Higgs production (see, \emph{e.g.}, Ref.\tcite{Ball:2016neh}).

Besides a pure QCD viewpoint, Higgs-boson radiative decays to quarkonia offer us the engaging opportunity of accessing and constraining the Yukawa coupling of the charm quark\tcite{Bodwin:2013gca,Bodwin:2014bpa,Kagan:2014ila,Konig:2015qat}.
A similar result can be achieved \emph{via} the measurement of the production cross section for the Higgs in association with a charmed jet\tcite{Brivio:2015fxa}.
Higgs decays to charmonia at large transverse momenta can be investigated on the basis of the $c$-quark fragmentation approximation\tcite{Han:2022rwq}.
Observables sensitive to the emission of charmed $B$-mesons are thought to be quite good probes for rare Higgs decays\tcite{Jiang:2015pah,Karyasov:2016hfm,Liao:2018nab}.

Coming back to strong interactions in ultraforward regimes, the hea\-vi\-ly asymmetric ranges in transverse momenta and rapidities accessible \emph{via} a FPF~$+$~A\-TLAS setup represent a fertile ground where to disengage the pure high-energy dynamics from DGLAP contaminations\tcite{Ducloue:2013bva,Celiberto:2015yba,Celiberto:2015mpa,Celiberto:2020wpk}.
Clues of a reached stability of our hybrid factorization in the ultraforward rapidity regime would re\-pre\-sent a striking advancement in semi-hard phe\-no\-me\-no\-lo\-gy and a core step forward toward assessing the feasibility of QCD precision studies at high energies.

\section{Heavy-flavor plus Higgs production in hybrid factorization}
\label{sec:theory}

We consider the ultraforward inclusive charmed-hadron plus Higgs production in unpolarized proton-proton ($pp$) collisions, diagrammatically represented in Fig.\tref{fig:process},
\begin{equation}
\label{process}
p(P_a) + p(P_b) \to  {\cal C}(\vec p_{\cal C}, y_{\cal C}) + {\cal X} + H(\vec p_H, y_H),
\end{equation}
both of the two emitted objects featuring high transverse momenta, $\vec p_{{\cal C},H} \gg \Lambda_{\rm QCD}$, and a large rapidity separation, $\DY = y_{\cal C} - y_H$.
The bound state labeled as ${\cal C}$ represents either a $\Lambda_c^\pm$ baryon or a $D^{*\pm}$ meson, whereas $H$ is a scalar Higgs boson.
The four-momenta of the parent protons, $P_{a,b}$, are taken as Sudakov vectors
satisfying $P_{a,b}^2 = 0$ and $P_a \scalarp P_b = s/2$, so that the final-state
transverse momenta can be decomposed in the following way
\begin{eqnarray}
p_{\cal C} &=& x_{\cal C} P_b + \frac{|\vec p_{\cal C}|^2}{x_{\cal C} s}P_a + p_{{\cal C}\perp} \,,
 \\ \nonumber
p_H &=& x_H P_a + \frac{m_{H \perp}^2}{x_H s}P_b + p_{H\perp} \,,
\label{Sudakov}
\end{eqnarray}
with $p_{{\cal C},H \perp}^2 = - |\vec p_{{\cal C},H}|^2$, and the spacial part of the four-vector $P_{a\parallel}$ being taken positive. Then, $m_{H \perp} = \sqrt{m_{H}^2+|\vec p_{H}|^2}$ is the Higgs-boson transverse mass.
The longitudinal momentum fractions, $x_{{\cal C},H}$, are related to the corresponding rapidities in the center-of-mass frame \emph{via} the relations
\begin{equation}
\label{rapidities}
y_{{\cal C}} = \ln \left( \frac{x_{{\cal C}} \sqrt{s}}{|\vec p_{\cal C}|} \right) \; ,
 \quad
y_H = - \ln \left( \frac{x_H \sqrt{s}}{m_{H \perp}} \right) \; ,
\end{equation}
thus having $\drv y_{{\cal C},H} = \pm \drv x_{{\cal C},H}/x_{{\cal C},H}$.
As for the rapidity distance, one has
\begin{equation}\label{rapidity_interval}
\DY = y_{\cal C} - y_H = \ln \left( \frac{x_{\cal C} x_H s}{|\Vec{p_{\cal C}}| m_{H \perp}} \right) \;.
\end{equation}

\begin{widetext}
The cross section is given as the Fourier series of the so-called \emph{azimuthal coefficients}
\begin{equation}
\frac{\drv \sigma}{\drv y_{\cal C} \, \drv y_H \, \drv |\vec p_{\cal C}| \, \drv |\vec p_H| \, \drv \varphi_{\cal C} \, \drv \varphi_H}
=\frac{1}{(2\pi)^2}\left[{\hat C}_0 + \sum_{\kappa=1}^\infty  2\cos (\kappa\varphi )\,
{\hat C}_\kappa\right]\; ,
\end{equation}
where $\varphi=\varphi_{\cal C}-\varphi_H-\pi$, with $\varphi_{{\cal C},H}$ the ${\cal C}$-hadron and Higgs azimuthal angles. A NLA formula for the $\varphi$-summed cross section, ${\hat C}_0$, and the other coefficients, ${\hat C}_{\kappa > 0}$,
reads
\begin{equation}\label{Cn}
\begin{split}
{\hat C}_\kappa &\,\equiv\, \int_0^{2\pi} \drv \varphi_{\cal C} \int_0^{2\pi} \drv \varphi_H\,
\cos(\kappa \varphi) \,
\frac{d\sigma}{\drv y_{\cal C} \, \drv y_H \, \drv |\vec p_{\cal C}| \, \drv |\vec p_H| \, \drv \varphi_{\cal C} \, \drv \varphi_H} \\
&\,=\, \frac{e^{\DY}}{s} \frac{m_{H \perp}}{|\vec p_H|} \int_{-\infty}^{+\infty} \drv \nu \, e^{\DY \bar \alpha_s(\mu_{R_c})\left\{\chi(\kappa,\nu)+\bar\alpha_s(\mu_{R_c})
	\left[\bar\chi(\kappa,\nu)+\frac{\beta_0}{8 N_c}\chi(\kappa,\nu)\left[-\chi(\kappa,\nu)
	+\frac{10}{3}+4\ln\left(\frac{\mu_{R_c}}{\sqrt{\vec p_{\cal C}\vec p_H}}\right)\right]\right]\right\}}
\\
&\,\times\, \left[ \alpha_s(\mu_{R_1}) c_{\cal C}(\kappa,\nu,|\vec p_{\cal C}|,x_{\cal C}) \right]
\left[ \alpha_s^2(\mu_{R_2}) [c_H(\kappa,\nu,|\vec p_H|,x_H)]^* \right] \,
\\
&\,\times\, \left\{1 
+ \alpha_s(\mu_{R_1}) \frac{c_{{\cal C}}^{(1)}(\kappa,\nu,|\vec p_{\cal C}|, x_{\cal C})}{c_{\cal C}(\kappa,\nu,|\vec p_{\cal C}|,
	x_{\cal C})}
+ \alpha_s(\mu_{R_2}) \left[\frac{c_H^{(1)}(\kappa,\nu,|\vec p_H|,x_H)}{c_H(\kappa,\nu,|\vec p_H|, x_H)}\right]^*
\right\} \; , 
\end{split}
\end{equation}
\end{widetext}
where $\bar \alpha_s \equiv N_c/\pi \, \alpha_s$, with $N_c$ the QCD color
number, $\beta_0$
the first coefficient in the expansion of the QCD $\beta$-function ($n_f$ is the active-flavor number), and $\chi\left(\kappa,\nu\right) $
the eigenvalue of the LO BFKL kernel, and $c_{{\cal C},H}(\kappa,\nu)$ are the ${\cal C}$-hadron and Higgs LO impact factors in the Mellin ($\kappa,\nu$)-space, given by
\begin{widetext}
\begin{equation}
\label{c-hadron}
\begin{split}
c_{{\cal C}}(\kappa,\nu,|\vec p_{\cal C}|,x_{\cal C})\,&=\,
2 \sqrt{\frac{C_F}{C_A}}
\left( \vec p_{\cal C}^{\: 2} \right)^{i\nu-1/2}\int_{x_{\cal C}}^{1}\frac{\drv\xi}{\xi}\left(\frac{\xi}{x_{\cal C}}\right)^{2i\nu-1}
\\
&\times\, \left[\frac{C_A}{C_F}f_g(\xi,\mu_{F_1})D_g^{{\cal C}}\left(\frac{x_{\cal C}}{\xi},\mu_{F_1}\right)
+\sum_{a=q,\bar q}f_a(\xi,\mu_{F_1})D_a^{{\cal C}}\left(\frac{x_{\cal C}}{\xi},\mu_{F_1}\right)\right]  
\end{split}
\end{equation}
for the ${\cal C}$-hadron, and
\begin{equation}
\label{c-higgs}
c_H(\kappa,\nu, |\vec p_H|, x_H) = \frac{1}{v^2} \frac{|\mathcal{H}_f(\vec p_H^{\: 2})|^2}
{128\pi^{3}\sqrt{2(N^2_{c}-1)}}
\left( \vec p_H^{\: 2} \right)^ {i\nu + 1/2} f_g(x_H,\mu_{F_2}) \; ,
\end{equation}
\end{widetext}
for the Higgs boson. The structure of equations presented above unambiguously shows how our hybrid high-energy and collinear factorization is built. To the high-energy convolution between the Green's function and impact factors (see Eq.\eref{Cn}), the collinear convolution between the hard factor, PDFs $f_a(x_{\cal C},\mu_{F_1})$ for the struck parton (gluon or quark) and FFs $D_a^{\cal C}\left(\frac{x_{\cal C}}{\xi},\mu_{F_1}\right)$ for the outgoing parton generating the ${\cal C}$-hadron corresponds (see Eq.\eref{c-hadron}).
The LO Higgs impact factor in the Mellin space was obtained in Eq.\tcite{Celiberto:2020tmb} and its expression contains the $f_g(x_H,\mu_{F_2})$ depicting the incoming gluon in the lower impact factor of Fig.\tref{fig:process} that takes part in the hard sub-process.
The $\mathcal{H}_f(\vec p_H^{\: 2})$ encodes the gluon-Reggeon-Higgs top-triangle vertex in the same figure. Its expression was originally calculated in Ref.\tcite{DelDuca:2003ba} and reads
\begin{widetext}
\begin{equation}
\label{IF_F}
 \hspace{-0.45cm}
 \mathcal{H}_f(\vec p_H^{\: 2}) = \frac{4 m_t^2}{m_{H \perp}^2}
 \left\{
 \left( \frac{1}{2} - \frac{2 m_t^2}{m_{H \perp}^2} \right) \left[ \Delta_h(\upsilon_2)^2 - \Delta_h(\upsilon_1)^2 \right]
  + \left( \frac{2 \vec p_H^{\: 2}}{m_{H \perp}^2} \right)
 \left[ \sqrt{\upsilon_1} \, \Delta_h(\upsilon_1) - \sqrt{\upsilon_2} \, \Delta_h(\upsilon_2) \right] + 2
  \right\} ,
\end{equation}
\end{widetext}
with $m_t = 173.21$ GeV the top-quark mass, $\upsilon_1 = 1
- 4 m_t^2/m_H^2$, $\upsilon_2 = 1 + 4 m_t^2/\vec p_H^{\: 2}$, the root
$\sqrt{\upsilon_1} = i \sqrt{|\upsilon_1|}$ holding for negative values of $\upsilon_1$.
Moreover one has
\begin{equation}
 \label{IF_W}
 \Delta_h(\upsilon) = \left\{
 \begin{aligned}
  &- 2 i \arcsin \frac{1}{\sqrt{1 - \upsilon}} \; , 
  &\upsilon < 0 \; ; \\ 
  &\ln \frac{1 + \sqrt{\upsilon}}{1 - \sqrt{\upsilon}} - i \pi \; ,
  &0 < \upsilon < 1 \; ; \\ 
  &\ln \frac{1 + \sqrt{\upsilon}}{\sqrt{\upsilon} - 1} \; ,
  &\upsilon > 1 \; . 
 \end{aligned}
 \right.
\end{equation}

We remark that our treatment for the LO Higgs vertex encodes finite top-mass contributions. The effect of taking the same calculation in the $(m_t \to + \infty)$ limit was gauged in Section~3.2 of Ref.\tcite{Celiberto:2020tmb}.
A result for the Higgs NLO impact factor was recently obtained in the infinite top-mass limit in the standard BFKL approach~\cite{Celiberto:2022fgx} as well as within the Lipatov’s high-energy effective action~\cite{Hentschinski:2020tbi,Nefedov:2019mrg}. However, its numerical implementation in the Mellin space, $c_H^{(1)}(\kappa,\nu,|\vec p_H|, x_H)$, is not yet available.
Therefore, in our study we consider a partial NLO implementation that includes some ``universal''  contributions to the Higgs impact factor, proportional to the corresponding LO impact factor.
These terms are obtained on the basis of a renormalization group analysis, namely \emph{via} the requirement of stability at NLO under variations of energy scales. Thus we have
\begin{widetext}
\begin{equation}\label{cH1}
c_H^{(1)}(\kappa,\nu,|\vec p_H|, x_H) \,=\,
c_H(\kappa, \nu, |\vec p_H|, x_H) \left\{ \frac{\beta_0}{2 \pi} 
\left(\ln \frac{\mu_{R_2}}{|\vec p_H|} + \frac{5}{6} \right) 
+ \chi\left(\kappa,\nu\right) \ln \frac{\sqrt{s_0}}{m_{H \perp}}
+ \frac{\beta_0}{2 \pi} \ln \frac{\mu_{R_2}}{m_{H \perp}}
\right.
\end{equation}
\[
\left.
\hspace{1.5cm} -\, \frac{1}{\pi f_g(x_H,\mu_{F_2})} \ln \frac{\mu_{F_2}}{m_{H \perp}} \int_{x_H}^1\frac{\drv \xi}{\xi}
\left[P_{gg}(\xi)f_g\left(\frac{x_H}{\xi},\mu_{F_2}\right)+\sum_{a={q,\bar q}} P_{ga}(\xi)
f_a\left(\frac{x_H}{\xi},\mu_{F_2}\right)\right]
\right\} \;.
\]
\end{widetext}

The NLO correction for the forward hadron impact factor was computed ten years ago~\cite{Ivanov:2012iv}.
For the sake of consistency with the treatment of the Higgs case, we take into account only the universal NLO corrections, \emph{i.e.} the ones proportional to the LO hadron impact factor, thus getting
\begin{widetext}
\begin{equation}
\label{cC1}
\begin{split}
c_{\cal C}^{(1)}(\kappa,\nu,&|\vec p_{\cal C}|, x_{\cal C}) \,=\,
c_{\cal C}(n, \nu, |\vec p_{\cal C}|, x_{\cal C}) \Biggl\{ \frac{\beta_0}{2 \pi} 
\left( \ln \frac{\mu_{R_1}}{|\vec p_{\cal C}|} + \frac{5}{6} \right)
+ \chi\left(\kappa,\nu\right) \ln \frac{\sqrt{s_0}}{|\vec p_{\cal C}|}
\Biggl.
\\
&\,-\, \frac{1}{\pi \tilde{f}(x_{\cal C},\mu_{F_1}) }
\ln \frac{\mu_{F_1}}{|\vec p_{\cal C}|}
\left[
\frac{C_A}{C_F} 
\int_{x_{\cal C}}^1\frac{\drv \xi}{\xi}
\left[P_{gg}(\xi)f_g\left(\frac{x_{\cal C}}{\xi},\mu_{F_1}\right)+\sum_{a={q,\bar q}} P_{ga}(\xi)
    f_a\left(\frac{x_{\cal C}}{\xi},\mu_{F_1}\right)\right]\right.
\\
&\,+\, 
\left.
\left.
\sum_{a = q,\bar{q}} \int_{x_{\cal C}}^1\frac{\drv \xi}{\xi}
\left[P_{ag}(\xi) f_g\left(\frac{x_{\cal C}}{\xi},\mu_{F_1}\right)
  + P_{aa}(\xi)f_a\left(\frac{x_{\cal C}}{\xi},\mu_{F_1}\right)\right]
\right]
\right\} \; ,
\end{split}
\end{equation}
where
\begin{equation}
\label{PDF_effective}
\tilde{f}(x,\mu_F) = \frac{C_A}{C_F} f_{g}(x,\mu_F) + \sum_{a = q,\bar{q}}f_{a}(x,\mu_F) \; .
\end{equation}
\end{widetext}
In Eqs.\eref{cH1} and\eref{cC1} $s_0$ is an energy-scale parameter, arbitrary within NLA accuracy, which we set to $s_0 = |\Vec{p_{\cal C}}| m_{H \perp}$.
Furthermore, $P_{ij}(\xi)$ depict the DGLAP splitting kernels at LO.

As it is well-known, any variation of the renormalization ($\mu_{R_{1,2,c}}$) and factorization ($\mu_{F_{1,2}}$) scales produces effects that are beyond the NLO accuracy. Therefore, in principle their choice can be arbitrary. It is convenient, however, to set these scales to the \emph{natural} values provided by the process kinematics. They are generally connected with physical hard scales.
From the inspection of the diagram in Fig.\tref{fig:process}, we can distinguish two sub-processes. The first one is the production of the ${\cal C}$-hadron in the upper fragmentation region, and the characteristic hard scale here is the $|\Vec{p_{\cal C}}|$ transverse momentum. The second one is the emission of the Higgs boson in the lower fragmentation region, featuring $m_{H \perp}$ as the typical hard scale.
Therefore, we go with a multi-scale strategy, namely we choose two distinct values for scales, depending on which sub-process they refer. We set $\mu_{R_1}=\mu_{F_1}=C_\mu  m_{\cal C \perp}$, with $m_{\cal C \perp} = \sqrt{m_{\cal C}^2+|\vec p_{\cal C}|^2}$ the charmed-hadron transverse mass, and $\mu_{R_2}=\mu_{F_2}=C_\mu m_{H \perp}$.
Moreover, since the two sub-processes are connected \emph{via} the gluon ladder (see Fig.\tref{fig:process}), we select a third value for the renormalization scale entering the exponentiated kernel in Eq.\eref{Cn}, given by the geometric mean of the former ones. We thus have $\mu_{R_c}= C_\mu \sqrt{\mu_{R_1}\mu_{R_2}}$.
The $C_\mu$ variation parameter will be set as explained in the next Section.

\section{Phenomenology}
\label{sec:pheno}

All the numerical studies presented in this Section were performed \emph{via} the {\tt JETHAD} multi-modular interface\tcite{Celiberto:2020wpk}.
In order to gauge the sensitivity of our predictions on the scale variation, we let renormalization and factorization scales stay around their \emph{natural} values provided by kinematics, up to a factor that ranges from 1/2 to two, controlled by the $C_\mu$ scale parameter.
We encode in uncertainty bands of our plots the comprehensive effect of scale variation and phase-space multi-dimensional integration, this latter being uniformly kept below 1\% by the {\tt JETHAD} integrator routines.

Collinear PDFs are calculated \emph{via} the {\tt MMHT14} NLO set~\cite{Harland-Lang:2014zoa} as provided by the {\tt LHAPDF} package~\cite{Buckley:2014ana}. As already mentioned, charmed hadrons are described in terms of VFNS collinear FFs. In particular, the {\tt KKKS08}\tcite{Kneesch:2007ey} and the {\tt KKSS19}\tcite{Kniehl:2020szu} NLO parametrizations are used to depict $D^{*\pm}$ meson and $\Lambda_c^\pm$ baryon emissions, respectively.
We remark that the use of these VFNS functions together with calculations for the hadron impact factor obtained in the light-quark limit\tcite{Ivanov:2012iv} is adequate, provided that the $\mu_{F_1}$ scale is larger than the threshold for DGLAP evolution of the charm quark, generally set to $m_c \simeq 1.275$ GeV.
For comparison, we present predictions for ultraforward detections of lighter hadron species, \emph{i.e.} charged pions and kaons, portrayed by the corresponding {\tt NNFF1.0}\tcite{Bertone:2017tyb} NLO FF sets.

Masses of the observed particles are set to: $m_{D^{*\pm}} = 2.01$ GeV, $m_{\Lambda_c^\pm} = 2.286$ GeV, $m_{\pi^\pm} = 139.57$ MeV, $m_{K^\pm} = 493.677$ MeV, and $m_H = 125.18$ GeV.
The center-of-mass energy is $\sqrt{s}= 14$ TeV.
All calculations are performed in the $\MSb$ renormalization scheme\tcite{PhysRevD.18.3998}.

\subsection{Observables and kinematics}
\label{ssec:observables}

The first class of observables included in our phenomenological study is represented by \emph{azimuthal-angle coefficients} integrated over the phase space of the two outgoing particles, at fixed values of their rapidity distance $\DY$,
\begin{widetext}
\begin{equation}
    \label{Cn_int}
    C_\kappa(s, \DY) =
    \int_{y^{\rm min}_{\cal C}}^{y^{\rm max}_{\cal C}} \hspace{-0.18cm} \drv y_{\cal C}
    \int_{y^{\rm min}_H}^{y^{\rm max}_H} \hspace{-0.18cm} \drv y_H
    \int_{p^{\rm min}_{\cal C}}^{p^{\rm max}_{\cal C}} \hspace{-0.18cm} \drv |\vec p_{\cal C}|
    \int_{p^{\rm min}_H}^{{p^{\rm max}_H}} \hspace{-0.18cm} \drv |\vec p_H|
    \, \delta \left(y_{\cal C} - y_H - \DY \right)
    \, {\cal C}_\kappa \left(y_{\cal C}, y_H, |\vec p_{\cal C}|, |\vec p_H|) \right)
    \, ,
\end{equation}
\end{widetext}
the ${\cal C}_\kappa$ phase-space differential coefficients being defined in Section\tref{sec:theory}. Among them, $C_0$ is the so-called $\DY$-distribution, which corresponds to the cross section differential in $\DY$ and summed over the azimuthal-angle distance $\varphi$.

The second class of observables matter of our investigation are the \emph{azimuthal-correlation moments}, $R_{\kappa0} = C_\kappa/C_0 \equiv \langle \cos (n \varphi) \rangle$, built as ratios between a given $C_{\kappa>0}$ coefficient and $C_0$, and their ratios\tcite{Vera:2006un,Vera:2007kn}, $R_{\kappa\lambda} = R_{\kappa0}/R_{\lambda0} \equiv \langle \cos (\kappa \varphi) \rangle / \langle \cos (\lambda \varphi) \rangle$.
Their behavior as functions of $\DY$ provides us with quantitative information about the weight of undetected gluon radiation predicted by the high-energy resummation.

The last observable considered in our analysis is the \emph{azimuthal distribution}\tcite{Marquet:2007xx} of the charmed hadron and the jet, as a function of $\varphi$ and at fixed values of $\DY$,
\begin{widetext}
\begin{equation}
    \label{phi_distribution}
    \frac{\drv \sigma (s, \DY, \varphi)}{\sigma \, \drv \varphi} \equiv
    \frac{1}{2 \pi} \left\{ 1 + 2 \sum_{\kappa=1}^{\infty} \cos(\kappa \varphi) \, \frac{C_\kappa}{C_0} \right\}
    = \frac{1}{\pi} \left\{ \frac{1}{2} + \sum_{\kappa = 1}^{\infty} \cos(\kappa \varphi) \langle \cos(\kappa \varphi) \rangle \right\} \; .
\end{equation}
\end{widetext}
From the experimental point of view, the azimuthal distribution is one of the most directly accessible observables. Indeed, since experimental measurements do not cover the entire azimuthal-angle plane due to apparatus limitations, the comparison of predictions with data of $\varphi$-differential distributions is easier than that for azimuthal correlations.
From the theory point of view, our distribution embodies signals of high-energy dynamics contained in all the $C_\kappa$ coefficients. Therefore, it is one of the best observables where to hunt for high-energy resummation effects.
We checked that an excellent stability of numerical computations of Eq.\eref{phi_distribution} is reached when the $\kappa$-sum is truncated at $\kappa_{\sup} \equiv 50$.

We perform phenomenological studies of all the considered observables by imposing realistic kinematic cuts of forthcoming experimental analyses at the LHC. 
In particular, we allow the charmed-hadron transverse momentum to be in the range 8~GeV~$< p_{\cal C} <$~20~GeV and its rapidity in the ultraforward rapidity window $6 < y_{\cal C} < 7.5$. This choice is in line with nominal acceptances of FPF detector plans\tcite{Anchordoqui:2021ghd,Feng:2022inv}. 
On one hand, there is a clear evidence that, when two particles are inclusively emitted in semi-hard configurations and with large rapidity separation\tcite{Celiberto:2020wpk}, the use of disjoint windows helps to single out the pure high-energy signal from DGLAP contaminations. On the other hand, as pointed out in Ref.\tcite{Celiberto:2020tmb}, Higgs emissions in $p_T$-ranges close to the ones of the other outgoing object give us the chance of quenching contributions coming from other resummation mechanisms (among them, the \emph{transverse-momentum} and the \emph{threshold} one). Therefore, we opt for a solution that stays in between the two options, namely we set 20~GeV~$< p_{H} <$~60~GeV. Concerning the Higgs rapidity, we allow for its detection in the ATLAS barrel, namely we set $|y_H| < 2.5$.

\begin{figure*}[!t]

   \includegraphics[scale=0.53,clip]{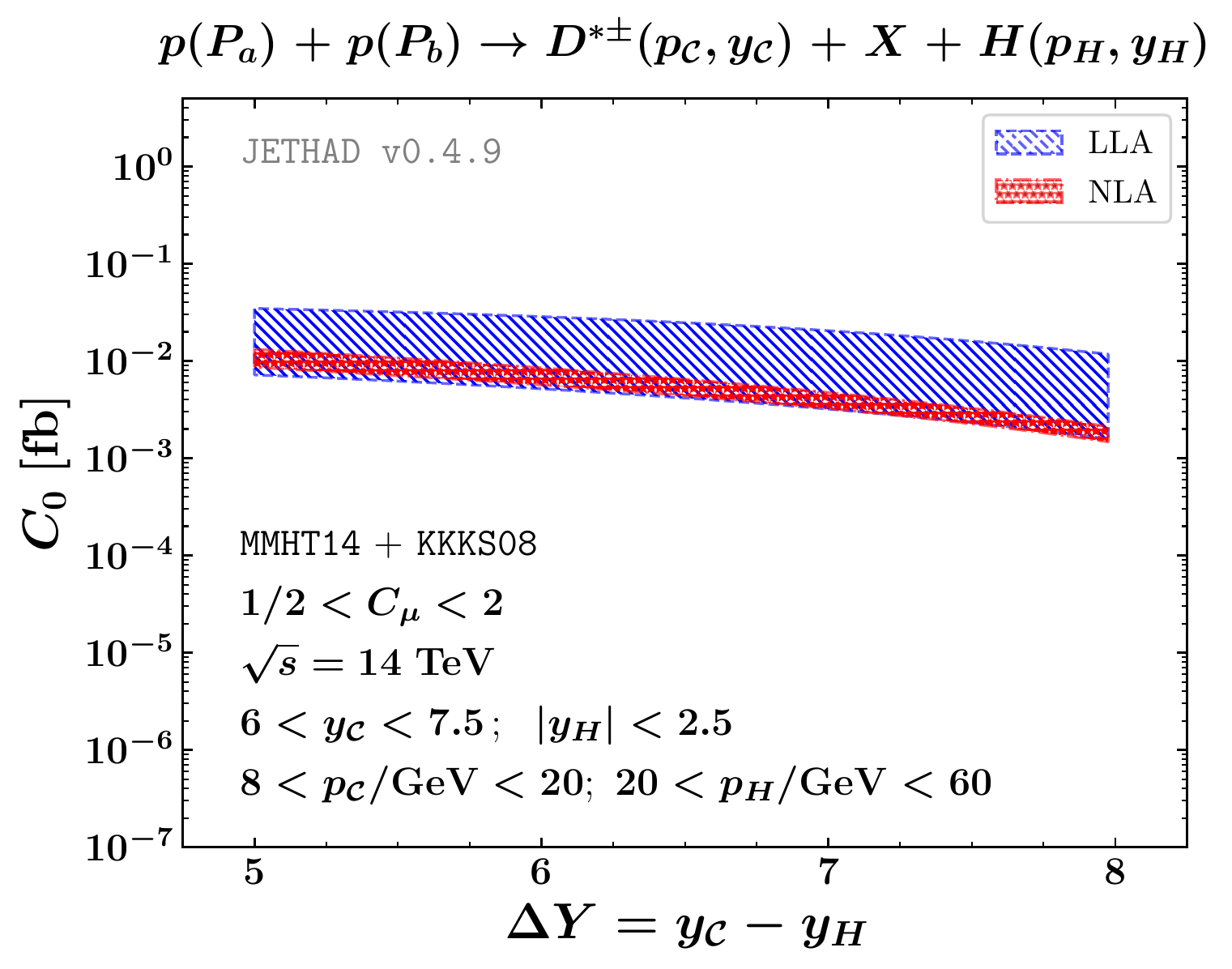}
   \includegraphics[scale=0.53,clip]{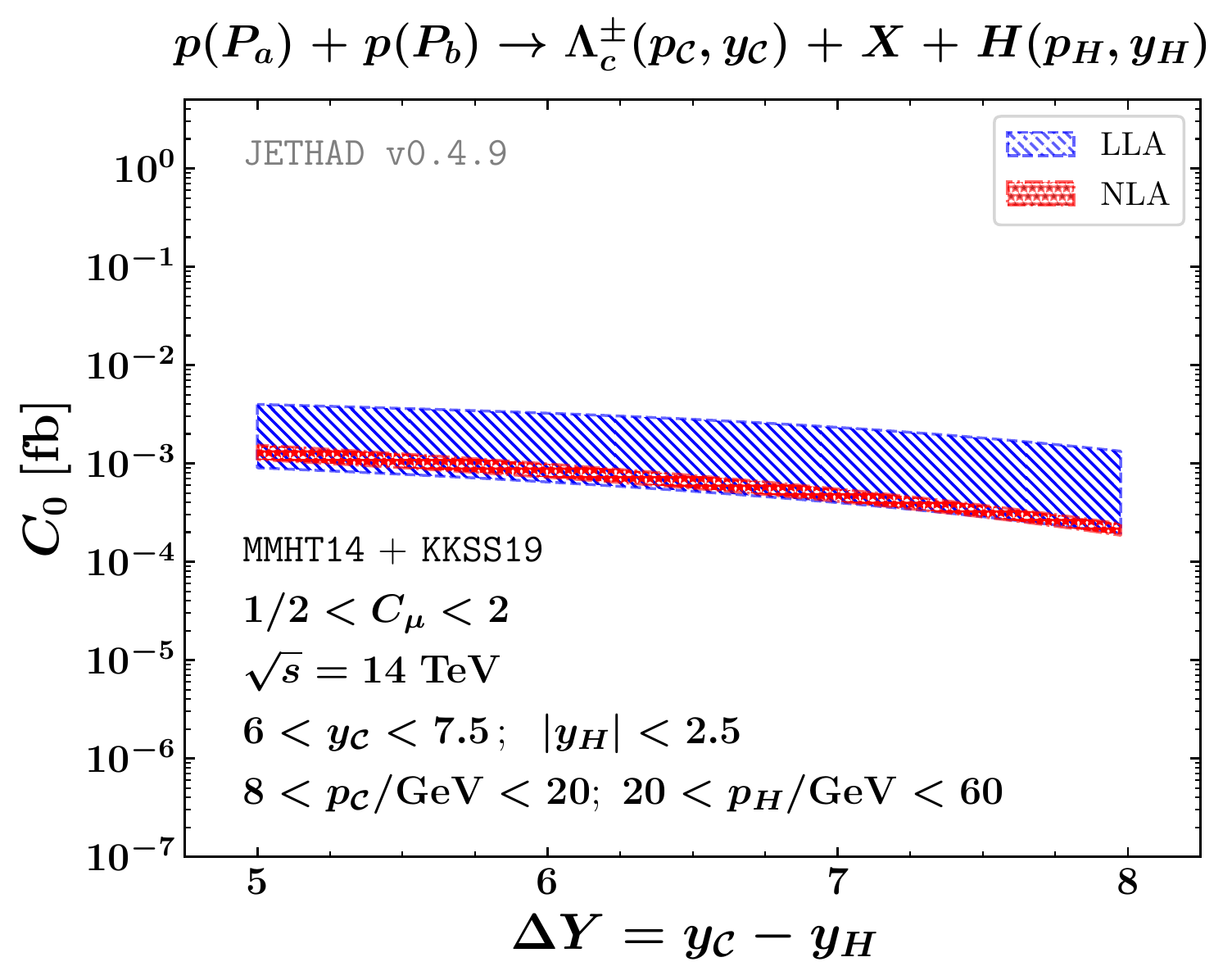}

   \includegraphics[scale=0.53,clip]{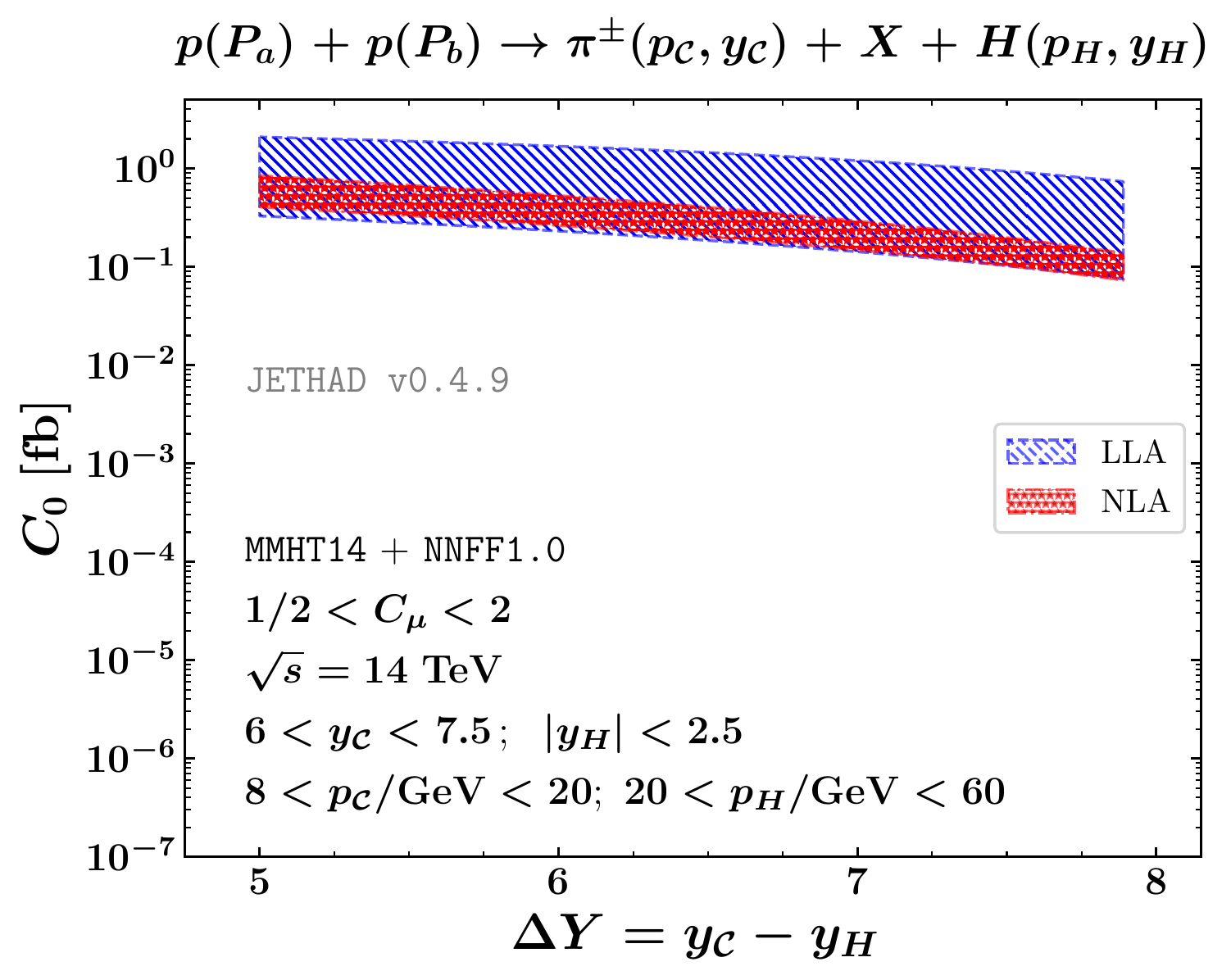}
   \hspace{0.10cm}
   \includegraphics[scale=0.53,clip]{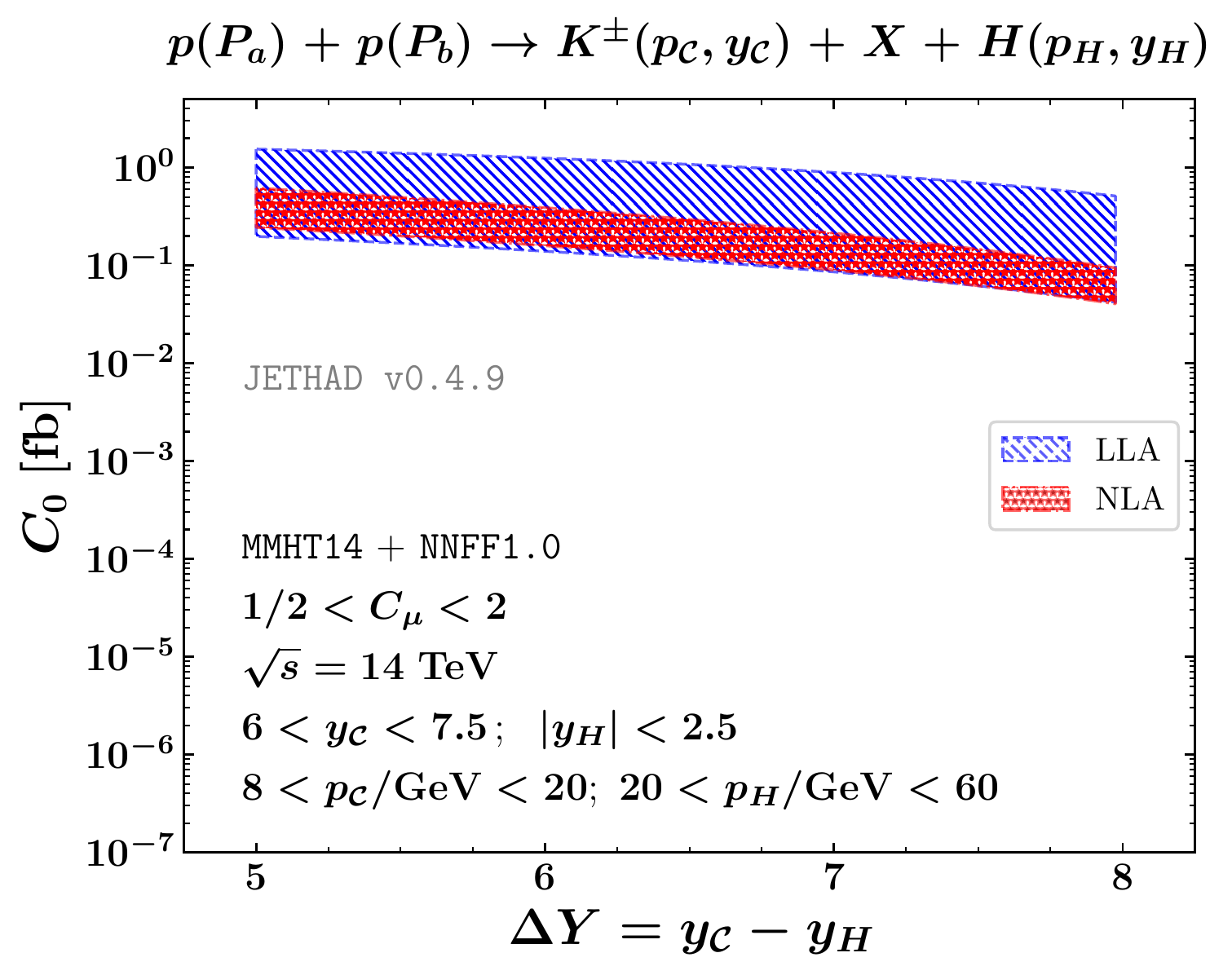}

\caption{$\DY$-distribution for the ultraforward inclusive hadron plus Higgs production at $\sqrt{s} = 14$ TeV. Predictions for charmed-hadron emissions (upper panels) are compared with light-hadron detections (lower panels). Text boxes inside panels refer to final-state kinematic cuts. Uncertainty bands embody the combined effect of scale variation and phase-space multi-dimensional integration.}
\label{fig:C0}
\end{figure*}

\begin{figure*}[!t]
\centering

\includegraphics[scale=0.53,clip]{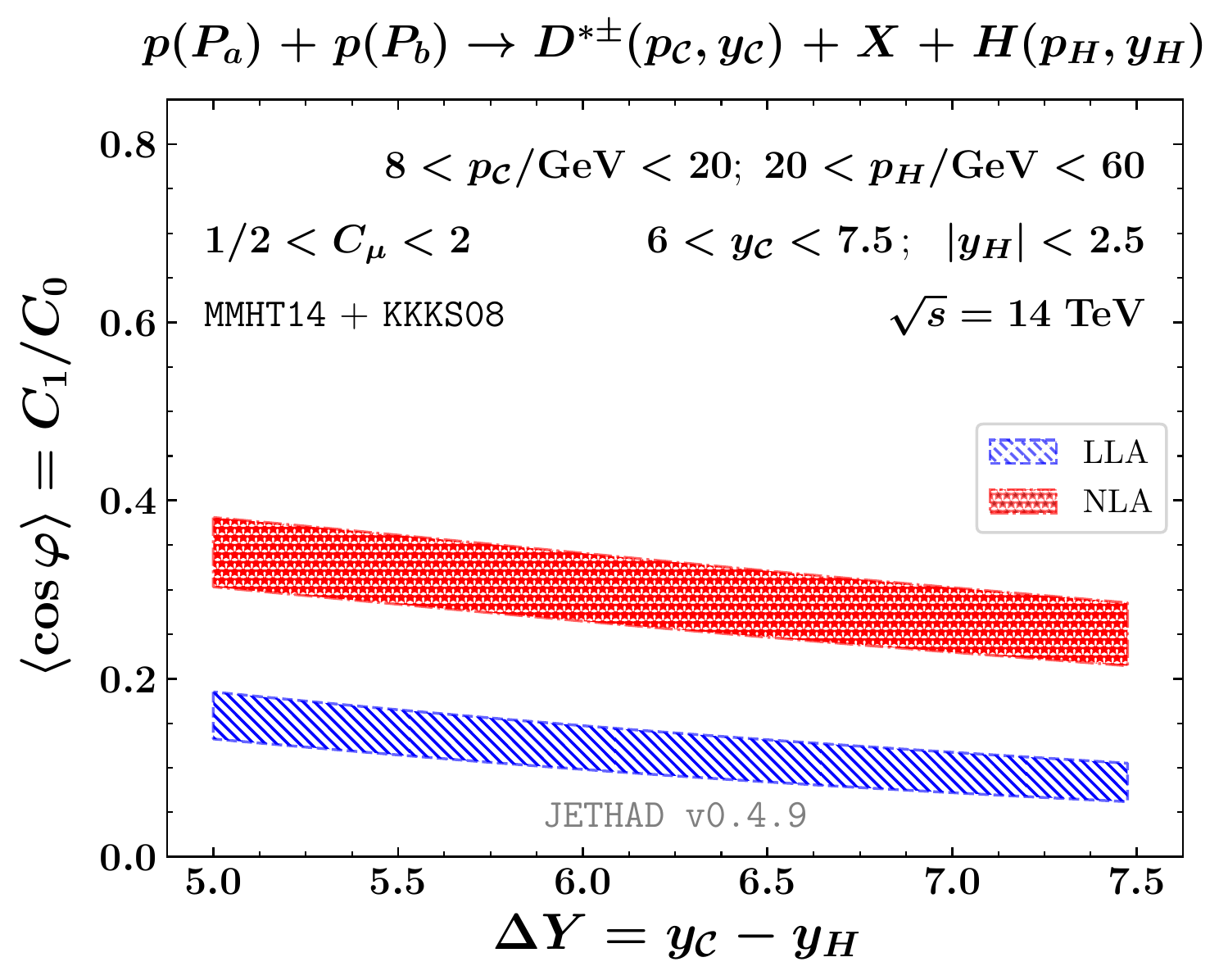}
\includegraphics[scale=0.53,clip]{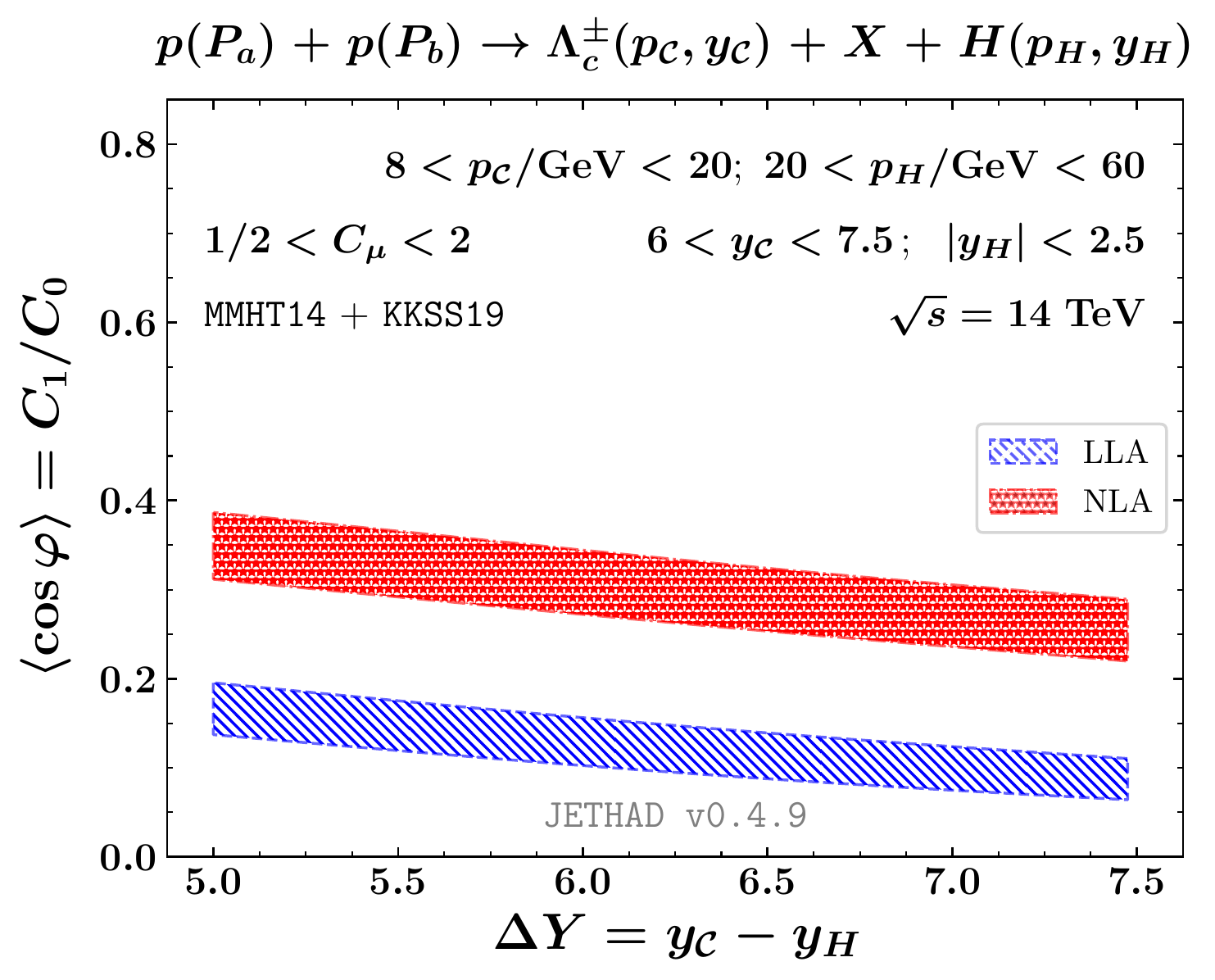}

\includegraphics[scale=0.53,clip]{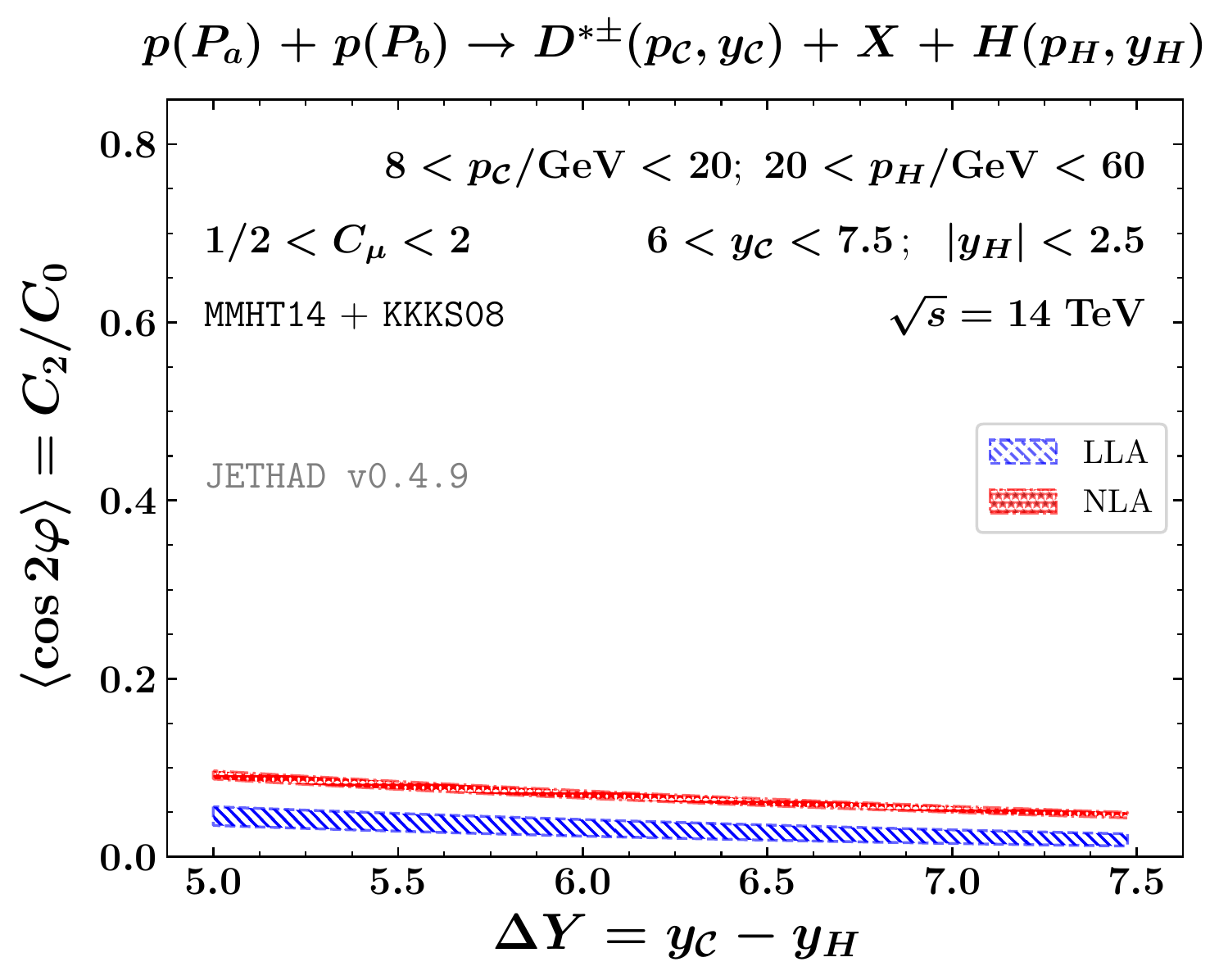}
\includegraphics[scale=0.53,clip]{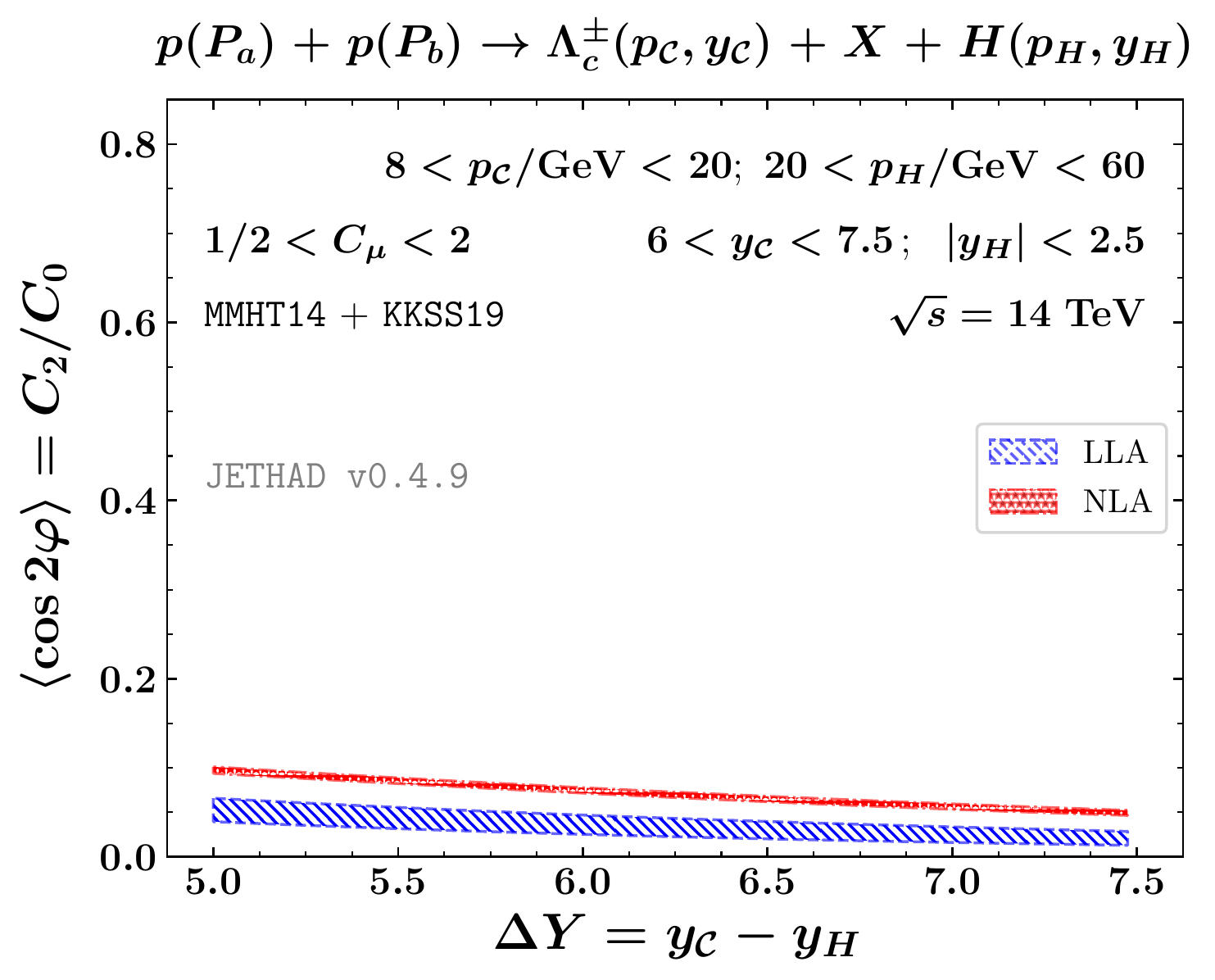}

\includegraphics[scale=0.53,clip]{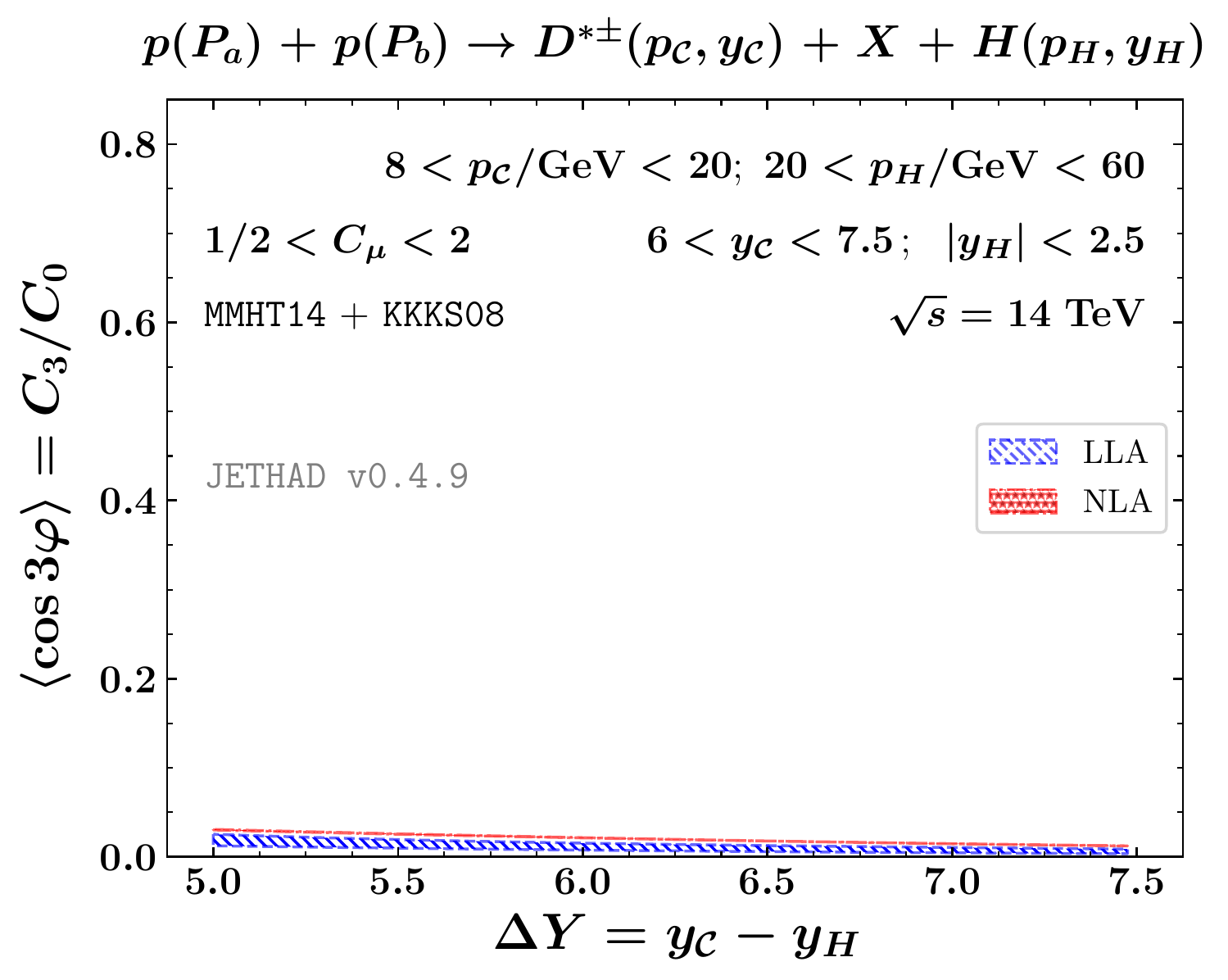}
\includegraphics[scale=0.53,clip]{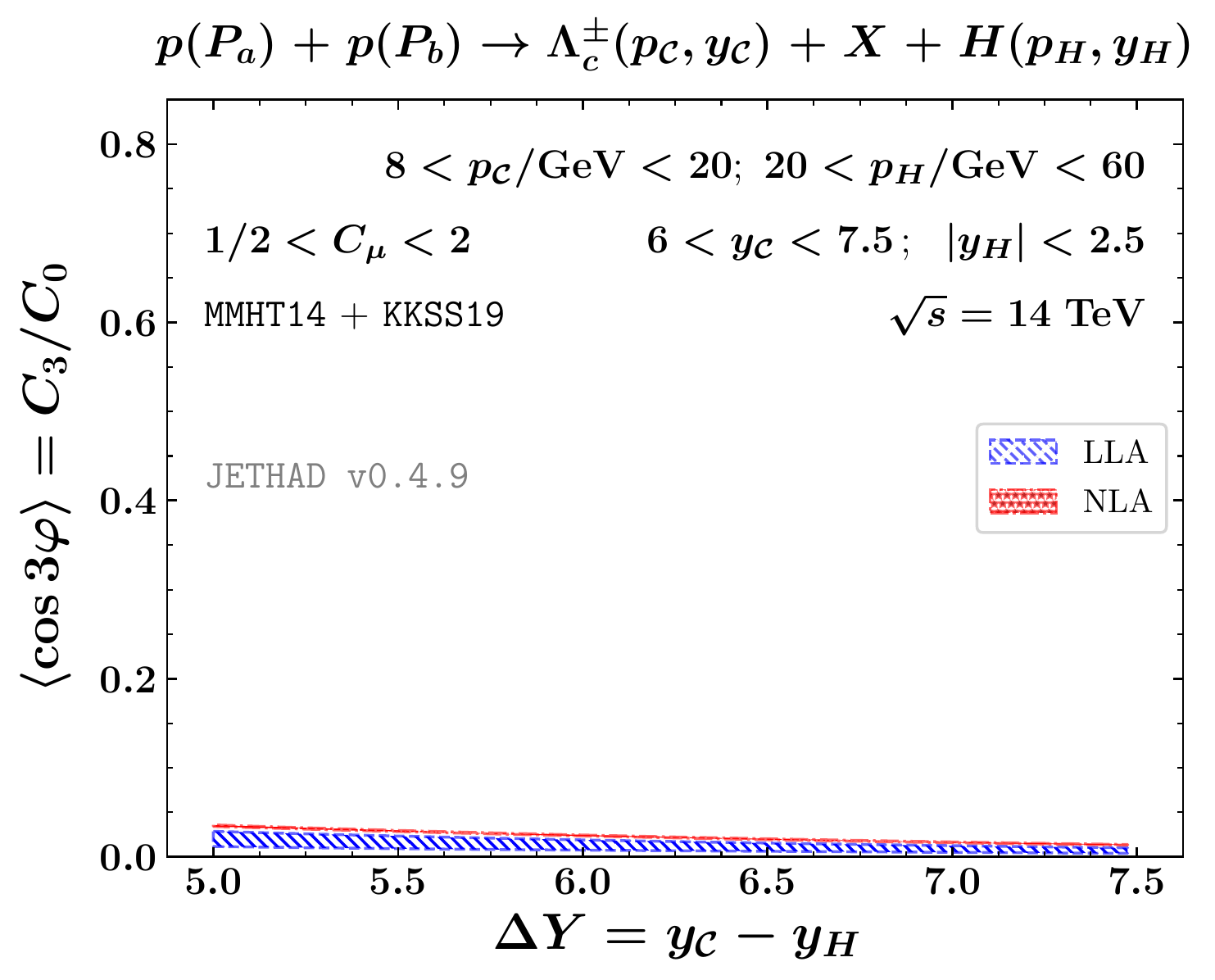}

\caption{$\DY$-pattern of azimuthal-correlation moments, $R_{\kappa0} \equiv C_{\kappa}/C_{0}$, for the ultraforward inclusive $D^{*^\pm}$ plus Higgs channel (left plots) and the inclusive $\Lambda_c^\pm$ plus Higgs channel (right panels) at $\sqrt{s} = 14$ TeV. Text boxes inside panels refer to final-state kinematic cuts. Uncertainty bands embody the combined effect of scale variation and phase-space multi-dimensional integration.}
\label{fig:Rn0}
\end{figure*}

\subsection{Results}
\label{ssec:results}
In this Section, we present our results for the observables previously introduced. In Fig.~\ref{fig:C0}, the total cross section summed over azimuthal angles and differential
in the final-state rapidity distance, $C_0$, is shown for our two single-charmed hadron species, $D^{* \pm}$ and $\Lambda_c^{\pm}$ (upper panels), and compared with the corresponding one for two lighter mesons, $\pi^{\pm}$ and $K^{\pm}$ (lower panels).

The stabilization pattern can be immediately observed from these plots. Here, NLA predictions are systematically contained inside LLA ones and the width of uncertainty bands considerably decreases when moving to the higher order. We also observe that the sensitivity of NLA predictions under scale variations lowers when charmed species are detected. This effect is due to the smoothly-behaved, non-decreasing with $\mu_F$, ${\cal C}$-hadron gluon FFs (see Section~3.4 of Ref.\tcite{Celiberto:2021dzy} for technical details).

Furthermore, statistics associated with heavy-particle emissions is one to two orders of magnitude lower than that lighter-meson one, but it still remains promising. As usually observed in processes described within the hybrid factorization, the net result of two competing effects, namely the growth of the partonic cross sections with energy and the simultaneous falloff of gluon PDFs, is the well known decreasing pattern of $C_0$ when $\DY$ grows.

\begin{figure*}[!t]
\centering

\includegraphics[scale=0.53,clip]{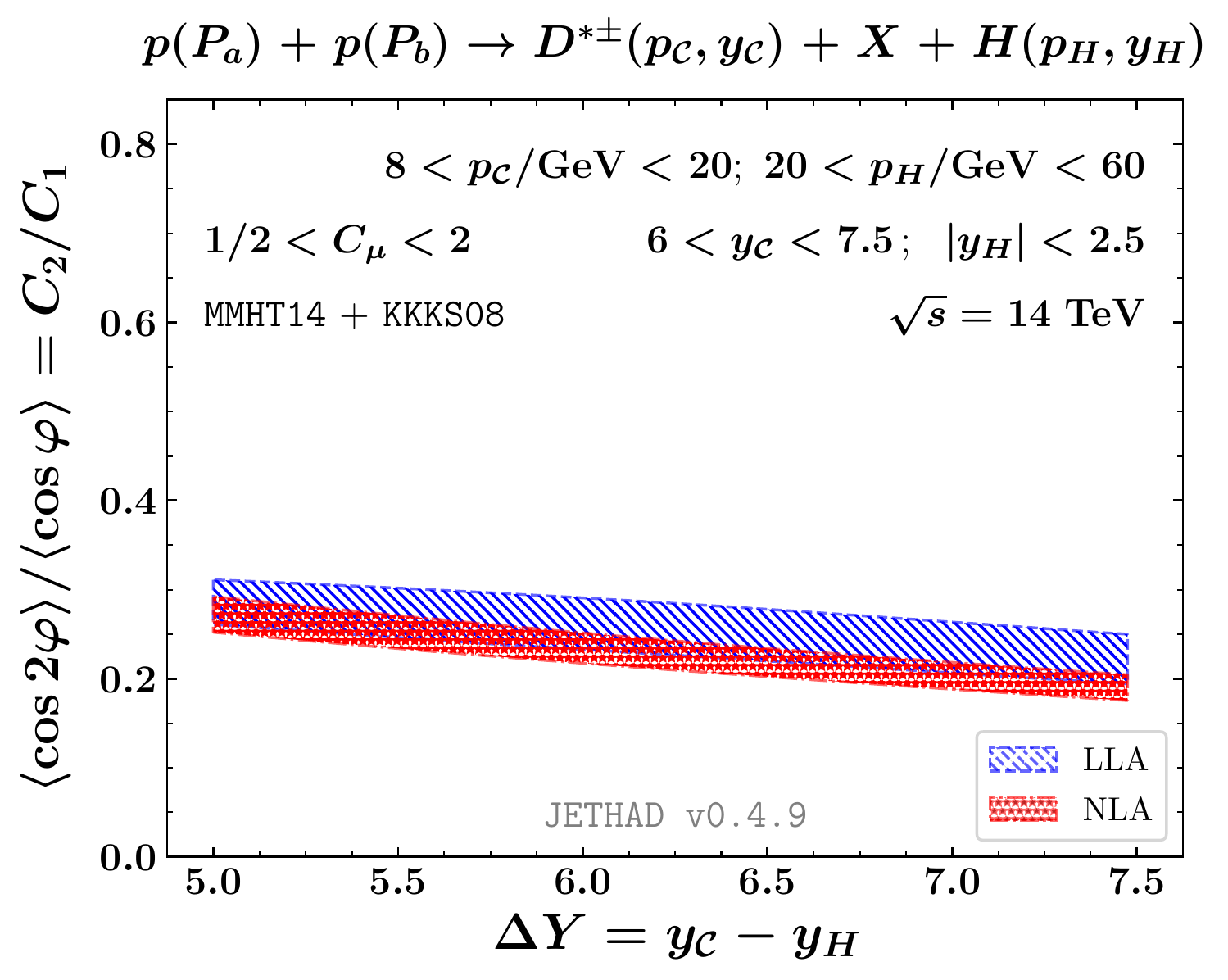}
\includegraphics[scale=0.53,clip]{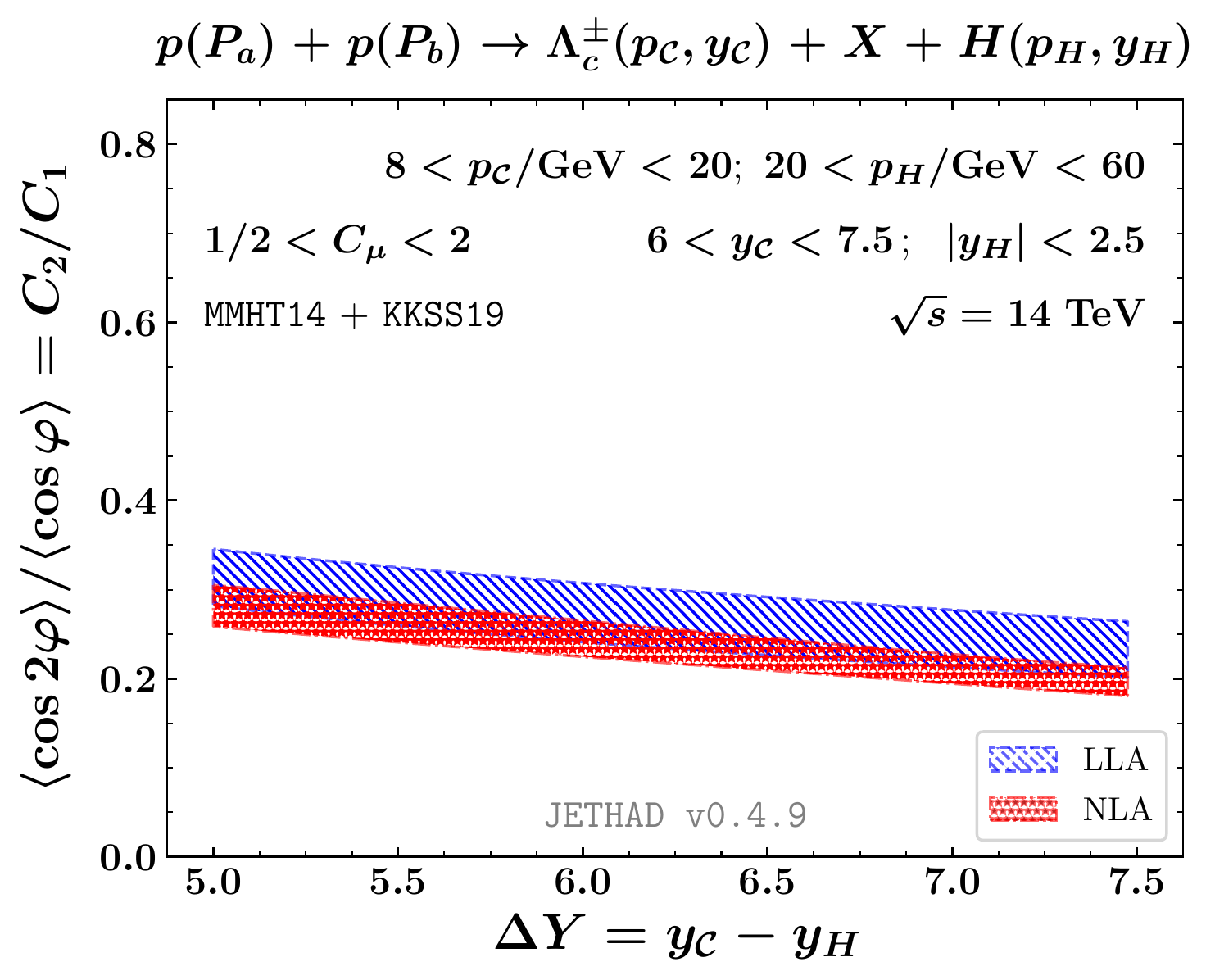}

\includegraphics[scale=0.53,clip]{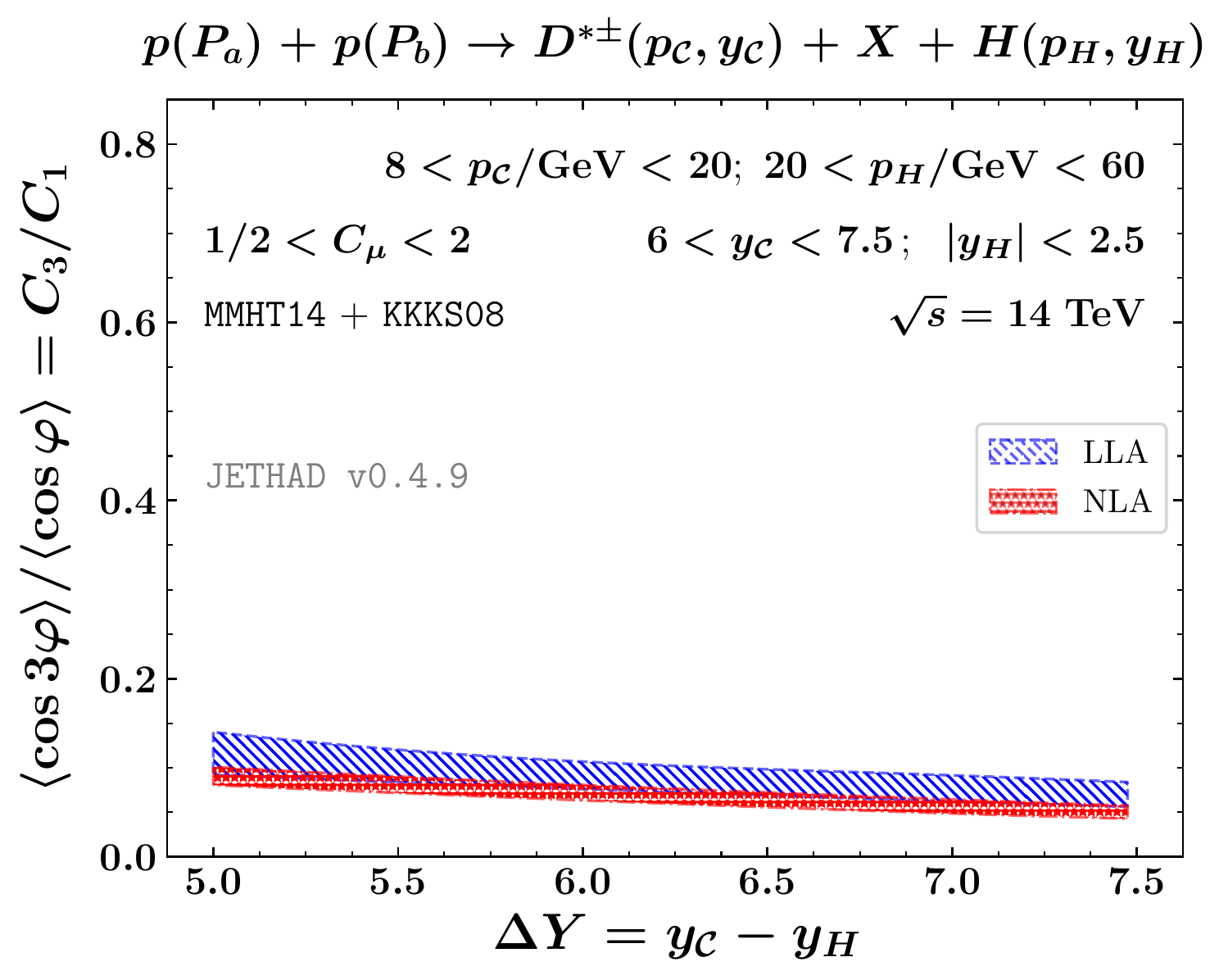}
\includegraphics[scale=0.53,clip]{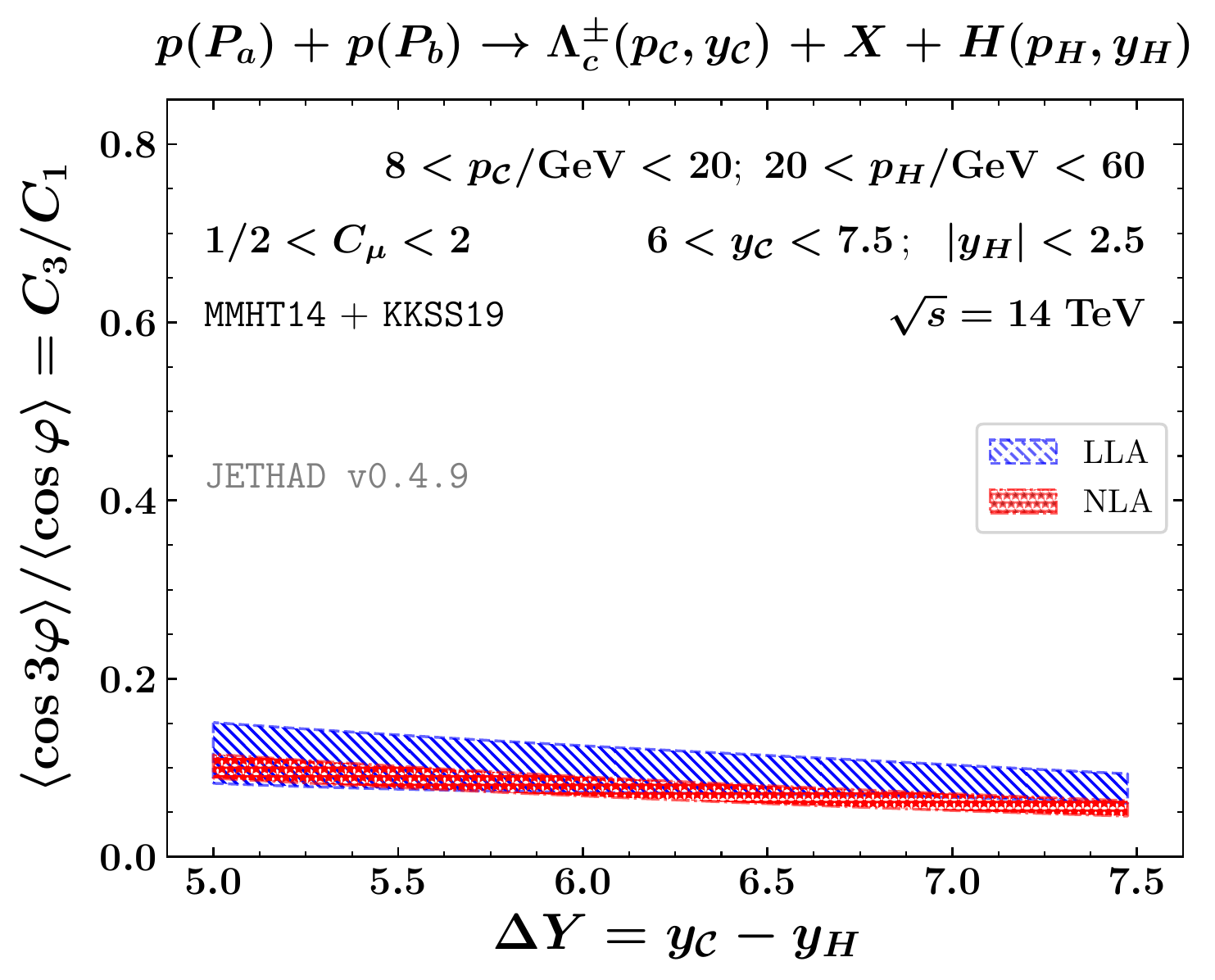}

\includegraphics[scale=0.53,clip]{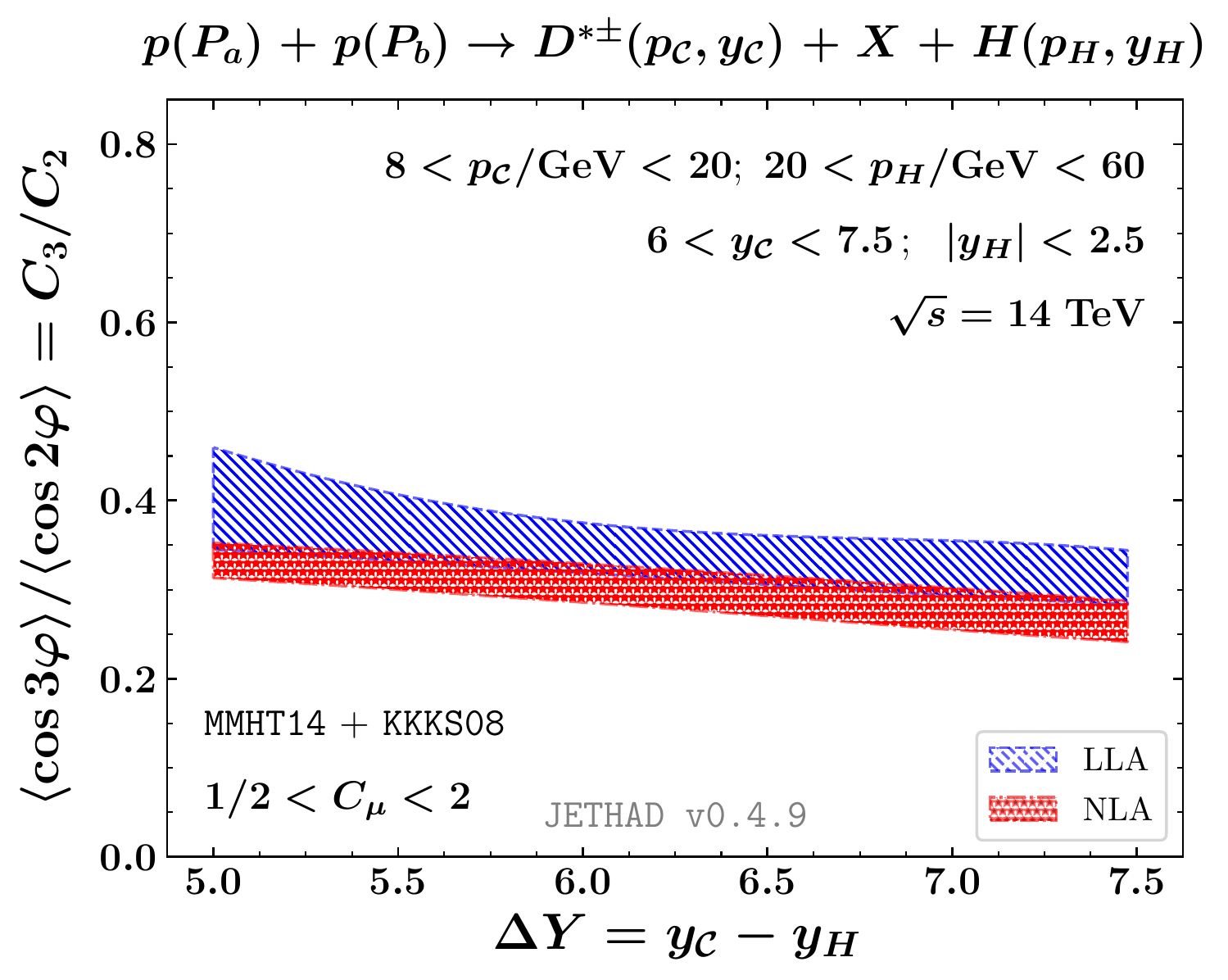}
\includegraphics[scale=0.53,clip]{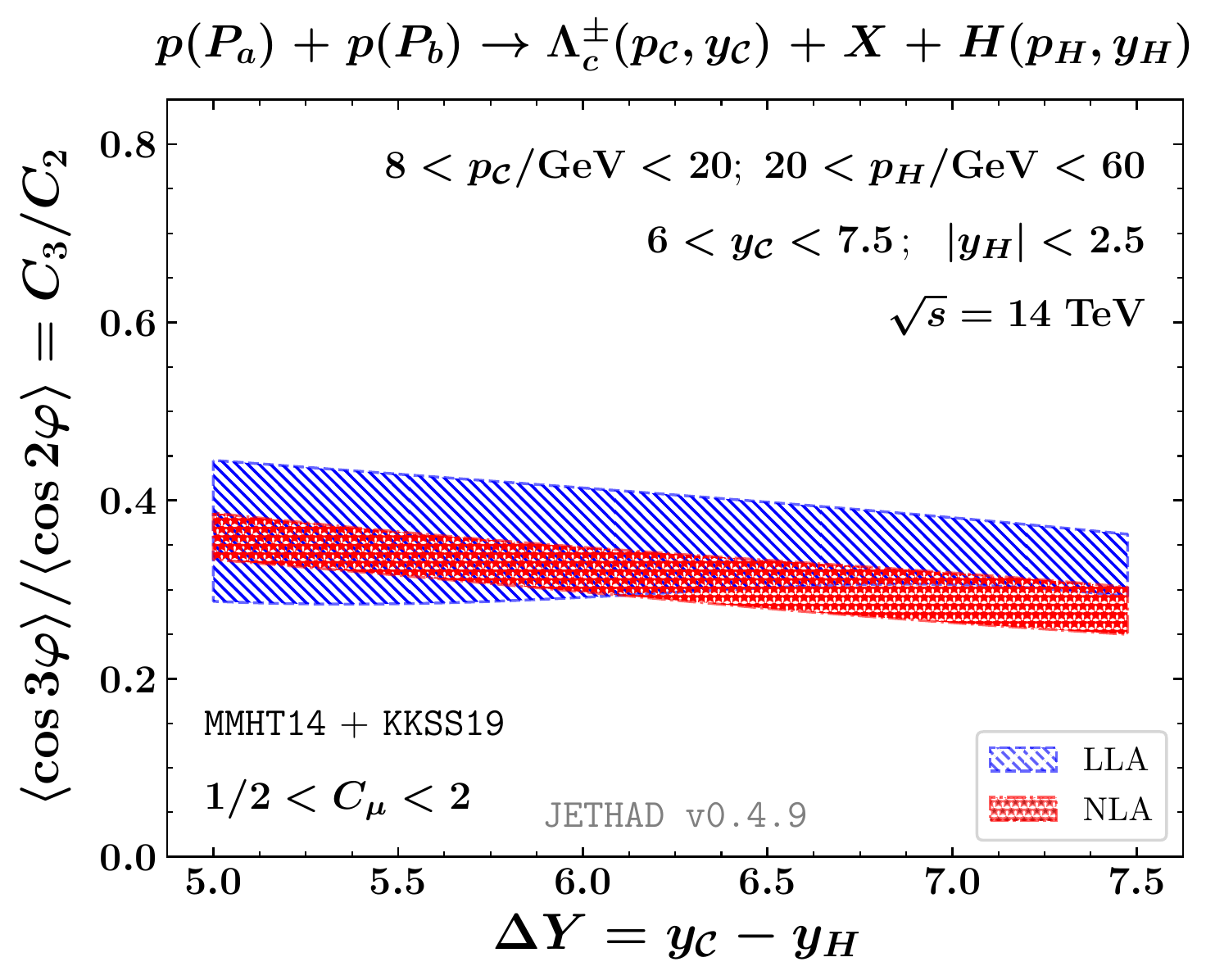}

\caption{$\DY$-pattern of azimuthal-correlation ratios, $R_{\kappa\lambda} \equiv C_{\kappa}/C_{\lambda}$, for the ultraforward inclusive $D^{*^\pm}$ plus Higgs channel (left panels) and the inclusive $\Lambda_c^\pm$ plus Higgs channel (right panels) at $\sqrt{s} = 14$ TeV. Text boxes inside panels refer to final-state kinematic cuts. Uncertainty bands embody the combined effect of scale variation and phase-space multi-dimensional integration.}
\label{fig:Rnm}
\end{figure*}

The azimuthal moments $R_{\kappa0}$ are shown in Fig.~\ref{fig:Rn0}, whereas their ratios $R_{\kappa\lambda}$ are presented in Fig.~\ref{fig:Rnm}. All these correlations fall off when $\DY$ increases, thus validating the assumption that the weight of the inclusive system of gluons emitted in the final state, labeled as $X$ in Eq.\eref{process} and in titles of our plots, grows with $\DY$. This leads to the observed decorrelation pattern, which is evidently stronger in the LLA case with respect to the NLA one. As it is well known from phenomenological studies of other semi-hard reactions, azimuthal-correlation moments are among the most sensitive observables to high-energy dynamics.
This brings to a missing or incomplete overlap between leading and next-to-leading predictions. It happens in particular for the $R_{10}$ and $R_{20}$ ratios, where LLA and NLA bands are disjoint, thus indicating that NLA corrections still have a sizable impact.

Nonetheless, at variance with Mueller--Navelet, light-hadron~$+$~jet and double light-hadron correlations, studies for charmed-hadron~$+$~Higgs around natural values of renormalization and factorization scales are possible. This reinforces our statement that final states proposed in the present work are particularly stable. On one hand, the ($\kappa = 0$) component for the all eigenvalues associated to the NLO BFKL kernel is more sensitive to contaminations coming from the collinear region. Thus, it is not surprising that ratios of correlations, \emph{i.e.} $R_{\kappa\lambda}$ moments with $\kappa$ and $\lambda$ larger than zero exhibit a milder sensitivity on higher-order corrections. On the other hand, however, it is well known that those higher components are less sensitive to BFKL. In other words, they carry less information about high-energy dynamics.

\begin{figure*}[!t]
\centering
\includegraphics[scale=0.53,clip]{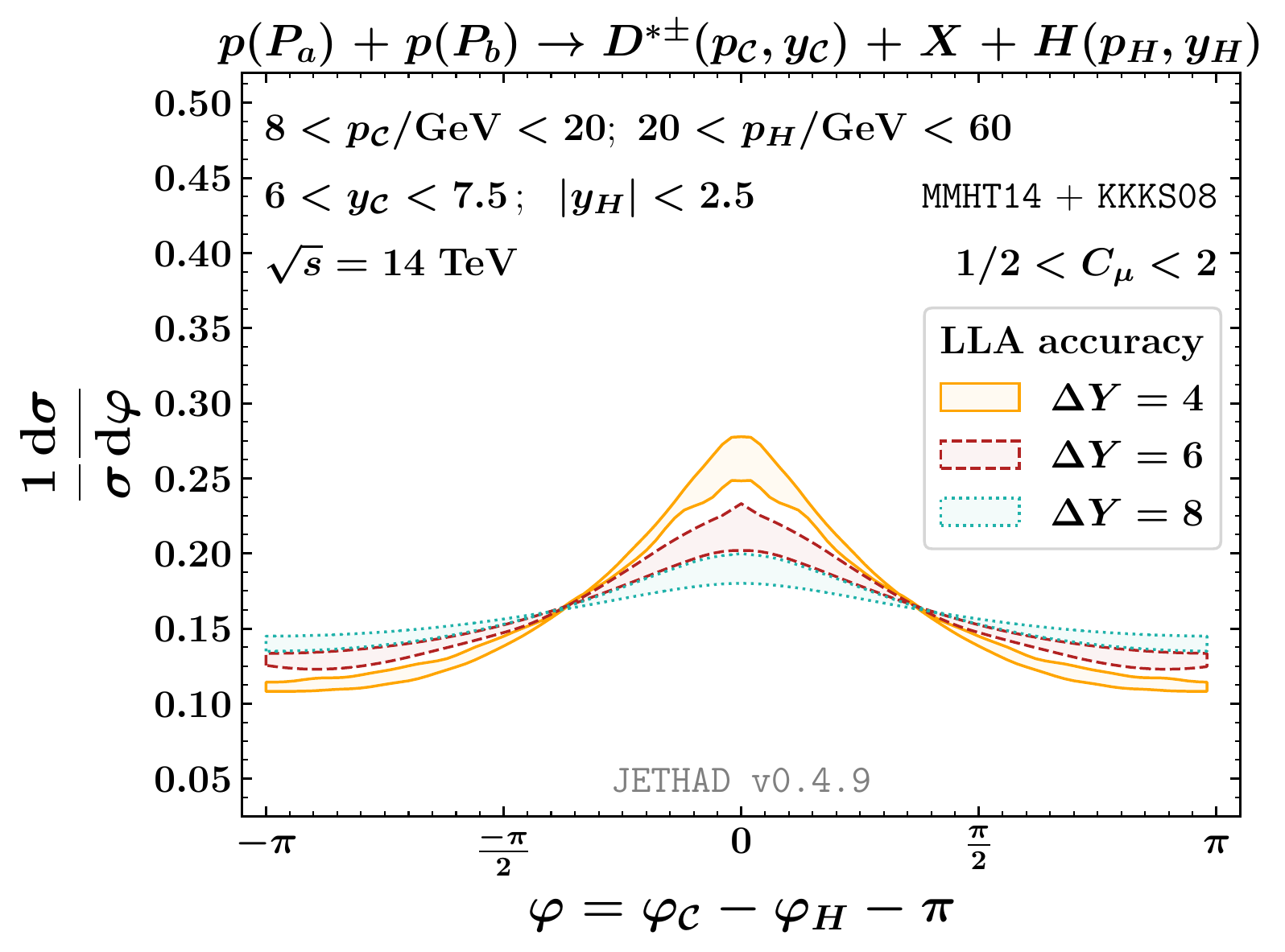}
\includegraphics[scale=0.53,clip]{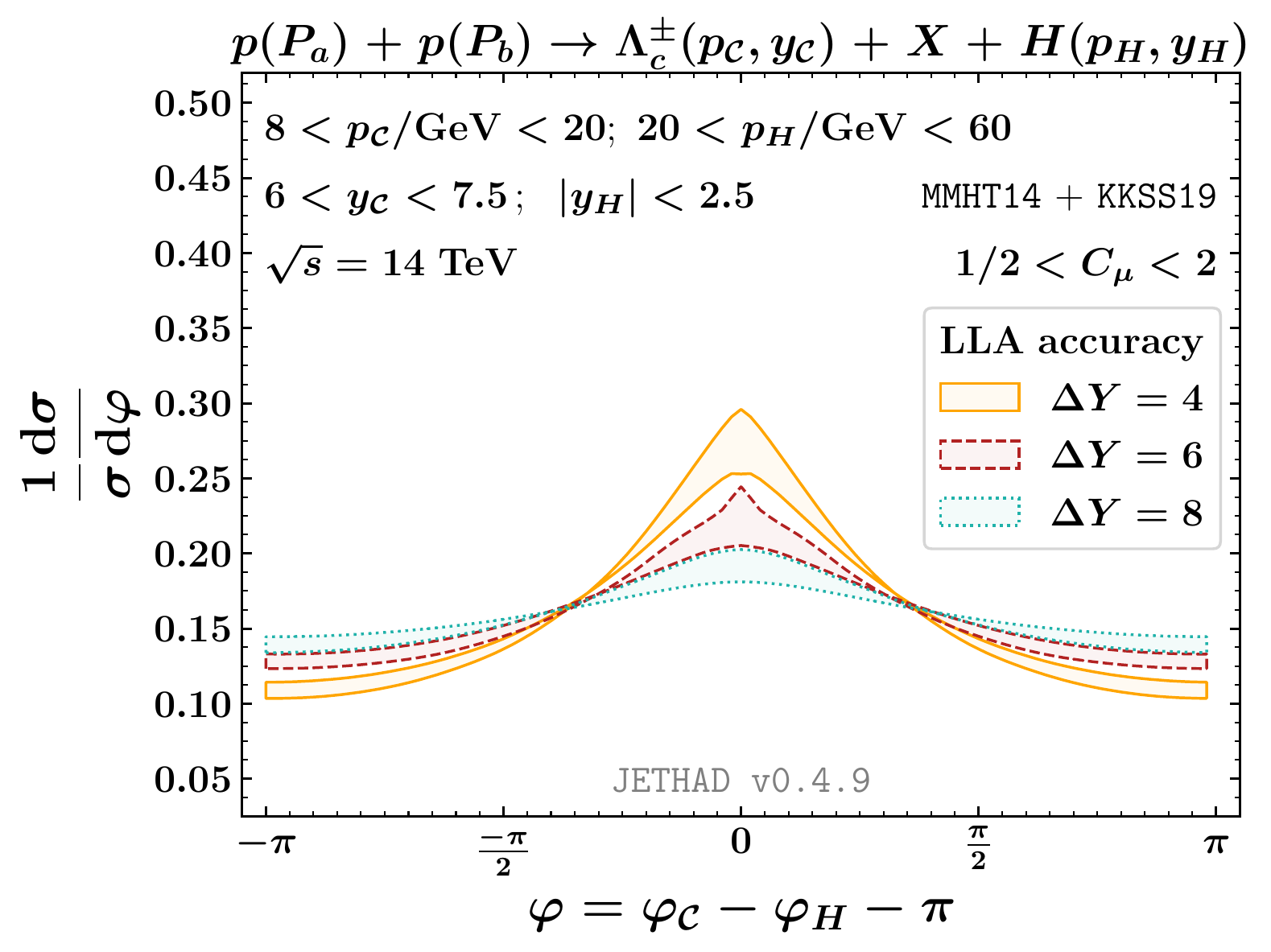}

\includegraphics[scale=0.53,clip]{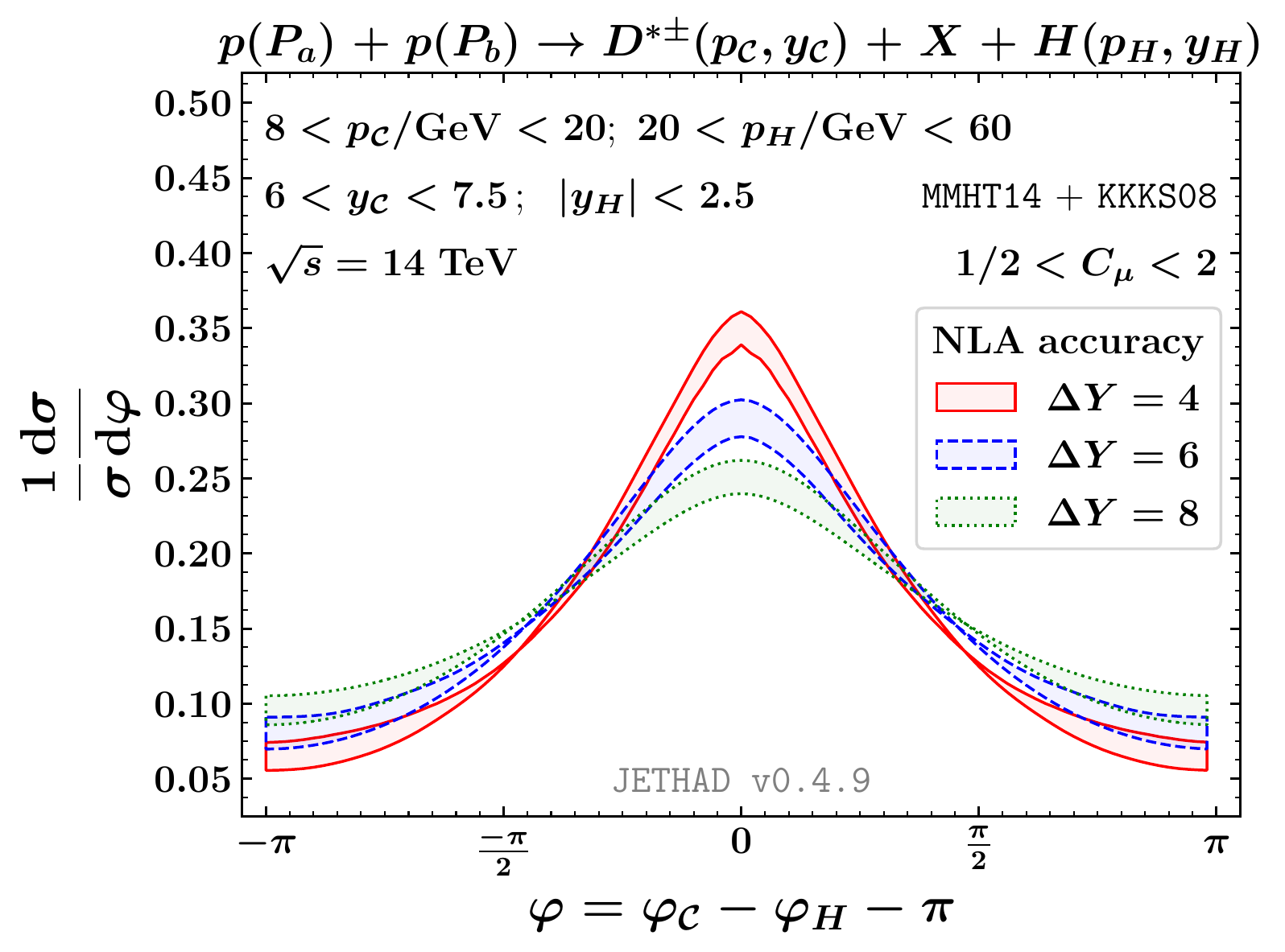}
\includegraphics[scale=0.53,clip]{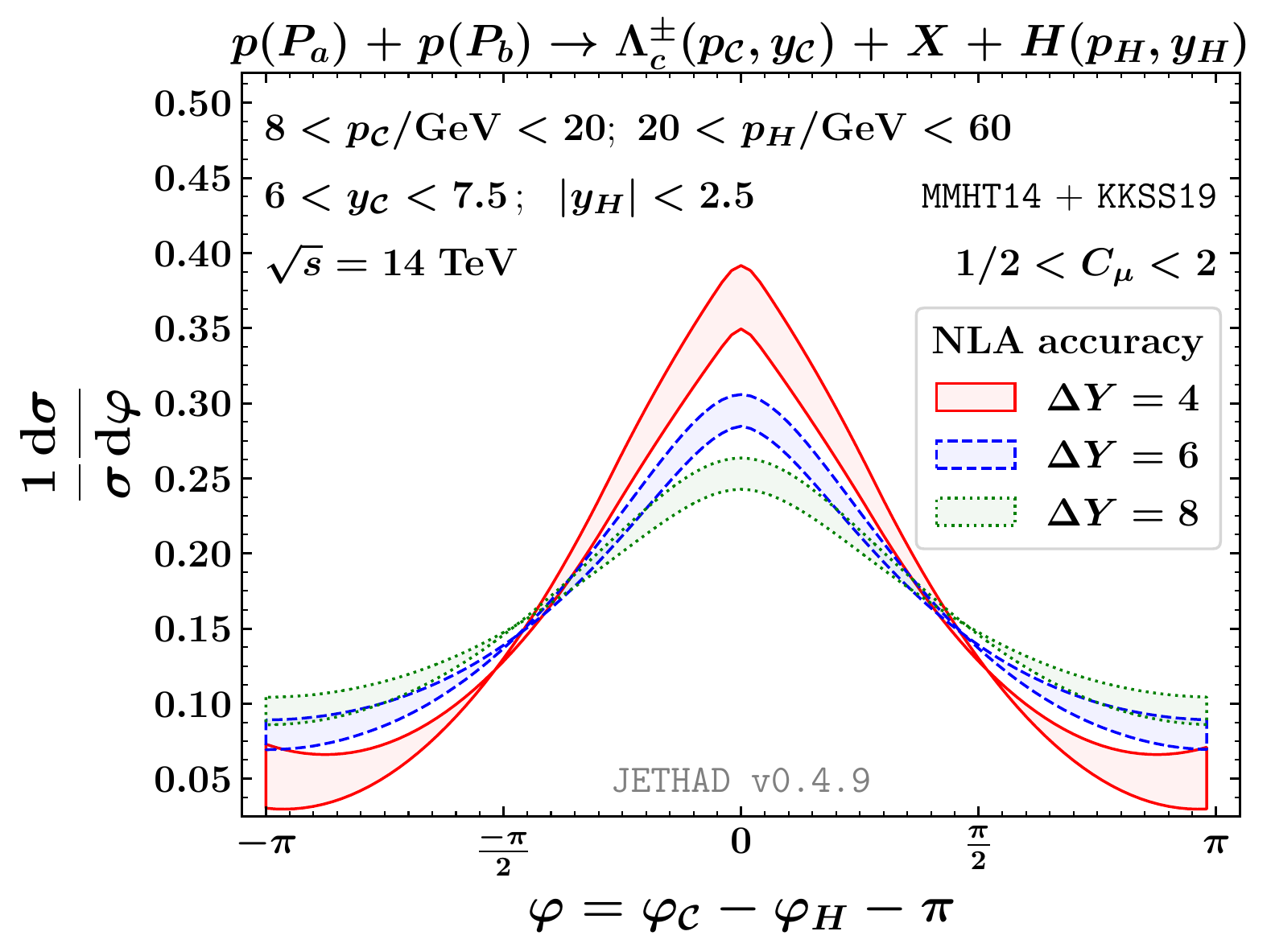}

\caption{LLA (upper panels) and NLA (lower panels) azimuthal distribution for the ultraforward inclusive $D^{*^\pm}$ plus Higgs channel (left panels) and the inclusive $\Lambda_c^\pm$ plus Higgs channel (right panels) at $\sqrt{s} = 14$ TeV, and for three distinct values of $\DY$. Text boxes inside panels refer to final-state kinematic cuts. Uncertainty bands embody the combined effect of scale variation and phase-space multi-dimensional integration.}
\label{fig:dsigma_dphi}
\end{figure*}

What has been observed for the azimuthal correlations is confirmed by the inspection of our results for the $\varphi$-distribution in plots of Fig.\tref{fig:dsigma_dphi}. Here, predictions for different values of $\DY$ are shown in the LLA (upper panels) as well as in the NLA (lower panels) accuracy. The general trend of spectra for final states featuring $D^{* \pm}$ (left panels) and $\Lambda_c^\pm$ (right panels) emissions is the presence of a clear peak around $\varphi=0$, namely when the ${\cal C}$-hadron and the Higgs boson are close to be emitted back-to-back. With the increase of $\DY$, the peak height decreases and the distribution width broadens. This reflects the fact that larger rapidity intervals lead to a more significant decorrelation of the hadron~$+$~Higgs system. Therefore the number of back-to-back events diminishes, while the rate of uncorrelated events grows.

The stronger decorrelation of LLA series observed in the $R_{\kappa\lambda}$ patterns (Figs.\tref{fig:Rn0} and\tref{fig:Rnm}) consistently translates into smaller peaks of the corresponding LLA azimuthal distributions with respect to NLA ones. We note that, both at LLA and NLA, peaks for $D^{* \pm}$ emissions are lower than $\Lambda_c^{\pm}$ ones. This could descend from to the different dependence on $\mu_F$ of the gluon FFs of the two heavy-flavor species. Preliminary tests have confirmed that the {\tt KKKS08} function depicting the gluon fragmentation to $D^{* \pm}$ smoothly increases with $\mu_F$, whereas the {\tt KKSS19} one for the gluon to $\Lambda_c^{\pm}$ production exhibits a non-decreasing \emph{plateau} (see upper left panel in Fig.~8 of Ref.\tcite{Celiberto:2021dzy}). Further studies are needed to unveil a possible connection between these patterns and the observed differences in the peaks of corresponding plots in Fig.\tref{fig:dsigma_dphi}.

Azimuthal distributions are very promising observables where to hunt for clear manifestations of the emergence of high-energy dynamics, and to discriminate between BFKL and other resummation approaches.
We notice the presence of intersection points in plots of Fig.~\ref{fig:dsigma_dphi} at which predictions for the $\varphi$-distribution overlap, regardless of $\DY$. These intersection points could come out as a consequence of the characteristic behavior the BFKL exponentiated kernel, and hence no such behavior can be present within the DGLAP resummation. More in particular, they could be sentinels of the existence of particular kinematic configurations at which the amount of the undetected-gluon radiation in central-rapidity regions accounted for by BFKL Green's function is stable under variations of the rapidity interval. Future studies are needed to reveal the connection between these intersection points and BFKL Green's function.
In this regards, a prospective NLA BFKL versus NLO fixed-order analysis, doable after the numerical implementation of the forward Higgs impact factor, would clarify if the aforementioned pattern genuinely descends from the high-energy resummation or not.

\section{Conclusions and outlook}
\label{sec:conclusions}

We have proposed the simultaneous inclusive emission of a single-charmed hadron and of a Higgs boson as a novel channel to access the high-energy spectrum of QCD.

By making predictions for rapidity and azimuthal-angle distributions at the hands of the hybrid high-energy and collinear factorization, we have provided evidence that distinctive signals of a \emph{natural stabilization} under higher-order corrections fairly emerge and allow for a consistent description of the con\-si\-de\-red observables at the energy scales suggested by process kinematics.
This motivates our interest in making use of the hybrid factorization as a useful tool to assess the feasibility of precision studies of high-energy QCD dynamics.

The predicted statistics for cross sections in a FPF~$+$~ATLAS coincidence setup is encouraging. 
We believe that the study of our heavy-flavor~$+$~Higgs-boson final states could be included in future experimental analyses jointly done \emph{via} FPF ultraforward detectors and the ATLAS barrel.

Two future developments will extend this work.
On one hand, we plan to enhance the accuracy of our analysis by accounting for the full NLO expression of the impact factors, when the numerical implementation of the forward Higgs one\tcite{Celiberto:2022fgx} will be available.
On the other hand, we propose to consider single inclusive emissions of ${\cal C}$-hadrons tagged at FPF rapidity ranges.
This will open a window opportunities to ($i$) explore the proton structure in the ultraforward regime and ($ii$) offer a theoretical common basis for studies on charm production and decays at the FPF, where the fair stability of our high-energy resummed predictions can represent a key ingredient to improve the standard fixed-order description for observables of interest.

\section*{Acknowledgements}

We thank members of the \textbf{FPF Collaboration} for inspiring discussions and for the warm atmosphere of the related meetings. We would like to express our gratitude to Dmitry Yu. Ivanov for a critical reading of the manuscript, for useful suggestions and for encouragement.

F.G.C. acknowledges support from the INFN/NINPHA project and thanks the Universit\`a degli Studi di Pavia for the warm hospitality.
M.F., M.M.A.M. and A.P. acknowledge support from the INFN/QFT@COL\-LI\-DERS project.

\bibliography{references}

\begin{thebibliography}{195}
\expandafter\ifx\csname natexlab\endcsname\relax\def\natexlab#1{#1}\fi
\expandafter\ifx\csname bibnamefont\endcsname\relax
  \def\bibnamefont#1{#1}\fi
\expandafter\ifx\csname bibfnamefont\endcsname\relax
  \def\bibfnamefont#1{#1}\fi
\expandafter\ifx\csname citenamefont\endcsname\relax
  \def\citenamefont#1{#1}\fi
\expandafter\ifx\csname url\endcsname\relax
  \def\url#1{\texttt{#1}}\fi
\expandafter\ifx\csname urlprefix\endcsname\relax\def\urlprefix{URL }\fi
\providecommand{\bibinfo}[2]{#2}
\providecommand{\eprint}[2][]{\url{#2}}

\bibitem[{\citenamefont{Gribov et~al.}(1983)\citenamefont{Gribov, Levin, and
  Ryskin}}]{Gribov:1983ivg}
\bibinfo{author}{\bibfnamefont{L.~V.} \bibnamefont{Gribov}},
  \bibinfo{author}{\bibfnamefont{E.~M.} \bibnamefont{Levin}}, \bibnamefont{and}
  \bibinfo{author}{\bibfnamefont{M.~G.} \bibnamefont{Ryskin}},
  \bibinfo{journal}{Phys. Rept.} \textbf{\bibinfo{volume}{100}},
  \bibinfo{pages}{1} (\bibinfo{year}{1983}).

\bibitem[{\citenamefont{Fadin et~al.}(1975)\citenamefont{Fadin, Kuraev, and
  Lipatov}}]{Fadin:1975cb}
\bibinfo{author}{\bibfnamefont{V.~S.} \bibnamefont{Fadin}},
  \bibinfo{author}{\bibfnamefont{E.}~\bibnamefont{Kuraev}}, \bibnamefont{and}
  \bibinfo{author}{\bibfnamefont{L.}~\bibnamefont{Lipatov}},
  \bibinfo{journal}{Phys. Lett. B} \textbf{\bibinfo{volume}{60}},
  \bibinfo{pages}{50} (\bibinfo{year}{1975}).

\bibitem[{\citenamefont{Kuraev et~al.}(1976)\citenamefont{Kuraev, Lipatov, and
  Fadin}}]{Kuraev:1976ge}
\bibinfo{author}{\bibfnamefont{E.~A.} \bibnamefont{Kuraev}},
  \bibinfo{author}{\bibfnamefont{L.~N.} \bibnamefont{Lipatov}},
  \bibnamefont{and} \bibinfo{author}{\bibfnamefont{V.~S.} \bibnamefont{Fadin}},
  \bibinfo{journal}{Sov. Phys. JETP} \textbf{\bibinfo{volume}{44}},
  \bibinfo{pages}{443} (\bibinfo{year}{1976}).

\bibitem[{\citenamefont{Kuraev et~al.}(1977)\citenamefont{Kuraev, Lipatov, and
  Fadin}}]{Kuraev:1977fs}
\bibinfo{author}{\bibfnamefont{E.}~\bibnamefont{Kuraev}},
  \bibinfo{author}{\bibfnamefont{L.}~\bibnamefont{Lipatov}}, \bibnamefont{and}
  \bibinfo{author}{\bibfnamefont{V.~S.} \bibnamefont{Fadin}},
  \bibinfo{journal}{Sov.\ Phys.\ JETP} \textbf{\bibinfo{volume}{45}},
  \bibinfo{pages}{199} (\bibinfo{year}{1977}).

\bibitem[{\citenamefont{Balitsky and Lipatov}(1978)}]{Balitsky:1978ic}
\bibinfo{author}{\bibfnamefont{I.}~\bibnamefont{Balitsky}} \bibnamefont{and}
  \bibinfo{author}{\bibfnamefont{L.}~\bibnamefont{Lipatov}},
  \bibinfo{journal}{Sov.\ J.\ Nucl.\ Phys.} \textbf{\bibinfo{volume}{28}},
  \bibinfo{pages}{822} (\bibinfo{year}{1978}).

\bibitem[{\citenamefont{Fadin and Lipatov}(1998)}]{Fadin:1998py}
\bibinfo{author}{\bibfnamefont{V.~S.} \bibnamefont{Fadin}} \bibnamefont{and}
  \bibinfo{author}{\bibfnamefont{L.~N.} \bibnamefont{Lipatov}},
  \bibinfo{journal}{Phys. Lett. B} \textbf{\bibinfo{volume}{429}},
  \bibinfo{pages}{127} (\bibinfo{year}{1998}), \eprint{hep-ph/9802290}.

\bibitem[{\citenamefont{Ciafaloni and Camici}(1998)}]{Ciafaloni:1998gs}
\bibinfo{author}{\bibfnamefont{M.}~\bibnamefont{Ciafaloni}} \bibnamefont{and}
  \bibinfo{author}{\bibfnamefont{G.}~\bibnamefont{Camici}},
  \bibinfo{journal}{Phys. Lett. B} \textbf{\bibinfo{volume}{430}},
  \bibinfo{pages}{349} (\bibinfo{year}{1998}), \eprint{hep-ph/9803389}.

\bibitem[{\citenamefont{Fadin et~al.}(1999)\citenamefont{Fadin, Fiore, and
  Papa}}]{Fadin:1998jv}
\bibinfo{author}{\bibfnamefont{V.~S.} \bibnamefont{Fadin}},
  \bibinfo{author}{\bibfnamefont{R.}~\bibnamefont{Fiore}}, \bibnamefont{and}
  \bibinfo{author}{\bibfnamefont{A.}~\bibnamefont{Papa}},
  \bibinfo{journal}{Phys. Rev. D} \textbf{\bibinfo{volume}{60}},
  \bibinfo{pages}{074025} (\bibinfo{year}{1999}), \eprint{hep-ph/9812456}.

\bibitem[{\citenamefont{Fadin and
  Gorbachev}(2000{\natexlab{a}})}]{Fadin:2000kx}
\bibinfo{author}{\bibfnamefont{V.~S.} \bibnamefont{Fadin}} \bibnamefont{and}
  \bibinfo{author}{\bibfnamefont{D.~A.} \bibnamefont{Gorbachev}},
  \bibinfo{journal}{JETP Lett.} \textbf{\bibinfo{volume}{71}},
  \bibinfo{pages}{222} (\bibinfo{year}{2000}{\natexlab{a}}).

\bibitem[{\citenamefont{Fadin and
  Gorbachev}(2000{\natexlab{b}})}]{Fadin:2000hu}
\bibinfo{author}{\bibfnamefont{V.~S.} \bibnamefont{Fadin}} \bibnamefont{and}
  \bibinfo{author}{\bibfnamefont{D.~A.} \bibnamefont{Gorbachev}},
  \bibinfo{journal}{Phys. Atom. Nucl.} \textbf{\bibinfo{volume}{63}},
  \bibinfo{pages}{2157} (\bibinfo{year}{2000}{\natexlab{b}}).

\bibitem[{\citenamefont{Fadin and Fiore}(2005{\natexlab{a}})}]{Fadin:2004zq}
\bibinfo{author}{\bibfnamefont{V.~S.} \bibnamefont{Fadin}} \bibnamefont{and}
  \bibinfo{author}{\bibfnamefont{R.}~\bibnamefont{Fiore}},
  \bibinfo{journal}{Phys. Lett. B} \textbf{\bibinfo{volume}{610}},
  \bibinfo{pages}{61} (\bibinfo{year}{2005}{\natexlab{a}}),
  \bibinfo{note}{[Erratum: Phys.Lett.B 621, 320 (2005)]},
  \eprint{hep-ph/0412386}.

\bibitem[{\citenamefont{Fadin and Fiore}(2005{\natexlab{b}})}]{Fadin:2005zj}
\bibinfo{author}{\bibfnamefont{V.~S.} \bibnamefont{Fadin}} \bibnamefont{and}
  \bibinfo{author}{\bibfnamefont{R.}~\bibnamefont{Fiore}},
  \bibinfo{journal}{Phys. Rev. D} \textbf{\bibinfo{volume}{72}},
  \bibinfo{pages}{014018} (\bibinfo{year}{2005}{\natexlab{b}}),
  \eprint{hep-ph/0502045}.

\bibitem[{\citenamefont{Amoroso et~al.}(2022)}]{Amoroso:2022eow}
\bibinfo{author}{\bibfnamefont{S.}~\bibnamefont{Amoroso}} \bibnamefont{et~al.},
  in \emph{\bibinfo{booktitle}{{2022 Snowmass Summer Study}}}
  (\bibinfo{year}{2022}), \eprint{2203.13923}.

\bibitem[{\citenamefont{Colferai et~al.}(2010)\citenamefont{Colferai,
  Schwennsen, Szymanowski, and Wallon}}]{Colferai:2010wu}
\bibinfo{author}{\bibfnamefont{D.}~\bibnamefont{Colferai}},
  \bibinfo{author}{\bibfnamefont{F.}~\bibnamefont{Schwennsen}},
  \bibinfo{author}{\bibfnamefont{L.}~\bibnamefont{Szymanowski}},
  \bibnamefont{and} \bibinfo{author}{\bibfnamefont{S.}~\bibnamefont{Wallon}},
  \bibinfo{journal}{JHEP} \textbf{\bibinfo{volume}{12}}, \bibinfo{pages}{026}
  (\bibinfo{year}{2010}), \eprint{1002.1365}.

\bibitem[{\citenamefont{Deak et~al.}(2009)\citenamefont{Deak, Hautmann, Jung,
  and Kutak}}]{Deak:2009xt}
\bibinfo{author}{\bibfnamefont{M.}~\bibnamefont{Deak}},
  \bibinfo{author}{\bibfnamefont{F.}~\bibnamefont{Hautmann}},
  \bibinfo{author}{\bibfnamefont{H.}~\bibnamefont{Jung}}, \bibnamefont{and}
  \bibinfo{author}{\bibfnamefont{K.}~\bibnamefont{Kutak}},
  \bibinfo{journal}{JHEP} \textbf{\bibinfo{volume}{09}}, \bibinfo{pages}{121}
  (\bibinfo{year}{2009}), \eprint{0908.0538}.

\bibitem[{\citenamefont{van Hameren et~al.}(2015)\citenamefont{van Hameren,
  Kotko, and Kutak}}]{vanHameren:2015uia}
\bibinfo{author}{\bibfnamefont{A.}~\bibnamefont{van Hameren}},
  \bibinfo{author}{\bibfnamefont{P.}~\bibnamefont{Kotko}}, \bibnamefont{and}
  \bibinfo{author}{\bibfnamefont{K.}~\bibnamefont{Kutak}},
  \bibinfo{journal}{Phys. Rev. D} \textbf{\bibinfo{volume}{92}},
  \bibinfo{pages}{054007} (\bibinfo{year}{2015}), \eprint{1505.02763}.

\bibitem[{\citenamefont{Deak et~al.}(2019)\citenamefont{Deak, van Hameren,
  Jung, Kusina, Kutak, and Serino}}]{Deak:2018obv}
\bibinfo{author}{\bibfnamefont{M.}~\bibnamefont{Deak}},
  \bibinfo{author}{\bibfnamefont{A.}~\bibnamefont{van Hameren}},
  \bibinfo{author}{\bibfnamefont{H.}~\bibnamefont{Jung}},
  \bibinfo{author}{\bibfnamefont{A.}~\bibnamefont{Kusina}},
  \bibinfo{author}{\bibfnamefont{K.}~\bibnamefont{Kutak}}, \bibnamefont{and}
  \bibinfo{author}{\bibfnamefont{M.}~\bibnamefont{Serino}},
  \bibinfo{journal}{Phys. Rev. D} \textbf{\bibinfo{volume}{99}},
  \bibinfo{pages}{094011} (\bibinfo{year}{2019}), \eprint{1809.03854}.

\bibitem[{\citenamefont{Van~Haevermaet
  et~al.}(2020)\citenamefont{Van~Haevermaet, Van~Hameren, Kotko, Kutak, and
  Van~Mechelen}}]{VanHaevermaet:2020rro}
\bibinfo{author}{\bibfnamefont{H.}~\bibnamefont{Van~Haevermaet}},
  \bibinfo{author}{\bibfnamefont{A.}~\bibnamefont{Van~Hameren}},
  \bibinfo{author}{\bibfnamefont{P.}~\bibnamefont{Kotko}},
  \bibinfo{author}{\bibfnamefont{K.}~\bibnamefont{Kutak}}, \bibnamefont{and}
  \bibinfo{author}{\bibfnamefont{P.}~\bibnamefont{Van~Mechelen}},
  \bibinfo{journal}{Eur. Phys. J. C} \textbf{\bibinfo{volume}{80}},
  \bibinfo{pages}{610} (\bibinfo{year}{2020}), \eprint{2004.07551}.

\bibitem[{\citenamefont{Blanco et~al.}(2020)\citenamefont{Blanco, van Hameren,
  Kotko, and Kutak}}]{Blanco:2020akb}
\bibinfo{author}{\bibfnamefont{E.}~\bibnamefont{Blanco}},
  \bibinfo{author}{\bibfnamefont{A.}~\bibnamefont{van Hameren}},
  \bibinfo{author}{\bibfnamefont{P.}~\bibnamefont{Kotko}}, \bibnamefont{and}
  \bibinfo{author}{\bibfnamefont{K.}~\bibnamefont{Kutak}},
  \bibinfo{journal}{JHEP} \textbf{\bibinfo{volume}{12}}, \bibinfo{pages}{158}
  (\bibinfo{year}{2020}), \eprint{2008.07916}.

\bibitem[{\citenamefont{van Hameren et~al.}(2021)\citenamefont{van Hameren,
  Kotko, Kutak, and Sapeta}}]{vanHameren:2020rqt}
\bibinfo{author}{\bibfnamefont{A.}~\bibnamefont{van Hameren}},
  \bibinfo{author}{\bibfnamefont{P.}~\bibnamefont{Kotko}},
  \bibinfo{author}{\bibfnamefont{K.}~\bibnamefont{Kutak}}, \bibnamefont{and}
  \bibinfo{author}{\bibfnamefont{S.}~\bibnamefont{Sapeta}},
  \bibinfo{journal}{Phys. Lett. B} \textbf{\bibinfo{volume}{814}},
  \bibinfo{pages}{136078} (\bibinfo{year}{2021}), \eprint{2010.13066}.

\bibitem[{\citenamefont{Guiot and van Hameren}(2021)}]{Guiot:2021vnp}
\bibinfo{author}{\bibfnamefont{B.}~\bibnamefont{Guiot}} \bibnamefont{and}
  \bibinfo{author}{\bibfnamefont{A.}~\bibnamefont{van Hameren}},
  \bibinfo{journal}{Phys. Rev. D} \textbf{\bibinfo{volume}{104}},
  \bibinfo{pages}{094038} (\bibinfo{year}{2021}), \eprint{2108.06419}.

\bibitem[{\citenamefont{van Hameren et~al.}(2022)\citenamefont{van Hameren,
  Motyka, and Ziarko}}]{vanHameren:2022mtk}
\bibinfo{author}{\bibfnamefont{A.}~\bibnamefont{van Hameren}},
  \bibinfo{author}{\bibfnamefont{L.}~\bibnamefont{Motyka}}, \bibnamefont{and}
  \bibinfo{author}{\bibfnamefont{G.}~\bibnamefont{Ziarko}}
  (\bibinfo{year}{2022}), \eprint{2205.09585}.

\bibitem[{\citenamefont{Celiberto}(2017)}]{Celiberto:2017ius}
\bibinfo{author}{\bibfnamefont{F.~G.} \bibnamefont{Celiberto}}, Ph.D. thesis,
  \bibinfo{school}{Universit\`a della Calabria and INFN-Cosenza}
  (\bibinfo{year}{2017}), \eprint{1707.04315}.

\bibitem[{\citenamefont{Hentschinski et~al.}(2022)}]{Hentschinski:2022xnd}
\bibinfo{author}{\bibfnamefont{M.}~\bibnamefont{Hentschinski}}
  \bibnamefont{et~al.}, in \emph{\bibinfo{booktitle}{{2022 Snowmass Summer
  Study}}} (\bibinfo{year}{2022}), \eprint{2203.08129}.

\bibitem[{\citenamefont{Adachi et~al.}(2022)}]{AlexanderAryshev:2022pkx}
\bibinfo{author}{\bibfnamefont{I.}~\bibnamefont{Adachi}} \bibnamefont{et~al.}
  (\bibinfo{collaboration}{ILC International Development Team and ILC
  Community}) (\bibinfo{year}{2022}), \eprint{2203.07622}.

\bibitem[{\citenamefont{Mueller and Navelet}(1987)}]{Mueller:1986ey}
\bibinfo{author}{\bibfnamefont{A.~H.} \bibnamefont{Mueller}} \bibnamefont{and}
  \bibinfo{author}{\bibfnamefont{H.}~\bibnamefont{Navelet}},
  \bibinfo{journal}{Nucl. Phys. B} \textbf{\bibinfo{volume}{282}},
  \bibinfo{pages}{727} (\bibinfo{year}{1987}).

\bibitem[{\citenamefont{Caporale
  et~al.}(2013{\natexlab{a}})\citenamefont{Caporale, Ivanov, Murdaca, and
  Papa}}]{Caporale:2012ih}
\bibinfo{author}{\bibfnamefont{F.}~\bibnamefont{Caporale}},
  \bibinfo{author}{\bibfnamefont{D.}~\bibnamefont{Ivanov}},
  \bibinfo{author}{\bibfnamefont{B.}~\bibnamefont{Murdaca}}, \bibnamefont{and}
  \bibinfo{author}{\bibfnamefont{A.}~\bibnamefont{Papa}},
  \bibinfo{journal}{Nucl. Phys. B} \textbf{\bibinfo{volume}{877}},
  \bibinfo{pages}{73} (\bibinfo{year}{2013}{\natexlab{a}}), \eprint{1211.7225}.

\bibitem[{\citenamefont{Duclou\'e et~al.}(2013)\citenamefont{Duclou\'e,
  Szymanowski, and Wallon}}]{Ducloue:2013hia}
\bibinfo{author}{\bibfnamefont{B.}~\bibnamefont{Duclou\'e}},
  \bibinfo{author}{\bibfnamefont{L.}~\bibnamefont{Szymanowski}},
  \bibnamefont{and} \bibinfo{author}{\bibfnamefont{S.}~\bibnamefont{Wallon}},
  \bibinfo{journal}{JHEP} \textbf{\bibinfo{volume}{05}}, \bibinfo{pages}{096}
  (\bibinfo{year}{2013}), \eprint{1302.7012}.

\bibitem[{\citenamefont{Duclou\'e et~al.}(2014)\citenamefont{Duclou\'e,
  Szymanowski, and Wallon}}]{Ducloue:2013bva}
\bibinfo{author}{\bibfnamefont{B.}~\bibnamefont{Duclou\'e}},
  \bibinfo{author}{\bibfnamefont{L.}~\bibnamefont{Szymanowski}},
  \bibnamefont{and} \bibinfo{author}{\bibfnamefont{S.}~\bibnamefont{Wallon}},
  \bibinfo{journal}{Phys. Rev. Lett.} \textbf{\bibinfo{volume}{112}},
  \bibinfo{pages}{082003} (\bibinfo{year}{2014}), \eprint{1309.3229}.

\bibitem[{\citenamefont{Caporale
  et~al.}(2013{\natexlab{b}})\citenamefont{Caporale, Murdaca, Sabio~Vera, and
  Salas}}]{Caporale:2013uva}
\bibinfo{author}{\bibfnamefont{F.}~\bibnamefont{Caporale}},
  \bibinfo{author}{\bibfnamefont{B.}~\bibnamefont{Murdaca}},
  \bibinfo{author}{\bibfnamefont{A.}~\bibnamefont{Sabio~Vera}},
  \bibnamefont{and} \bibinfo{author}{\bibfnamefont{C.}~\bibnamefont{Salas}},
  \bibinfo{journal}{Nucl. Phys. B} \textbf{\bibinfo{volume}{875}},
  \bibinfo{pages}{134} (\bibinfo{year}{2013}{\natexlab{b}}),
  \eprint{1305.4620}.

\bibitem[{\citenamefont{Caporale et~al.}(2014)\citenamefont{Caporale, Ivanov,
  Murdaca, and Papa}}]{Caporale:2014gpa}
\bibinfo{author}{\bibfnamefont{F.}~\bibnamefont{Caporale}},
  \bibinfo{author}{\bibfnamefont{D.~{\relax Yu}.} \bibnamefont{Ivanov}},
  \bibinfo{author}{\bibfnamefont{B.}~\bibnamefont{Murdaca}}, \bibnamefont{and}
  \bibinfo{author}{\bibfnamefont{A.}~\bibnamefont{Papa}},
  \bibinfo{journal}{Eur. Phys. J. C} \textbf{\bibinfo{volume}{74}},
  \bibinfo{pages}{3084} (\bibinfo{year}{2014}), \bibinfo{note}{[Erratum:
  Eur.Phys.J.C 75, 535 (2015)]}, \eprint{1407.8431}.

\bibitem[{\citenamefont{Colferai and Niccoli}(2015)}]{Colferai:2015zfa}
\bibinfo{author}{\bibfnamefont{D.}~\bibnamefont{Colferai}} \bibnamefont{and}
  \bibinfo{author}{\bibfnamefont{A.}~\bibnamefont{Niccoli}},
  \bibinfo{journal}{JHEP} \textbf{\bibinfo{volume}{04}}, \bibinfo{pages}{071}
  (\bibinfo{year}{2015}), \eprint{1501.07442}.

\bibitem[{\citenamefont{Caporale et~al.}(2015)\citenamefont{Caporale, Ivanov,
  Murdaca, and Papa}}]{Caporale:2015uva}
\bibinfo{author}{\bibfnamefont{F.}~\bibnamefont{Caporale}},
  \bibinfo{author}{\bibfnamefont{D.~{\relax Yu}.} \bibnamefont{Ivanov}},
  \bibinfo{author}{\bibfnamefont{B.}~\bibnamefont{Murdaca}}, \bibnamefont{and}
  \bibinfo{author}{\bibfnamefont{A.}~\bibnamefont{Papa}},
  \bibinfo{journal}{Phys. Rev. D} \textbf{\bibinfo{volume}{91}},
  \bibinfo{pages}{114009} (\bibinfo{year}{2015}), \eprint{1504.06471}.

\bibitem[{\citenamefont{Duclou\'e et~al.}(2015)\citenamefont{Duclou\'e,
  Szymanowski, and Wallon}}]{Ducloue:2015jba}
\bibinfo{author}{\bibfnamefont{B.}~\bibnamefont{Duclou\'e}},
  \bibinfo{author}{\bibfnamefont{L.}~\bibnamefont{Szymanowski}},
  \bibnamefont{and} \bibinfo{author}{\bibfnamefont{S.}~\bibnamefont{Wallon}},
  \bibinfo{journal}{Phys. Rev. D} \textbf{\bibinfo{volume}{92}},
  \bibinfo{pages}{076002} (\bibinfo{year}{2015}), \eprint{1507.04735}.

\bibitem[{\citenamefont{Celiberto
  et~al.}(2015{\natexlab{a}})\citenamefont{Celiberto, Ivanov, Murdaca, and
  Papa}}]{Celiberto:2015yba}
\bibinfo{author}{\bibfnamefont{F.~G.} \bibnamefont{Celiberto}},
  \bibinfo{author}{\bibfnamefont{D.~{\relax Yu}.} \bibnamefont{Ivanov}},
  \bibinfo{author}{\bibfnamefont{B.}~\bibnamefont{Murdaca}}, \bibnamefont{and}
  \bibinfo{author}{\bibfnamefont{A.}~\bibnamefont{Papa}},
  \bibinfo{journal}{Eur. Phys. J. C} \textbf{\bibinfo{volume}{75}},
  \bibinfo{pages}{292} (\bibinfo{year}{2015}{\natexlab{a}}),
  \eprint{1504.08233}.

\bibitem[{\citenamefont{Celiberto
  et~al.}(2015{\natexlab{b}})\citenamefont{Celiberto, Ivanov, Murdaca, and
  Papa}}]{Celiberto:2015mpa}
\bibinfo{author}{\bibfnamefont{F.~G.} \bibnamefont{Celiberto}},
  \bibinfo{author}{\bibfnamefont{D.~{\relax Yu}.} \bibnamefont{Ivanov}},
  \bibinfo{author}{\bibfnamefont{B.}~\bibnamefont{Murdaca}}, \bibnamefont{and}
  \bibinfo{author}{\bibfnamefont{A.}~\bibnamefont{Papa}},
  \bibinfo{journal}{Acta Phys. Polon. Supp.} \textbf{\bibinfo{volume}{8}},
  \bibinfo{pages}{935} (\bibinfo{year}{2015}{\natexlab{b}}),
  \eprint{1510.01626}.

\bibitem[{\citenamefont{Celiberto
  et~al.}(2016{\natexlab{a}})\citenamefont{Celiberto, Ivanov, Murdaca, and
  Papa}}]{Celiberto:2016ygs}
\bibinfo{author}{\bibfnamefont{F.~G.} \bibnamefont{Celiberto}},
  \bibinfo{author}{\bibfnamefont{D.~{\relax Yu}.} \bibnamefont{Ivanov}},
  \bibinfo{author}{\bibfnamefont{B.}~\bibnamefont{Murdaca}}, \bibnamefont{and}
  \bibinfo{author}{\bibfnamefont{A.}~\bibnamefont{Papa}},
  \bibinfo{journal}{Eur. Phys. J. C} \textbf{\bibinfo{volume}{76}},
  \bibinfo{pages}{224} (\bibinfo{year}{2016}{\natexlab{a}}),
  \eprint{1601.07847}.

\bibitem[{\citenamefont{Celiberto
  et~al.}(2016{\natexlab{b}})\citenamefont{Celiberto, Ivanov, Murdaca, and
  Papa}}]{Celiberto:2016vva}
\bibinfo{author}{\bibfnamefont{F.~G.} \bibnamefont{Celiberto}},
  \bibinfo{author}{\bibfnamefont{D.~{\relax Yu}.} \bibnamefont{Ivanov}},
  \bibinfo{author}{\bibfnamefont{B.}~\bibnamefont{Murdaca}}, \bibnamefont{and}
  \bibinfo{author}{\bibfnamefont{A.}~\bibnamefont{Papa}},
  \bibinfo{journal}{PoS} \textbf{\bibinfo{volume}{DIS2016}},
  \bibinfo{pages}{176} (\bibinfo{year}{2016}{\natexlab{b}}),
  \eprint{1606.08892}.

\bibitem[{\citenamefont{Caporale et~al.}(2018)\citenamefont{Caporale,
  Celiberto, Chachamis, Gordo~G\'omez, and Sabio~Vera}}]{Caporale:2018qnm}
\bibinfo{author}{\bibfnamefont{F.}~\bibnamefont{Caporale}},
  \bibinfo{author}{\bibfnamefont{F.~G.} \bibnamefont{Celiberto}},
  \bibinfo{author}{\bibfnamefont{G.}~\bibnamefont{Chachamis}},
  \bibinfo{author}{\bibfnamefont{D.}~\bibnamefont{Gordo~G\'omez}},
  \bibnamefont{and}
  \bibinfo{author}{\bibfnamefont{A.}~\bibnamefont{Sabio~Vera}},
  \bibinfo{journal}{Nucl. Phys. B} \textbf{\bibinfo{volume}{935}},
  \bibinfo{pages}{412} (\bibinfo{year}{2018}), \eprint{1806.06309}.

\bibitem[{\citenamefont{Celiberto
  et~al.}(2016{\natexlab{c}})\citenamefont{Celiberto, Ivanov, Murdaca, and
  Papa}}]{Celiberto:2016hae}
\bibinfo{author}{\bibfnamefont{F.~G.} \bibnamefont{Celiberto}},
  \bibinfo{author}{\bibfnamefont{D.~{\relax Yu}.} \bibnamefont{Ivanov}},
  \bibinfo{author}{\bibfnamefont{B.}~\bibnamefont{Murdaca}}, \bibnamefont{and}
  \bibinfo{author}{\bibfnamefont{A.}~\bibnamefont{Papa}},
  \bibinfo{journal}{Phys. Rev. D} \textbf{\bibinfo{volume}{94}},
  \bibinfo{pages}{034013} (\bibinfo{year}{2016}{\natexlab{c}}),
  \eprint{1604.08013}.

\bibitem[{\citenamefont{Celiberto
  et~al.}(2017{\natexlab{a}})\citenamefont{Celiberto, Ivanov, Murdaca, and
  Papa}}]{Celiberto:2016zgb}
\bibinfo{author}{\bibfnamefont{F.~G.} \bibnamefont{Celiberto}},
  \bibinfo{author}{\bibfnamefont{D.~{\relax Yu}.} \bibnamefont{Ivanov}},
  \bibinfo{author}{\bibfnamefont{B.}~\bibnamefont{Murdaca}}, \bibnamefont{and}
  \bibinfo{author}{\bibfnamefont{A.}~\bibnamefont{Papa}}, \bibinfo{journal}{AIP
  Conf. Proc.} \textbf{\bibinfo{volume}{1819}}, \bibinfo{pages}{060005}
  (\bibinfo{year}{2017}{\natexlab{a}}), \eprint{1611.04811}.

\bibitem[{\citenamefont{Celiberto
  et~al.}(2017{\natexlab{b}})\citenamefont{Celiberto, Ivanov, Murdaca, and
  Papa}}]{Celiberto:2017ptm}
\bibinfo{author}{\bibfnamefont{F.~G.} \bibnamefont{Celiberto}},
  \bibinfo{author}{\bibfnamefont{D.~{\relax Yu}.} \bibnamefont{Ivanov}},
  \bibinfo{author}{\bibfnamefont{B.}~\bibnamefont{Murdaca}}, \bibnamefont{and}
  \bibinfo{author}{\bibfnamefont{A.}~\bibnamefont{Papa}},
  \bibinfo{journal}{Eur. Phys. J. C} \textbf{\bibinfo{volume}{77}},
  \bibinfo{pages}{382} (\bibinfo{year}{2017}{\natexlab{b}}),
  \eprint{1701.05077}.

\bibitem[{\citenamefont{Celiberto
  et~al.}(2017{\natexlab{c}})\citenamefont{Celiberto, Ivanov, Murdaca, and
  Papa}}]{Celiberto:2017uae}
\bibinfo{author}{\bibfnamefont{F.~G.} \bibnamefont{Celiberto}},
  \bibinfo{author}{\bibfnamefont{D.~{\relax Yu}.} \bibnamefont{Ivanov}},
  \bibinfo{author}{\bibfnamefont{B.}~\bibnamefont{Murdaca}}, \bibnamefont{and}
  \bibinfo{author}{\bibfnamefont{A.}~\bibnamefont{Papa}}, in
  \emph{\bibinfo{booktitle}{{25th Low-x Meeting}}}
  (\bibinfo{year}{2017}{\natexlab{c}}), \eprint{1709.01128}.

\bibitem[{\citenamefont{Celiberto
  et~al.}(2017{\natexlab{d}})\citenamefont{Celiberto, Ivanov, Murdaca, and
  Papa}}]{Celiberto:2017ydk}
\bibinfo{author}{\bibfnamefont{F.~G.} \bibnamefont{Celiberto}},
  \bibinfo{author}{\bibfnamefont{D.~{\relax Yu}.} \bibnamefont{Ivanov}},
  \bibinfo{author}{\bibfnamefont{B.}~\bibnamefont{Murdaca}}, \bibnamefont{and}
  \bibinfo{author}{\bibfnamefont{A.}~\bibnamefont{Papa}}, in
  \emph{\bibinfo{booktitle}{{17th conference on Elastic and Diffractive
  Scattering}}} (\bibinfo{year}{2017}{\natexlab{d}}), \eprint{1709.04758}.

\bibitem[{\citenamefont{Caporale
  et~al.}(2016{\natexlab{a}})\citenamefont{Caporale, Chachamis, Murdaca, and
  Sabio~Vera}}]{Caporale:2015vya}
\bibinfo{author}{\bibfnamefont{F.}~\bibnamefont{Caporale}},
  \bibinfo{author}{\bibfnamefont{G.}~\bibnamefont{Chachamis}},
  \bibinfo{author}{\bibfnamefont{B.}~\bibnamefont{Murdaca}}, \bibnamefont{and}
  \bibinfo{author}{\bibfnamefont{A.}~\bibnamefont{Sabio~Vera}},
  \bibinfo{journal}{Phys. Rev. Lett.} \textbf{\bibinfo{volume}{116}},
  \bibinfo{pages}{012001} (\bibinfo{year}{2016}{\natexlab{a}}),
  \eprint{1508.07711}.

\bibitem[{\citenamefont{Caporale
  et~al.}(2016{\natexlab{b}})\citenamefont{Caporale, Celiberto, Chachamis, and
  Sabio~Vera}}]{Caporale:2015int}
\bibinfo{author}{\bibfnamefont{F.}~\bibnamefont{Caporale}},
  \bibinfo{author}{\bibfnamefont{F.~G.} \bibnamefont{Celiberto}},
  \bibinfo{author}{\bibfnamefont{G.}~\bibnamefont{Chachamis}},
  \bibnamefont{and}
  \bibinfo{author}{\bibfnamefont{A.}~\bibnamefont{Sabio~Vera}},
  \bibinfo{journal}{Eur. Phys. J. C} \textbf{\bibinfo{volume}{76}},
  \bibinfo{pages}{165} (\bibinfo{year}{2016}{\natexlab{b}}),
  \eprint{1512.03364}.

\bibitem[{\citenamefont{Caporale
  et~al.}(2016{\natexlab{c}})\citenamefont{Caporale, Celiberto, Chachamis,
  Gordo~G\'omez, and Sabio~Vera}}]{Caporale:2016soq}
\bibinfo{author}{\bibfnamefont{F.}~\bibnamefont{Caporale}},
  \bibinfo{author}{\bibfnamefont{F.~G.} \bibnamefont{Celiberto}},
  \bibinfo{author}{\bibfnamefont{G.}~\bibnamefont{Chachamis}},
  \bibinfo{author}{\bibfnamefont{D.}~\bibnamefont{Gordo~G\'omez}},
  \bibnamefont{and}
  \bibinfo{author}{\bibfnamefont{A.}~\bibnamefont{Sabio~Vera}},
  \bibinfo{journal}{Nucl. Phys. B} \textbf{\bibinfo{volume}{910}},
  \bibinfo{pages}{374} (\bibinfo{year}{2016}{\natexlab{c}}),
  \eprint{1603.07785}.

\bibitem[{\citenamefont{Caporale
  et~al.}(2016{\natexlab{d}})\citenamefont{Caporale, Celiberto, Chachamis, and
  Sabio~Vera}}]{Caporale:2016vxt}
\bibinfo{author}{\bibfnamefont{F.}~\bibnamefont{Caporale}},
  \bibinfo{author}{\bibfnamefont{F.~G.} \bibnamefont{Celiberto}},
  \bibinfo{author}{\bibfnamefont{G.}~\bibnamefont{Chachamis}},
  \bibnamefont{and}
  \bibinfo{author}{\bibfnamefont{A.}~\bibnamefont{Sabio~Vera}},
  \bibinfo{journal}{PoS} \textbf{\bibinfo{volume}{DIS2016}},
  \bibinfo{pages}{177} (\bibinfo{year}{2016}{\natexlab{d}}),
  \eprint{1610.01880}.

\bibitem[{\citenamefont{Caporale
  et~al.}(2017{\natexlab{a}})\citenamefont{Caporale, Celiberto, Chachamis,
  Gordo~G\'omez, and Sabio~Vera}}]{Caporale:2016xku}
\bibinfo{author}{\bibfnamefont{F.}~\bibnamefont{Caporale}},
  \bibinfo{author}{\bibfnamefont{F.~G.} \bibnamefont{Celiberto}},
  \bibinfo{author}{\bibfnamefont{G.}~\bibnamefont{Chachamis}},
  \bibinfo{author}{\bibfnamefont{D.}~\bibnamefont{Gordo~G\'omez}},
  \bibnamefont{and}
  \bibinfo{author}{\bibfnamefont{A.}~\bibnamefont{Sabio~Vera}},
  \bibinfo{journal}{Eur. Phys. J. C} \textbf{\bibinfo{volume}{77}},
  \bibinfo{pages}{5} (\bibinfo{year}{2017}{\natexlab{a}}), \eprint{1606.00574}.

\bibitem[{\citenamefont{Celiberto}(2016)}]{Celiberto:2016vhn}
\bibinfo{author}{\bibfnamefont{F.~G.} \bibnamefont{Celiberto}},
  \bibinfo{journal}{Frascati Phys. Ser.} \textbf{\bibinfo{volume}{63}},
  \bibinfo{pages}{43} (\bibinfo{year}{2016}), \eprint{1606.07327}.

\bibitem[{\citenamefont{Caporale
  et~al.}(2017{\natexlab{b}})\citenamefont{Caporale, Celiberto, Chachamis,
  Gordo~G\'omez, and Sabio~Vera}}]{Caporale:2016djm}
\bibinfo{author}{\bibfnamefont{F.}~\bibnamefont{Caporale}},
  \bibinfo{author}{\bibfnamefont{F.~G.} \bibnamefont{Celiberto}},
  \bibinfo{author}{\bibfnamefont{G.}~\bibnamefont{Chachamis}},
  \bibinfo{author}{\bibfnamefont{D.}~\bibnamefont{Gordo~G\'omez}},
  \bibnamefont{and}
  \bibinfo{author}{\bibfnamefont{A.}~\bibnamefont{Sabio~Vera}},
  \bibinfo{journal}{AIP Conf. Proc.} \textbf{\bibinfo{volume}{1819}},
  \bibinfo{pages}{060009} (\bibinfo{year}{2017}{\natexlab{b}}),
  \eprint{1611.04813}.

\bibitem[{\citenamefont{Chachamis
  et~al.}(2016{\natexlab{a}})\citenamefont{Chachamis, Caporale, Celiberto,
  Gomez, and Sabio~Vera}}]{Chachamis:2016lyi}
\bibinfo{author}{\bibfnamefont{G.}~\bibnamefont{Chachamis}},
  \bibinfo{author}{\bibfnamefont{F.}~\bibnamefont{Caporale}},
  \bibinfo{author}{\bibfnamefont{F.~G.} \bibnamefont{Celiberto}},
  \bibinfo{author}{\bibfnamefont{D.~G.} \bibnamefont{Gomez}}, \bibnamefont{and}
  \bibinfo{author}{\bibfnamefont{A.}~\bibnamefont{Sabio~Vera}}
  (\bibinfo{year}{2016}{\natexlab{a}}), \eprint{1610.01342}.

\bibitem[{\citenamefont{Chachamis
  et~al.}(2016{\natexlab{b}})\citenamefont{Chachamis, Caporale, Celiberto,
  Gomez~Gordo, and Sabio~Vera}}]{Chachamis:2016qct}
\bibinfo{author}{\bibfnamefont{G.}~\bibnamefont{Chachamis}},
  \bibinfo{author}{\bibfnamefont{F.}~\bibnamefont{Caporale}},
  \bibinfo{author}{\bibfnamefont{F.}~\bibnamefont{Celiberto}},
  \bibinfo{author}{\bibfnamefont{D.}~\bibnamefont{Gomez~Gordo}},
  \bibnamefont{and}
  \bibinfo{author}{\bibfnamefont{A.}~\bibnamefont{Sabio~Vera}},
  \bibinfo{journal}{PoS} \textbf{\bibinfo{volume}{DIS2016}},
  \bibinfo{pages}{178} (\bibinfo{year}{2016}{\natexlab{b}}).

\bibitem[{\citenamefont{Caporale
  et~al.}(2017{\natexlab{c}})\citenamefont{Caporale, Celiberto, Chachamis,
  Gordo~Gomez, Murdaca, and Sabio~Vera}}]{Caporale:2016pqe}
\bibinfo{author}{\bibfnamefont{F.}~\bibnamefont{Caporale}},
  \bibinfo{author}{\bibfnamefont{F.~G.} \bibnamefont{Celiberto}},
  \bibinfo{author}{\bibfnamefont{G.}~\bibnamefont{Chachamis}},
  \bibinfo{author}{\bibfnamefont{D.}~\bibnamefont{Gordo~Gomez}},
  \bibinfo{author}{\bibfnamefont{B.}~\bibnamefont{Murdaca}}, \bibnamefont{and}
  \bibinfo{author}{\bibfnamefont{A.}~\bibnamefont{Sabio~Vera}}, in
  \emph{\bibinfo{booktitle}{{24th Low-x Meeting}}}
  (\bibinfo{year}{2017}{\natexlab{c}}), vol.~\bibinfo{volume}{5},
  p.~\bibinfo{pages}{47}, \eprint{1610.04765}.

\bibitem[{\citenamefont{Caporale
  et~al.}(2017{\natexlab{d}})\citenamefont{Caporale, Celiberto, Chachamis,
  Gordo~G\'omez, and Sabio~Vera}}]{Caporale:2016lnh}
\bibinfo{author}{\bibfnamefont{F.}~\bibnamefont{Caporale}},
  \bibinfo{author}{\bibfnamefont{F.~G.} \bibnamefont{Celiberto}},
  \bibinfo{author}{\bibfnamefont{G.}~\bibnamefont{Chachamis}},
  \bibinfo{author}{\bibfnamefont{D.}~\bibnamefont{Gordo~G\'omez}},
  \bibnamefont{and}
  \bibinfo{author}{\bibfnamefont{A.}~\bibnamefont{Sabio~Vera}},
  \bibinfo{journal}{EPJ Web Conf.} \textbf{\bibinfo{volume}{164}},
  \bibinfo{pages}{07027} (\bibinfo{year}{2017}{\natexlab{d}}),
  \eprint{1612.02771}.

\bibitem[{\citenamefont{Caporale
  et~al.}(2017{\natexlab{e}})\citenamefont{Caporale, Celiberto, Chachamis,
  Gordo~G\'omez, and Sabio~Vera}}]{Caporale:2016zkc}
\bibinfo{author}{\bibfnamefont{F.}~\bibnamefont{Caporale}},
  \bibinfo{author}{\bibfnamefont{F.~G.} \bibnamefont{Celiberto}},
  \bibinfo{author}{\bibfnamefont{G.}~\bibnamefont{Chachamis}},
  \bibinfo{author}{\bibfnamefont{D.}~\bibnamefont{Gordo~G\'omez}},
  \bibnamefont{and}
  \bibinfo{author}{\bibfnamefont{A.}~\bibnamefont{Sabio~Vera}},
  \bibinfo{journal}{Phys. Rev. D} \textbf{\bibinfo{volume}{95}},
  \bibinfo{pages}{074007} (\bibinfo{year}{2017}{\natexlab{e}}),
  \eprint{1612.05428}.

\bibitem[{\citenamefont{Caporale
  et~al.}(2017{\natexlab{f}})\citenamefont{Caporale, Celiberto, Gordo~Gomez,
  Sabio~Vera, and Chachamis}}]{Caporale:2017jqj}
\bibinfo{author}{\bibfnamefont{F.}~\bibnamefont{Caporale}},
  \bibinfo{author}{\bibfnamefont{F.~G.} \bibnamefont{Celiberto}},
  \bibinfo{author}{\bibfnamefont{D.}~\bibnamefont{Gordo~Gomez}},
  \bibinfo{author}{\bibfnamefont{A.}~\bibnamefont{Sabio~Vera}},
  \bibnamefont{and}
  \bibinfo{author}{\bibfnamefont{G.}~\bibnamefont{Chachamis}}, in
  \emph{\bibinfo{booktitle}{{25th Low-x Meeting}}}
  (\bibinfo{year}{2017}{\natexlab{f}}), \eprint{1801.00014}.

\bibitem[{\citenamefont{Bolognino
  et~al.}(2018{\natexlab{a}})\citenamefont{Bolognino, Celiberto, Ivanov,
  Mohammed, and Papa}}]{Bolognino:2018oth}
\bibinfo{author}{\bibfnamefont{A.~D.} \bibnamefont{Bolognino}},
  \bibinfo{author}{\bibfnamefont{F.~G.} \bibnamefont{Celiberto}},
  \bibinfo{author}{\bibfnamefont{D.~{\relax Yu}.} \bibnamefont{Ivanov}},
  \bibinfo{author}{\bibfnamefont{M.~M.} \bibnamefont{Mohammed}},
  \bibnamefont{and} \bibinfo{author}{\bibfnamefont{A.}~\bibnamefont{Papa}},
  \bibinfo{journal}{Eur. Phys. J. C} \textbf{\bibinfo{volume}{78}},
  \bibinfo{pages}{772} (\bibinfo{year}{2018}{\natexlab{a}}),
  \eprint{1808.05483}.

\bibitem[{\citenamefont{Bolognino
  et~al.}(2019{\natexlab{a}})\citenamefont{Bolognino, Celiberto, Ivanov,
  Mohammed, and Papa}}]{Bolognino:2019cac}
\bibinfo{author}{\bibfnamefont{A.~D.} \bibnamefont{Bolognino}},
  \bibinfo{author}{\bibfnamefont{F.~G.} \bibnamefont{Celiberto}},
  \bibinfo{author}{\bibfnamefont{D.~{\relax Yu}.} \bibnamefont{Ivanov}},
  \bibinfo{author}{\bibfnamefont{M.~M.~A.} \bibnamefont{Mohammed}},
  \bibnamefont{and} \bibinfo{author}{\bibfnamefont{A.}~\bibnamefont{Papa}},
  \bibinfo{journal}{PoS} \textbf{\bibinfo{volume}{DIS2019}},
  \bibinfo{pages}{049} (\bibinfo{year}{2019}{\natexlab{a}}),
  \eprint{1906.11800}.

\bibitem[{\citenamefont{Bolognino
  et~al.}(2019{\natexlab{b}})\citenamefont{Bolognino, Celiberto, Ivanov,
  Mohammed, and Papa}}]{Bolognino:2019yqj}
\bibinfo{author}{\bibfnamefont{A.~D.} \bibnamefont{Bolognino}},
  \bibinfo{author}{\bibfnamefont{F.~G.} \bibnamefont{Celiberto}},
  \bibinfo{author}{\bibfnamefont{D.~{\relax Yu}.} \bibnamefont{Ivanov}},
  \bibinfo{author}{\bibfnamefont{M.~M.} \bibnamefont{Mohammed}},
  \bibnamefont{and} \bibinfo{author}{\bibfnamefont{A.}~\bibnamefont{Papa}},
  \bibinfo{journal}{Acta Phys. Polon. Supp.} \textbf{\bibinfo{volume}{12}},
  \bibinfo{pages}{773} (\bibinfo{year}{2019}{\natexlab{b}}),
  \eprint{1902.04511}.

\bibitem[{\citenamefont{Celiberto}(2021{\natexlab{a}})}]{Celiberto:2020wpk}
\bibinfo{author}{\bibfnamefont{F.~G.} \bibnamefont{Celiberto}},
  \bibinfo{journal}{Eur. Phys. J. C} \textbf{\bibinfo{volume}{81}},
  \bibinfo{pages}{691} (\bibinfo{year}{2021}{\natexlab{a}}),
  \eprint{2008.07378}.

\bibitem[{\citenamefont{Celiberto et~al.}(2020)\citenamefont{Celiberto, Ivanov,
  and Papa}}]{Celiberto:2020rxb}
\bibinfo{author}{\bibfnamefont{F.~G.} \bibnamefont{Celiberto}},
  \bibinfo{author}{\bibfnamefont{D.~{\relax Yu}.} \bibnamefont{Ivanov}},
  \bibnamefont{and} \bibinfo{author}{\bibfnamefont{A.}~\bibnamefont{Papa}},
  \bibinfo{journal}{Phys. Rev. D} \textbf{\bibinfo{volume}{102}},
  \bibinfo{pages}{094019} (\bibinfo{year}{2020}), \eprint{2008.10513}.

\bibitem[{\citenamefont{Celiberto
  et~al.}(2021{\natexlab{a}})\citenamefont{Celiberto, Ivanov, Mohammed, and
  Papa}}]{Celiberto:2020tmb}
\bibinfo{author}{\bibfnamefont{F.~G.} \bibnamefont{Celiberto}},
  \bibinfo{author}{\bibfnamefont{D.~{\relax Yu}.} \bibnamefont{Ivanov}},
  \bibinfo{author}{\bibfnamefont{M.~M.~A.} \bibnamefont{Mohammed}},
  \bibnamefont{and} \bibinfo{author}{\bibfnamefont{A.}~\bibnamefont{Papa}},
  \bibinfo{journal}{Eur. Phys. J. C} \textbf{\bibinfo{volume}{81}},
  \bibinfo{pages}{293} (\bibinfo{year}{2021}{\natexlab{a}}),
  \eprint{2008.00501}.

\bibitem[{\citenamefont{Mohammed}(2022)}]{Mohammed:2022gbk}
\bibinfo{author}{\bibfnamefont{M.~M.~A.} \bibnamefont{Mohammed}}, Ph.D. thesis,
  \bibinfo{school}{Universit\`a della Calabria and INFN-Cosenza}
  (\bibinfo{year}{2022}), \eprint{2204.11606}.

\bibitem[{\citenamefont{Celiberto
  et~al.}(2021{\natexlab{b}})\citenamefont{Celiberto, Ivanov, Mohammed, and
  Papa}}]{Celiberto:2021fjf}
\bibinfo{author}{\bibfnamefont{F.~G.} \bibnamefont{Celiberto}},
  \bibinfo{author}{\bibfnamefont{D.~{\relax Yu}.} \bibnamefont{Ivanov}},
  \bibinfo{author}{\bibfnamefont{M.~M.~A.} \bibnamefont{Mohammed}},
  \bibnamefont{and} \bibinfo{author}{\bibfnamefont{A.}~\bibnamefont{Papa}}
  (\bibinfo{year}{2021}{\natexlab{b}}), \eprint{2107.13037}.

\bibitem[{\citenamefont{Celiberto
  et~al.}(2022{\natexlab{a}})\citenamefont{Celiberto, Fucilla, Papa, Ivanov,
  and Mohammed}}]{Celiberto:2021tky}
\bibinfo{author}{\bibfnamefont{F.~G.} \bibnamefont{Celiberto}},
  \bibinfo{author}{\bibfnamefont{M.}~\bibnamefont{Fucilla}},
  \bibinfo{author}{\bibfnamefont{A.}~\bibnamefont{Papa}},
  \bibinfo{author}{\bibfnamefont{D.~{\relax Yu}.} \bibnamefont{Ivanov}},
  \bibnamefont{and} \bibinfo{author}{\bibfnamefont{M.~M.~A.}
  \bibnamefont{Mohammed}}, \bibinfo{journal}{PoS}
  \textbf{\bibinfo{volume}{EPS-HEP2021}}, \bibinfo{pages}{589}
  (\bibinfo{year}{2022}{\natexlab{a}}), \eprint{2110.09358}.

\bibitem[{\citenamefont{Celiberto
  et~al.}(2022{\natexlab{b}})\citenamefont{Celiberto, Fucilla, Ivanov,
  Mohammed, and Papa}}]{Celiberto:2021txb}
\bibinfo{author}{\bibfnamefont{F.~G.} \bibnamefont{Celiberto}},
  \bibinfo{author}{\bibfnamefont{M.}~\bibnamefont{Fucilla}},
  \bibinfo{author}{\bibfnamefont{D.~{\relax Yu}.} \bibnamefont{Ivanov}},
  \bibinfo{author}{\bibfnamefont{M.~M.~A.} \bibnamefont{Mohammed}},
  \bibnamefont{and} \bibinfo{author}{\bibfnamefont{A.}~\bibnamefont{Papa}},
  \bibinfo{journal}{PoS} \textbf{\bibinfo{volume}{PANIC2021}},
  \bibinfo{pages}{352} (\bibinfo{year}{2022}{\natexlab{b}}),
  \eprint{2111.13090}.

\bibitem[{\citenamefont{Celiberto
  et~al.}(2021{\natexlab{c}})\citenamefont{Celiberto, Fucilla, Ivanov,
  Mohammed, and Papa}}]{Celiberto:2021xpm}
\bibinfo{author}{\bibfnamefont{F.~G.} \bibnamefont{Celiberto}},
  \bibinfo{author}{\bibfnamefont{M.}~\bibnamefont{Fucilla}},
  \bibinfo{author}{\bibfnamefont{D.~{\relax Yu}.} \bibnamefont{Ivanov}},
  \bibinfo{author}{\bibfnamefont{M.~M.~A.} \bibnamefont{Mohammed}},
  \bibnamefont{and} \bibinfo{author}{\bibfnamefont{A.}~\bibnamefont{Papa}}
  (\bibinfo{year}{2021}{\natexlab{c}}), \eprint{2110.12649}.

\bibitem[{\citenamefont{Bolognino
  et~al.}(2021{\natexlab{a}})\citenamefont{Bolognino, Celiberto, Fucilla,
  Ivanov, and Papa}}]{Bolognino:2021mrc}
\bibinfo{author}{\bibfnamefont{A.~D.} \bibnamefont{Bolognino}},
  \bibinfo{author}{\bibfnamefont{F.~G.} \bibnamefont{Celiberto}},
  \bibinfo{author}{\bibfnamefont{M.}~\bibnamefont{Fucilla}},
  \bibinfo{author}{\bibfnamefont{D.~{\relax Yu}.} \bibnamefont{Ivanov}},
  \bibnamefont{and} \bibinfo{author}{\bibfnamefont{A.}~\bibnamefont{Papa}},
  \bibinfo{journal}{Phys. Rev. D} \textbf{\bibinfo{volume}{103}},
  \bibinfo{pages}{094004} (\bibinfo{year}{2021}{\natexlab{a}}),
  \eprint{2103.07396}.

\bibitem[{\citenamefont{Bolognino
  et~al.}(2021{\natexlab{b}})\citenamefont{Bolognino, Celiberto, Fucilla,
  Ivanov, and Papa}}]{Bolognino:2021hxx}
\bibinfo{author}{\bibfnamefont{A.~D.} \bibnamefont{Bolognino}},
  \bibinfo{author}{\bibfnamefont{F.~G.} \bibnamefont{Celiberto}},
  \bibinfo{author}{\bibfnamefont{M.}~\bibnamefont{Fucilla}},
  \bibinfo{author}{\bibfnamefont{D.~{\relax Yu}.} \bibnamefont{Ivanov}},
  \bibnamefont{and} \bibinfo{author}{\bibfnamefont{A.}~\bibnamefont{Papa}}, in
  \emph{\bibinfo{booktitle}{{DIS 2021}}} (\bibinfo{year}{2021}{\natexlab{b}}),
  \eprint{2107.12120}.

\bibitem[{\citenamefont{Golec-Biernat et~al.}(2018)\citenamefont{Golec-Biernat,
  Motyka, and Stebel}}]{Golec-Biernat:2018kem}
\bibinfo{author}{\bibfnamefont{K.}~\bibnamefont{Golec-Biernat}},
  \bibinfo{author}{\bibfnamefont{L.}~\bibnamefont{Motyka}}, \bibnamefont{and}
  \bibinfo{author}{\bibfnamefont{T.}~\bibnamefont{Stebel}},
  \bibinfo{journal}{JHEP} \textbf{\bibinfo{volume}{12}}, \bibinfo{pages}{091}
  (\bibinfo{year}{2018}), \eprint{1811.04361}.

\bibitem[{\citenamefont{Boussarie et~al.}(2018)\citenamefont{Boussarie,
  Duclou\'e, Szymanowski, and Wallon}}]{Boussarie:2017oae}
\bibinfo{author}{\bibfnamefont{R.}~\bibnamefont{Boussarie}},
  \bibinfo{author}{\bibfnamefont{B.}~\bibnamefont{Duclou\'e}},
  \bibinfo{author}{\bibfnamefont{L.}~\bibnamefont{Szymanowski}},
  \bibnamefont{and} \bibinfo{author}{\bibfnamefont{S.}~\bibnamefont{Wallon}},
  \bibinfo{journal}{Phys. Rev. D} \textbf{\bibinfo{volume}{97}},
  \bibinfo{pages}{014008} (\bibinfo{year}{2018}), \eprint{1709.01380}.

\bibitem[{\citenamefont{Celiberto
  et~al.}(2018{\natexlab{a}})\citenamefont{Celiberto, Ivanov, Murdaca, and
  Papa}}]{Celiberto:2017nyx}
\bibinfo{author}{\bibfnamefont{F.~G.} \bibnamefont{Celiberto}},
  \bibinfo{author}{\bibfnamefont{D.~{\relax Yu}.} \bibnamefont{Ivanov}},
  \bibinfo{author}{\bibfnamefont{B.}~\bibnamefont{Murdaca}}, \bibnamefont{and}
  \bibinfo{author}{\bibfnamefont{A.}~\bibnamefont{Papa}},
  \bibinfo{journal}{Phys. Lett. B} \textbf{\bibinfo{volume}{777}},
  \bibinfo{pages}{141} (\bibinfo{year}{2018}{\natexlab{a}}),
  \eprint{1709.10032}.

\bibitem[{\citenamefont{Bolognino
  et~al.}(2019{\natexlab{c}})\citenamefont{Bolognino, Celiberto, Fucilla,
  Ivanov, and Papa}}]{Bolognino:2019yls}
\bibinfo{author}{\bibfnamefont{A.~D.} \bibnamefont{Bolognino}},
  \bibinfo{author}{\bibfnamefont{F.~G.} \bibnamefont{Celiberto}},
  \bibinfo{author}{\bibfnamefont{M.}~\bibnamefont{Fucilla}},
  \bibinfo{author}{\bibfnamefont{D.~{\relax Yu}.} \bibnamefont{Ivanov}},
  \bibnamefont{and} \bibinfo{author}{\bibfnamefont{A.}~\bibnamefont{Papa}},
  \bibinfo{journal}{Eur. Phys. J. C} \textbf{\bibinfo{volume}{79}},
  \bibinfo{pages}{939} (\bibinfo{year}{2019}{\natexlab{c}}),
  \eprint{1909.03068}.

\bibitem[{\citenamefont{Bolognino
  et~al.}(2019{\natexlab{d}})\citenamefont{Bolognino, Celiberto, Fucilla,
  Ivanov, Murdaca, and Papa}}]{Bolognino:2019ccd}
\bibinfo{author}{\bibfnamefont{A.~D.} \bibnamefont{Bolognino}},
  \bibinfo{author}{\bibfnamefont{F.~G.} \bibnamefont{Celiberto}},
  \bibinfo{author}{\bibfnamefont{M.}~\bibnamefont{Fucilla}},
  \bibinfo{author}{\bibfnamefont{D.~{\relax Yu}.} \bibnamefont{Ivanov}},
  \bibinfo{author}{\bibfnamefont{B.}~\bibnamefont{Murdaca}}, \bibnamefont{and}
  \bibinfo{author}{\bibfnamefont{A.}~\bibnamefont{Papa}},
  \bibinfo{journal}{PoS} \textbf{\bibinfo{volume}{DIS2019}},
  \bibinfo{pages}{067} (\bibinfo{year}{2019}{\natexlab{d}}),
  \eprint{1906.05940}.

\bibitem[{\citenamefont{Celiberto
  et~al.}(2021{\natexlab{d}})\citenamefont{Celiberto, Fucilla, Ivanov, and
  Papa}}]{Celiberto:2021dzy}
\bibinfo{author}{\bibfnamefont{F.~G.} \bibnamefont{Celiberto}},
  \bibinfo{author}{\bibfnamefont{M.}~\bibnamefont{Fucilla}},
  \bibinfo{author}{\bibfnamefont{D.~{\relax Yu}.} \bibnamefont{Ivanov}},
  \bibnamefont{and} \bibinfo{author}{\bibfnamefont{A.}~\bibnamefont{Papa}},
  \bibinfo{journal}{Eur. Phys. J. C} \textbf{\bibinfo{volume}{81}},
  \bibinfo{pages}{780} (\bibinfo{year}{2021}{\natexlab{d}}),
  \eprint{2105.06432}.

\bibitem[{\citenamefont{Celiberto
  et~al.}(2021{\natexlab{e}})\citenamefont{Celiberto, Fucilla, Ivanov,
  Mohammed, and Papa}}]{Celiberto:2021fdp}
\bibinfo{author}{\bibfnamefont{F.~G.} \bibnamefont{Celiberto}},
  \bibinfo{author}{\bibfnamefont{M.}~\bibnamefont{Fucilla}},
  \bibinfo{author}{\bibfnamefont{D.~{\relax Yu}.} \bibnamefont{Ivanov}},
  \bibinfo{author}{\bibfnamefont{M.~M.~A.} \bibnamefont{Mohammed}},
  \bibnamefont{and} \bibinfo{author}{\bibfnamefont{A.}~\bibnamefont{Papa}},
  \bibinfo{journal}{Phys. Rev. D} \textbf{\bibinfo{volume}{104}},
  \bibinfo{pages}{114007} (\bibinfo{year}{2021}{\natexlab{e}}),
  \eprint{2109.11875}.

\bibitem[{\citenamefont{Celiberto}(2022{\natexlab{a}})}]{Celiberto:2022rfj}
\bibinfo{author}{\bibfnamefont{F.~G.} \bibnamefont{Celiberto}},
  \bibinfo{journal}{Phys. Rev. D} \textbf{\bibinfo{volume}{105}},
  \bibinfo{pages}{114008} (\bibinfo{year}{2022}{\natexlab{a}}),
  \eprint{2204.06497}.

\bibitem[{\citenamefont{Bolognino
  et~al.}(2022{\natexlab{a}})\citenamefont{Bolognino, Celiberto, Fucilla,
  Ivanov, and Papa}}]{Bolognino:2022wgl}
\bibinfo{author}{\bibfnamefont{A.~D.} \bibnamefont{Bolognino}},
  \bibinfo{author}{\bibfnamefont{F.~G.} \bibnamefont{Celiberto}},
  \bibinfo{author}{\bibfnamefont{M.}~\bibnamefont{Fucilla}},
  \bibinfo{author}{\bibfnamefont{D.~{\relax Yu}.} \bibnamefont{Ivanov}},
  \bibnamefont{and} \bibinfo{author}{\bibfnamefont{A.}~\bibnamefont{Papa}},
  \bibinfo{journal}{PoS} \textbf{\bibinfo{volume}{EPS-HEP2021}},
  \bibinfo{pages}{389} (\bibinfo{year}{2022}{\natexlab{a}}).

\bibitem[{\citenamefont{Celiberto and Fucilla}(2022)}]{Celiberto:2022dyf}
\bibinfo{author}{\bibfnamefont{F.~G.} \bibnamefont{Celiberto}}
  \bibnamefont{and} \bibinfo{author}{\bibfnamefont{M.}~\bibnamefont{Fucilla}},
  \bibinfo{journal}{under revision in Eur. Phys. J. C}  (\bibinfo{year}{2022}),
  \eprint{2202.12227}.

\bibitem[{\citenamefont{Celiberto}(2022{\natexlab{b}})}]{Celiberto:2022keu}
\bibinfo{author}{\bibfnamefont{F.~G.} \bibnamefont{Celiberto}}
  (\bibinfo{year}{2022}{\natexlab{b}}), \eprint{2206.09413}.

\bibitem[{\citenamefont{Anikin et~al.}(2011)\citenamefont{Anikin, Besse,
  Ivanov, Pire, Szymanowski, and Wallon}}]{Anikin:2011sa}
\bibinfo{author}{\bibfnamefont{I.}~\bibnamefont{Anikin}},
  \bibinfo{author}{\bibfnamefont{A.}~\bibnamefont{Besse}},
  \bibinfo{author}{\bibfnamefont{D.~{\relax Yu}.} \bibnamefont{Ivanov}},
  \bibinfo{author}{\bibfnamefont{B.}~\bibnamefont{Pire}},
  \bibinfo{author}{\bibfnamefont{L.}~\bibnamefont{Szymanowski}},
  \bibnamefont{and} \bibinfo{author}{\bibfnamefont{S.}~\bibnamefont{Wallon}},
  \bibinfo{journal}{Phys. Rev. D} \textbf{\bibinfo{volume}{84}},
  \bibinfo{pages}{054004} (\bibinfo{year}{2011}), \eprint{1105.1761}.

\bibitem[{\citenamefont{Besse et~al.}(2013)\citenamefont{Besse, Szymanowski,
  and Wallon}}]{Besse:2013muy}
\bibinfo{author}{\bibfnamefont{A.}~\bibnamefont{Besse}},
  \bibinfo{author}{\bibfnamefont{L.}~\bibnamefont{Szymanowski}},
  \bibnamefont{and} \bibinfo{author}{\bibfnamefont{S.}~\bibnamefont{Wallon}},
  \bibinfo{journal}{JHEP} \textbf{\bibinfo{volume}{11}}, \bibinfo{pages}{062}
  (\bibinfo{year}{2013}), \eprint{1302.1766}.

\bibitem[{\citenamefont{Bolognino
  et~al.}(2018{\natexlab{b}})\citenamefont{Bolognino, Celiberto, Ivanov, and
  Papa}}]{Bolognino:2018rhb}
\bibinfo{author}{\bibfnamefont{A.~D.} \bibnamefont{Bolognino}},
  \bibinfo{author}{\bibfnamefont{F.~G.} \bibnamefont{Celiberto}},
  \bibinfo{author}{\bibfnamefont{D.~{\relax Yu}.} \bibnamefont{Ivanov}},
  \bibnamefont{and} \bibinfo{author}{\bibfnamefont{A.}~\bibnamefont{Papa}},
  \bibinfo{journal}{Eur. Phys. J.} \textbf{\bibinfo{volume}{C78}},
  \bibinfo{pages}{1023} (\bibinfo{year}{2018}{\natexlab{b}}),
  \eprint{1808.02395}.

\bibitem[{\citenamefont{Bolognino
  et~al.}(2018{\natexlab{c}})\citenamefont{Bolognino, Celiberto, Ivanov, and
  Papa}}]{Bolognino:2018mlw}
\bibinfo{author}{\bibfnamefont{A.~D.} \bibnamefont{Bolognino}},
  \bibinfo{author}{\bibfnamefont{F.~G.} \bibnamefont{Celiberto}},
  \bibinfo{author}{\bibfnamefont{D.~{\relax Yu}.} \bibnamefont{Ivanov}},
  \bibnamefont{and} \bibinfo{author}{\bibfnamefont{A.}~\bibnamefont{Papa}},
  \bibinfo{journal}{Frascati Phys. Ser.} \textbf{\bibinfo{volume}{67}},
  \bibinfo{pages}{76} (\bibinfo{year}{2018}{\natexlab{c}}),
  \eprint{1808.02958}.

\bibitem[{\citenamefont{Bolognino
  et~al.}(2019{\natexlab{e}})\citenamefont{Bolognino, Celiberto, Ivanov, and
  Papa}}]{Bolognino:2019bko}
\bibinfo{author}{\bibfnamefont{A.~D.} \bibnamefont{Bolognino}},
  \bibinfo{author}{\bibfnamefont{F.~G.} \bibnamefont{Celiberto}},
  \bibinfo{author}{\bibfnamefont{D.~{\relax Yu}.} \bibnamefont{Ivanov}},
  \bibnamefont{and} \bibinfo{author}{\bibfnamefont{A.}~\bibnamefont{Papa}},
  \bibinfo{journal}{Acta Phys. Polon. Supp.} \textbf{\bibinfo{volume}{12}},
  \bibinfo{pages}{891} (\bibinfo{year}{2019}{\natexlab{e}}),
  \eprint{1902.04520}.

\bibitem[{\citenamefont{Bolognino et~al.}(2020)\citenamefont{Bolognino,
  Szczurek, and Schaefer}}]{Bolognino:2019pba}
\bibinfo{author}{\bibfnamefont{A.~D.} \bibnamefont{Bolognino}},
  \bibinfo{author}{\bibfnamefont{A.}~\bibnamefont{Szczurek}}, \bibnamefont{and}
  \bibinfo{author}{\bibfnamefont{W.}~\bibnamefont{Schaefer}},
  \bibinfo{journal}{Phys. Rev. D} \textbf{\bibinfo{volume}{101}},
  \bibinfo{pages}{054041} (\bibinfo{year}{2020}), \eprint{1912.06507}.

\bibitem[{\citenamefont{Celiberto}(2019)}]{Celiberto:2019slj}
\bibinfo{author}{\bibfnamefont{F.~G.} \bibnamefont{Celiberto}},
  \bibinfo{journal}{Nuovo Cim.} \textbf{\bibinfo{volume}{C42}},
  \bibinfo{pages}{220} (\bibinfo{year}{2019}), \eprint{1912.11313}.

\bibitem[{\citenamefont{Bolognino
  et~al.}(2021{\natexlab{c}})\citenamefont{Bolognino, Celiberto, Ivanov, Papa,
  Sch\"afer, and Szczurek}}]{Bolognino:2021niq}
\bibinfo{author}{\bibfnamefont{A.~D.} \bibnamefont{Bolognino}},
  \bibinfo{author}{\bibfnamefont{F.~G.} \bibnamefont{Celiberto}},
  \bibinfo{author}{\bibfnamefont{D.~{\relax Yu}.} \bibnamefont{Ivanov}},
  \bibinfo{author}{\bibfnamefont{A.}~\bibnamefont{Papa}},
  \bibinfo{author}{\bibfnamefont{W.}~\bibnamefont{Sch\"afer}},
  \bibnamefont{and} \bibinfo{author}{\bibfnamefont{A.}~\bibnamefont{Szczurek}},
  \bibinfo{journal}{Eur. Phys. J. C} \textbf{\bibinfo{volume}{81}},
  \bibinfo{pages}{846} (\bibinfo{year}{2021}{\natexlab{c}}),
  \eprint{2107.13415}.

\bibitem[{\citenamefont{Bolognino
  et~al.}(2021{\natexlab{d}})\citenamefont{Bolognino, Celiberto, Ivanov, and
  Papa}}]{Bolognino:2021gjm}
\bibinfo{author}{\bibfnamefont{A.~D.} \bibnamefont{Bolognino}},
  \bibinfo{author}{\bibfnamefont{F.~G.} \bibnamefont{Celiberto}},
  \bibinfo{author}{\bibfnamefont{D.~{\relax Yu}.} \bibnamefont{Ivanov}},
  \bibnamefont{and} \bibinfo{author}{\bibfnamefont{A.}~\bibnamefont{Papa}}, in
  \emph{\bibinfo{booktitle}{{DIS 2021}}} (\bibinfo{year}{2021}{\natexlab{d}}),
  \eprint{2107.12725}.

\bibitem[{\citenamefont{Bolognino
  et~al.}(2022{\natexlab{b}})\citenamefont{Bolognino, Celiberto, Fucilla,
  Ivanov, Papa, Sch\"afer, and Szczurek}}]{Bolognino:2022uty}
\bibinfo{author}{\bibfnamefont{A.~D.} \bibnamefont{Bolognino}},
  \bibinfo{author}{\bibfnamefont{F.~G.} \bibnamefont{Celiberto}},
  \bibinfo{author}{\bibfnamefont{M.}~\bibnamefont{Fucilla}},
  \bibinfo{author}{\bibfnamefont{D.~{\relax Yu}.} \bibnamefont{Ivanov}},
  \bibinfo{author}{\bibfnamefont{A.}~\bibnamefont{Papa}},
  \bibinfo{author}{\bibfnamefont{W.}~\bibnamefont{Sch\"afer}},
  \bibnamefont{and} \bibinfo{author}{\bibfnamefont{A.}~\bibnamefont{Szczurek}},
  in \emph{\bibinfo{booktitle}{{19th International Conference on Hadron
  Spectroscopy and Structure}}} (\bibinfo{year}{2022}{\natexlab{b}}),
  \eprint{2202.02513}.

\bibitem[{\citenamefont{Celiberto}(2022{\natexlab{c}})}]{Celiberto:2022fam}
\bibinfo{author}{\bibfnamefont{F.~G.} \bibnamefont{Celiberto}}
  (\bibinfo{year}{2022}{\natexlab{c}}), \eprint{2202.04207}.

\bibitem[{\citenamefont{Motyka et~al.}(2015)\citenamefont{Motyka, Sadzikowski,
  and Stebel}}]{Motyka:2014lya}
\bibinfo{author}{\bibfnamefont{L.}~\bibnamefont{Motyka}},
  \bibinfo{author}{\bibfnamefont{M.}~\bibnamefont{Sadzikowski}},
  \bibnamefont{and} \bibinfo{author}{\bibfnamefont{T.}~\bibnamefont{Stebel}},
  \bibinfo{journal}{JHEP} \textbf{\bibinfo{volume}{05}}, \bibinfo{pages}{087}
  (\bibinfo{year}{2015}), \eprint{1412.4675}.

\bibitem[{\citenamefont{Motyka et~al.}(2017)\citenamefont{Motyka, Sadzikowski,
  and Stebel}}]{Motyka:2016lta}
\bibinfo{author}{\bibfnamefont{L.}~\bibnamefont{Motyka}},
  \bibinfo{author}{\bibfnamefont{M.}~\bibnamefont{Sadzikowski}},
  \bibnamefont{and} \bibinfo{author}{\bibfnamefont{T.}~\bibnamefont{Stebel}},
  \bibinfo{journal}{Phys. Rev.} \textbf{\bibinfo{volume}{D95}},
  \bibinfo{pages}{114025} (\bibinfo{year}{2017}), \eprint{1609.04300}.

\bibitem[{\citenamefont{Brzeminski et~al.}(2017)\citenamefont{Brzeminski,
  Motyka, Sadzikowski, and Stebel}}]{Brzeminski:2016lwh}
\bibinfo{author}{\bibfnamefont{D.}~\bibnamefont{Brzeminski}},
  \bibinfo{author}{\bibfnamefont{L.}~\bibnamefont{Motyka}},
  \bibinfo{author}{\bibfnamefont{M.}~\bibnamefont{Sadzikowski}},
  \bibnamefont{and} \bibinfo{author}{\bibfnamefont{T.}~\bibnamefont{Stebel}},
  \bibinfo{journal}{JHEP} \textbf{\bibinfo{volume}{01}}, \bibinfo{pages}{005}
  (\bibinfo{year}{2017}), \eprint{1611.04449}.

\bibitem[{\citenamefont{Celiberto
  et~al.}(2018{\natexlab{b}})\citenamefont{Celiberto, Gordo~G\'omez, and
  Sabio~Vera}}]{Celiberto:2018muu}
\bibinfo{author}{\bibfnamefont{F.~G.} \bibnamefont{Celiberto}},
  \bibinfo{author}{\bibfnamefont{D.}~\bibnamefont{Gordo~G\'omez}},
  \bibnamefont{and}
  \bibinfo{author}{\bibfnamefont{A.}~\bibnamefont{Sabio~Vera}},
  \bibinfo{journal}{Phys. Lett.} \textbf{\bibinfo{volume}{B786}},
  \bibinfo{pages}{201} (\bibinfo{year}{2018}{\natexlab{b}}),
  \eprint{1808.09511}.

\bibitem[{\citenamefont{Ball et~al.}(2018)\citenamefont{Ball, Bertone, Bonvini,
  Marzani, Rojo, and Rottoli}}]{Ball:2017otu}
\bibinfo{author}{\bibfnamefont{R.~D.} \bibnamefont{Ball}},
  \bibinfo{author}{\bibfnamefont{V.}~\bibnamefont{Bertone}},
  \bibinfo{author}{\bibfnamefont{M.}~\bibnamefont{Bonvini}},
  \bibinfo{author}{\bibfnamefont{S.}~\bibnamefont{Marzani}},
  \bibinfo{author}{\bibfnamefont{J.}~\bibnamefont{Rojo}}, \bibnamefont{and}
  \bibinfo{author}{\bibfnamefont{L.}~\bibnamefont{Rottoli}},
  \bibinfo{journal}{Eur. Phys. J.} \textbf{\bibinfo{volume}{C78}},
  \bibinfo{pages}{321} (\bibinfo{year}{2018}), \eprint{1710.05935}.

\bibitem[{\citenamefont{Abdolmaleki et~al.}(2018)}]{Abdolmaleki:2018jln}
\bibinfo{author}{\bibfnamefont{H.}~\bibnamefont{Abdolmaleki}}
  \bibnamefont{et~al.} (\bibinfo{collaboration}{xFitter Developers' Team}),
  \bibinfo{journal}{Eur. Phys. J. C} \textbf{\bibinfo{volume}{78}},
  \bibinfo{pages}{621} (\bibinfo{year}{2018}), \eprint{1802.00064}.

\bibitem[{\citenamefont{Bonvini and Giuli}(2019)}]{Bonvini:2019wxf}
\bibinfo{author}{\bibfnamefont{M.}~\bibnamefont{Bonvini}} \bibnamefont{and}
  \bibinfo{author}{\bibfnamefont{F.}~\bibnamefont{Giuli}},
  \bibinfo{journal}{Eur. Phys. J. Plus} \textbf{\bibinfo{volume}{134}},
  \bibinfo{pages}{531} (\bibinfo{year}{2019}), \eprint{1902.11125}.

\bibitem[{\citenamefont{Mulders and Rodrigues}(2001)}]{Mulders:2000sh}
\bibinfo{author}{\bibfnamefont{P.~J.} \bibnamefont{Mulders}} \bibnamefont{and}
  \bibinfo{author}{\bibfnamefont{J.}~\bibnamefont{Rodrigues}},
  \bibinfo{journal}{Phys. Rev.} \textbf{\bibinfo{volume}{D63}},
  \bibinfo{pages}{094021} (\bibinfo{year}{2001}), \eprint{hep-ph/0009343}.

\bibitem[{\citenamefont{Meissner et~al.}(2007)\citenamefont{Meissner, Metz, and
  Goeke}}]{Meissner:2007rx}
\bibinfo{author}{\bibfnamefont{S.}~\bibnamefont{Meissner}},
  \bibinfo{author}{\bibfnamefont{A.}~\bibnamefont{Metz}}, \bibnamefont{and}
  \bibinfo{author}{\bibfnamefont{K.}~\bibnamefont{Goeke}},
  \bibinfo{journal}{Phys. Rev.} \textbf{\bibinfo{volume}{D76}},
  \bibinfo{pages}{034002} (\bibinfo{year}{2007}), \eprint{hep-ph/0703176}.

\bibitem[{\citenamefont{Lorce' and Pasquini}(2013)}]{Lorce:2013pza}
\bibinfo{author}{\bibfnamefont{C.}~\bibnamefont{Lorce'}} \bibnamefont{and}
  \bibinfo{author}{\bibfnamefont{B.}~\bibnamefont{Pasquini}},
  \bibinfo{journal}{JHEP} \textbf{\bibinfo{volume}{09}}, \bibinfo{pages}{138}
  (\bibinfo{year}{2013}), \eprint{1307.4497}.

\bibitem[{\citenamefont{Boer et~al.}(2016)\citenamefont{Boer, Cotogno, van
  Daal, Mulders, Signori, and Zhou}}]{Boer:2016xqr}
\bibinfo{author}{\bibfnamefont{D.}~\bibnamefont{Boer}},
  \bibinfo{author}{\bibfnamefont{S.}~\bibnamefont{Cotogno}},
  \bibinfo{author}{\bibfnamefont{T.}~\bibnamefont{van Daal}},
  \bibinfo{author}{\bibfnamefont{P.~J.} \bibnamefont{Mulders}},
  \bibinfo{author}{\bibfnamefont{A.}~\bibnamefont{Signori}}, \bibnamefont{and}
  \bibinfo{author}{\bibfnamefont{Y.-J.} \bibnamefont{Zhou}},
  \bibinfo{journal}{JHEP} \textbf{\bibinfo{volume}{10}}, \bibinfo{pages}{013}
  (\bibinfo{year}{2016}), \eprint{1607.01654}.

\bibitem[{\citenamefont{Bacchetta et~al.}(2020)\citenamefont{Bacchetta,
  Celiberto, Radici, and Taels}}]{Bacchetta:2020vty}
\bibinfo{author}{\bibfnamefont{A.}~\bibnamefont{Bacchetta}},
  \bibinfo{author}{\bibfnamefont{F.~G.} \bibnamefont{Celiberto}},
  \bibinfo{author}{\bibfnamefont{M.}~\bibnamefont{Radici}}, \bibnamefont{and}
  \bibinfo{author}{\bibfnamefont{P.}~\bibnamefont{Taels}},
  \bibinfo{journal}{Eur. Phys. J. C} \textbf{\bibinfo{volume}{80}},
  \bibinfo{pages}{733} (\bibinfo{year}{2020}), \eprint{2005.02288}.

\bibitem[{\citenamefont{Celiberto}(2021{\natexlab{b}})}]{Celiberto:2021zww}
\bibinfo{author}{\bibfnamefont{F.~G.} \bibnamefont{Celiberto}},
  \bibinfo{journal}{Nuovo Cim.} \textbf{\bibinfo{volume}{C44}},
  \bibinfo{pages}{36} (\bibinfo{year}{2021}{\natexlab{b}}),
  \eprint{2101.04630}.

\bibitem[{\citenamefont{Bacchetta et~al.}(2021)\citenamefont{Bacchetta,
  Celiberto, Radici, and Taels}}]{Bacchetta:2021oht}
\bibinfo{author}{\bibfnamefont{A.}~\bibnamefont{Bacchetta}},
  \bibinfo{author}{\bibfnamefont{F.~G.} \bibnamefont{Celiberto}},
  \bibinfo{author}{\bibfnamefont{M.}~\bibnamefont{Radici}}, \bibnamefont{and}
  \bibinfo{author}{\bibfnamefont{P.}~\bibnamefont{Taels}}, in
  \emph{\bibinfo{booktitle}{{DIS 2021}}} (\bibinfo{year}{2021}),
  \eprint{2107.13446}.

\bibitem[{\citenamefont{Bacchetta
  et~al.}(2022{\natexlab{a}})\citenamefont{Bacchetta, Celiberto, and
  Radici}}]{Bacchetta:2021lvw}
\bibinfo{author}{\bibfnamefont{A.}~\bibnamefont{Bacchetta}},
  \bibinfo{author}{\bibfnamefont{F.~G.} \bibnamefont{Celiberto}},
  \bibnamefont{and} \bibinfo{author}{\bibfnamefont{M.}~\bibnamefont{Radici}},
  \bibinfo{journal}{PoS} \textbf{\bibinfo{volume}{EPS-HEP2021}},
  \bibinfo{pages}{376} (\bibinfo{year}{2022}{\natexlab{a}}),
  \eprint{2111.01686}.

\bibitem[{\citenamefont{Bacchetta
  et~al.}(2022{\natexlab{b}})\citenamefont{Bacchetta, Celiberto, and
  Radici}}]{Bacchetta:2021twk}
\bibinfo{author}{\bibfnamefont{A.}~\bibnamefont{Bacchetta}},
  \bibinfo{author}{\bibfnamefont{F.~G.} \bibnamefont{Celiberto}},
  \bibnamefont{and} \bibinfo{author}{\bibfnamefont{M.}~\bibnamefont{Radici}},
  \bibinfo{journal}{PoS} \textbf{\bibinfo{volume}{PANIC2021}},
  \bibinfo{pages}{378} (\bibinfo{year}{2022}{\natexlab{b}}),
  \eprint{2111.03567}.

\bibitem[{\citenamefont{Bacchetta
  et~al.}(2022{\natexlab{c}})\citenamefont{Bacchetta, Celiberto, and
  Radici}}]{Bacchetta:2022esb}
\bibinfo{author}{\bibfnamefont{A.}~\bibnamefont{Bacchetta}},
  \bibinfo{author}{\bibfnamefont{F.~G.} \bibnamefont{Celiberto}},
  \bibnamefont{and} \bibinfo{author}{\bibfnamefont{M.}~\bibnamefont{Radici}}
  (\bibinfo{year}{2022}{\natexlab{c}}), \eprint{2201.10508}.

\bibitem[{\citenamefont{Arbuzov et~al.}(2021)}]{Arbuzov:2020cqg}
\bibinfo{author}{\bibfnamefont{A.}~\bibnamefont{Arbuzov}} \bibnamefont{et~al.},
  \bibinfo{journal}{Prog. Part. Nucl. Phys.} \textbf{\bibinfo{volume}{119}},
  \bibinfo{pages}{103858} (\bibinfo{year}{2021}), \eprint{2011.15005}.

\bibitem[{\citenamefont{Abdul~Khalek et~al.}(2021)}]{AbdulKhalek:2021gbh}
\bibinfo{author}{\bibfnamefont{R.}~\bibnamefont{Abdul~Khalek}}
  \bibnamefont{et~al.} (\bibinfo{year}{2021}), \eprint{2103.05419}.

\bibitem[{\citenamefont{Abdul~Khalek et~al.}(2022)}]{Khalek:2022bzd}
\bibinfo{author}{\bibfnamefont{R.}~\bibnamefont{Abdul~Khalek}}
  \bibnamefont{et~al.}, in \emph{\bibinfo{booktitle}{{2022 Snowmass Summer
  Study}}} (\bibinfo{year}{2022}), \eprint{2203.13199}.

\bibitem[{\citenamefont{Bacchetta
  et~al.}(2022{\natexlab{d}})\citenamefont{Bacchetta, Celiberto, and
  Radici}}]{Bacchetta:2022crh}
\bibinfo{author}{\bibfnamefont{A.}~\bibnamefont{Bacchetta}},
  \bibinfo{author}{\bibfnamefont{F.~G.} \bibnamefont{Celiberto}},
  \bibnamefont{and} \bibinfo{author}{\bibfnamefont{M.}~\bibnamefont{Radici}}
  (\bibinfo{year}{2022}{\natexlab{d}}), \eprint{2206.07815}.

\bibitem[{\citenamefont{Brodsky
  et~al.}(1997{\natexlab{a}})\citenamefont{Brodsky, Hautmann, and
  Soper}}]{Brodsky:1996sg}
\bibinfo{author}{\bibfnamefont{S.~J.} \bibnamefont{Brodsky}},
  \bibinfo{author}{\bibfnamefont{F.}~\bibnamefont{Hautmann}}, \bibnamefont{and}
  \bibinfo{author}{\bibfnamefont{D.~E.} \bibnamefont{Soper}},
  \bibinfo{journal}{Phys. Rev. Lett.} \textbf{\bibinfo{volume}{78}},
  \bibinfo{pages}{803} (\bibinfo{year}{1997}{\natexlab{a}}),
  \bibinfo{note}{[Erratum: Phys.Rev.Lett. 79, 3544 (1997)]},
  \eprint{hep-ph/9610260}.

\bibitem[{\citenamefont{Brodsky
  et~al.}(1997{\natexlab{b}})\citenamefont{Brodsky, Hautmann, and
  Soper}}]{Brodsky:1997sd}
\bibinfo{author}{\bibfnamefont{S.~J.} \bibnamefont{Brodsky}},
  \bibinfo{author}{\bibfnamefont{F.}~\bibnamefont{Hautmann}}, \bibnamefont{and}
  \bibinfo{author}{\bibfnamefont{D.~E.} \bibnamefont{Soper}},
  \bibinfo{journal}{Phys. Rev. D} \textbf{\bibinfo{volume}{56}},
  \bibinfo{pages}{6957} (\bibinfo{year}{1997}{\natexlab{b}}),
  \eprint{hep-ph/9706427}.

\bibitem[{\citenamefont{Brodsky et~al.}(1999)\citenamefont{Brodsky, Fadin, Kim,
  Lipatov, and Pivovarov}}]{Brodsky:1998kn}
\bibinfo{author}{\bibfnamefont{S.~J.} \bibnamefont{Brodsky}},
  \bibinfo{author}{\bibfnamefont{V.~S.} \bibnamefont{Fadin}},
  \bibinfo{author}{\bibfnamefont{V.~T.} \bibnamefont{Kim}},
  \bibinfo{author}{\bibfnamefont{L.~N.} \bibnamefont{Lipatov}},
  \bibnamefont{and} \bibinfo{author}{\bibfnamefont{G.~B.}
  \bibnamefont{Pivovarov}}, \bibinfo{journal}{JETP Lett.}
  \textbf{\bibinfo{volume}{70}}, \bibinfo{pages}{155} (\bibinfo{year}{1999}),
  \eprint{hep-ph/9901229}.

\bibitem[{\citenamefont{Brodsky et~al.}(2002)\citenamefont{Brodsky, Fadin, Kim,
  Lipatov, and Pivovarov}}]{Brodsky:2002ka}
\bibinfo{author}{\bibfnamefont{S.~J.} \bibnamefont{Brodsky}},
  \bibinfo{author}{\bibfnamefont{V.~S.} \bibnamefont{Fadin}},
  \bibinfo{author}{\bibfnamefont{V.~T.} \bibnamefont{Kim}},
  \bibinfo{author}{\bibfnamefont{L.~N.} \bibnamefont{Lipatov}},
  \bibnamefont{and} \bibinfo{author}{\bibfnamefont{G.~B.}
  \bibnamefont{Pivovarov}}, \bibinfo{journal}{JETP Lett.}
  \textbf{\bibinfo{volume}{76}}, \bibinfo{pages}{249} (\bibinfo{year}{2002}),
  \eprint{hep-ph/0207297}.

\bibitem[{\citenamefont{Kniehl et~al.}(2008)\citenamefont{Kniehl, Kramer,
  Schienbein, and Spiesberger}}]{Kniehl:2008zza}
\bibinfo{author}{\bibfnamefont{B.~A.} \bibnamefont{Kniehl}},
  \bibinfo{author}{\bibfnamefont{G.}~\bibnamefont{Kramer}},
  \bibinfo{author}{\bibfnamefont{I.}~\bibnamefont{Schienbein}},
  \bibnamefont{and}
  \bibinfo{author}{\bibfnamefont{H.}~\bibnamefont{Spiesberger}},
  \bibinfo{journal}{Phys. Rev. D} \textbf{\bibinfo{volume}{77}},
  \bibinfo{pages}{014011} (\bibinfo{year}{2008}), \eprint{0705.4392}.

\bibitem[{\citenamefont{Kramer and Spiesberger}(2018)}]{Kramer:2018vde}
\bibinfo{author}{\bibfnamefont{G.}~\bibnamefont{Kramer}} \bibnamefont{and}
  \bibinfo{author}{\bibfnamefont{H.}~\bibnamefont{Spiesberger}},
  \bibinfo{journal}{Phys. Rev. D} \textbf{\bibinfo{volume}{98}},
  \bibinfo{pages}{114010} (\bibinfo{year}{2018}), \eprint{1809.04297}.

\bibitem[{\citenamefont{Kniehl et~al.}(2020)\citenamefont{Kniehl, Kramer,
  Schienbein, and Spiesberger}}]{Kniehl:2020szu}
\bibinfo{author}{\bibfnamefont{B.~A.} \bibnamefont{Kniehl}},
  \bibinfo{author}{\bibfnamefont{G.}~\bibnamefont{Kramer}},
  \bibinfo{author}{\bibfnamefont{I.}~\bibnamefont{Schienbein}},
  \bibnamefont{and}
  \bibinfo{author}{\bibfnamefont{H.}~\bibnamefont{Spiesberger}},
  \bibinfo{journal}{Phys. Rev. D} \textbf{\bibinfo{volume}{101}},
  \bibinfo{pages}{114021} (\bibinfo{year}{2020}), \eprint{2004.04213}.

\bibitem[{\citenamefont{Mele and Nason}(1991)}]{Mele:1990cw}
\bibinfo{author}{\bibfnamefont{B.}~\bibnamefont{Mele}} \bibnamefont{and}
  \bibinfo{author}{\bibfnamefont{P.}~\bibnamefont{Nason}},
  \bibinfo{journal}{Nucl. Phys. B} \textbf{\bibinfo{volume}{361}},
  \bibinfo{pages}{626} (\bibinfo{year}{1991}), \bibinfo{note}{[Erratum:
  Nucl.Phys.B 921, 841--842 (2017)]}.

\bibitem[{\citenamefont{Cacciari and Greco}(1994)}]{Cacciari:1993mq}
\bibinfo{author}{\bibfnamefont{M.}~\bibnamefont{Cacciari}} \bibnamefont{and}
  \bibinfo{author}{\bibfnamefont{M.}~\bibnamefont{Greco}},
  \bibinfo{journal}{Nucl. Phys. B} \textbf{\bibinfo{volume}{421}},
  \bibinfo{pages}{530} (\bibinfo{year}{1994}), \eprint{hep-ph/9311260}.

\bibitem[{\citenamefont{Caswell and Lepage}(1986)}]{Caswell:1985ui}
\bibinfo{author}{\bibfnamefont{W.~E.} \bibnamefont{Caswell}} \bibnamefont{and}
  \bibinfo{author}{\bibfnamefont{G.~P.} \bibnamefont{Lepage}},
  \bibinfo{journal}{Phys. Lett. B} \textbf{\bibinfo{volume}{167}},
  \bibinfo{pages}{437} (\bibinfo{year}{1986}).

\bibitem[{\citenamefont{Thacker and Lepage}(1991)}]{Thacker:1990bm}
\bibinfo{author}{\bibfnamefont{B.~A.} \bibnamefont{Thacker}} \bibnamefont{and}
  \bibinfo{author}{\bibfnamefont{G.~P.} \bibnamefont{Lepage}},
  \bibinfo{journal}{Phys. Rev. D} \textbf{\bibinfo{volume}{43}},
  \bibinfo{pages}{196} (\bibinfo{year}{1991}).

\bibitem[{\citenamefont{Bodwin et~al.}(1995)\citenamefont{Bodwin, Braaten, and
  Lepage}}]{Bodwin:1994jh}
\bibinfo{author}{\bibfnamefont{G.~T.} \bibnamefont{Bodwin}},
  \bibinfo{author}{\bibfnamefont{E.}~\bibnamefont{Braaten}}, \bibnamefont{and}
  \bibinfo{author}{\bibfnamefont{G.~P.} \bibnamefont{Lepage}},
  \bibinfo{journal}{Phys. Rev. D} \textbf{\bibinfo{volume}{51}},
  \bibinfo{pages}{1125} (\bibinfo{year}{1995}), \bibinfo{note}{[Erratum:
  Phys.Rev.D 55, 5853 (1997)]}, \eprint{hep-ph/9407339}.

\bibitem[{\citenamefont{Braaten et~al.}(1993)\citenamefont{Braaten, Cheung, and
  Yuan}}]{Braaten:1993mp}
\bibinfo{author}{\bibfnamefont{E.}~\bibnamefont{Braaten}},
  \bibinfo{author}{\bibfnamefont{K.-m.} \bibnamefont{Cheung}},
  \bibnamefont{and} \bibinfo{author}{\bibfnamefont{T.~C.} \bibnamefont{Yuan}},
  \bibinfo{journal}{Phys. Rev. D} \textbf{\bibinfo{volume}{48}},
  \bibinfo{pages}{4230} (\bibinfo{year}{1993}), \eprint{hep-ph/9302307}.

\bibitem[{\citenamefont{Zheng et~al.}(2019)\citenamefont{Zheng, Chang, and
  Wu}}]{Zheng:2019dfk}
\bibinfo{author}{\bibfnamefont{X.-C.} \bibnamefont{Zheng}},
  \bibinfo{author}{\bibfnamefont{C.-H.} \bibnamefont{Chang}}, \bibnamefont{and}
  \bibinfo{author}{\bibfnamefont{X.-G.} \bibnamefont{Wu}},
  \bibinfo{journal}{Phys. Rev. D} \textbf{\bibinfo{volume}{100}},
  \bibinfo{pages}{014005} (\bibinfo{year}{2019}), \eprint{1905.09171}.

\bibitem[{\citenamefont{Binosi et~al.}(2009)\citenamefont{Binosi, Collins,
  Kaufhold, and Theussl}}]{Binosi:2008ig}
\bibinfo{author}{\bibfnamefont{D.}~\bibnamefont{Binosi}},
  \bibinfo{author}{\bibfnamefont{J.}~\bibnamefont{Collins}},
  \bibinfo{author}{\bibfnamefont{C.}~\bibnamefont{Kaufhold}}, \bibnamefont{and}
  \bibinfo{author}{\bibfnamefont{L.}~\bibnamefont{Theussl}},
  \bibinfo{journal}{Comput. Phys. Commun.} \textbf{\bibinfo{volume}{180}},
  \bibinfo{pages}{1709} (\bibinfo{year}{2009}), \eprint{0811.4113}.

\bibitem[{\citenamefont{Del~Duca and Schmidt}(1994)}]{DelDuca:1993ga}
\bibinfo{author}{\bibfnamefont{V.}~\bibnamefont{Del~Duca}} \bibnamefont{and}
  \bibinfo{author}{\bibfnamefont{C.~R.} \bibnamefont{Schmidt}},
  \bibinfo{journal}{Phys. Rev. D} \textbf{\bibinfo{volume}{49}},
  \bibinfo{pages}{177} (\bibinfo{year}{1994}), \eprint{hep-ph/9305346}.

\bibitem[{\citenamefont{Del~Duca et~al.}(2003)\citenamefont{Del~Duca, Kilgore,
  Oleari, Schmidt, and Zeppenfeld}}]{DelDuca:2003ba}
\bibinfo{author}{\bibfnamefont{V.}~\bibnamefont{Del~Duca}},
  \bibinfo{author}{\bibfnamefont{W.}~\bibnamefont{Kilgore}},
  \bibinfo{author}{\bibfnamefont{C.}~\bibnamefont{Oleari}},
  \bibinfo{author}{\bibfnamefont{C.~R.} \bibnamefont{Schmidt}},
  \bibnamefont{and}
  \bibinfo{author}{\bibfnamefont{D.}~\bibnamefont{Zeppenfeld}},
  \bibinfo{journal}{Phys. Rev. D} \textbf{\bibinfo{volume}{67}},
  \bibinfo{pages}{073003} (\bibinfo{year}{2003}), \eprint{hep-ph/0301013}.

\bibitem[{\citenamefont{Bartels et~al.}(2006)\citenamefont{Bartels, Bondarenko,
  Kutak, and Motyka}}]{Bartels:2006ea}
\bibinfo{author}{\bibfnamefont{J.}~\bibnamefont{Bartels}},
  \bibinfo{author}{\bibfnamefont{S.}~\bibnamefont{Bondarenko}},
  \bibinfo{author}{\bibfnamefont{K.}~\bibnamefont{Kutak}}, \bibnamefont{and}
  \bibinfo{author}{\bibfnamefont{L.}~\bibnamefont{Motyka}},
  \bibinfo{journal}{Phys. Rev. D} \textbf{\bibinfo{volume}{73}},
  \bibinfo{pages}{093004} (\bibinfo{year}{2006}), \eprint{hep-ph/0601128}.

\bibitem[{\citenamefont{Xiao and Yuan}(2018)}]{Xiao:2018esv}
\bibinfo{author}{\bibfnamefont{B.-W.} \bibnamefont{Xiao}} \bibnamefont{and}
  \bibinfo{author}{\bibfnamefont{F.}~\bibnamefont{Yuan}},
  \bibinfo{journal}{Phys. Lett. B} \textbf{\bibinfo{volume}{782}},
  \bibinfo{pages}{28} (\bibinfo{year}{2018}), \eprint{1801.05478}.

\bibitem[{\citenamefont{Pasechnik et~al.}(2006)\citenamefont{Pasechnik,
  Teryaev, and Szczurek}}]{Pasechnik:2006du}
\bibinfo{author}{\bibfnamefont{R.~S.} \bibnamefont{Pasechnik}},
  \bibinfo{author}{\bibfnamefont{O.~V.} \bibnamefont{Teryaev}},
  \bibnamefont{and} \bibinfo{author}{\bibfnamefont{A.}~\bibnamefont{Szczurek}},
  \bibinfo{journal}{Eur. Phys. J. C} \textbf{\bibinfo{volume}{47}},
  \bibinfo{pages}{429} (\bibinfo{year}{2006}), \eprint{hep-ph/0603258}.

\bibitem[{\citenamefont{Ball and Forte}(1995)}]{Ball:1995vc}
\bibinfo{author}{\bibfnamefont{R.~D.} \bibnamefont{Ball}} \bibnamefont{and}
  \bibinfo{author}{\bibfnamefont{S.}~\bibnamefont{Forte}},
  \bibinfo{journal}{Phys. Lett. B} \textbf{\bibinfo{volume}{351}},
  \bibinfo{pages}{313} (\bibinfo{year}{1995}), \eprint{hep-ph/9501231}.

\bibitem[{\citenamefont{Ball and Forte}(1997)}]{Ball:1997vf}
\bibinfo{author}{\bibfnamefont{R.~D.} \bibnamefont{Ball}} \bibnamefont{and}
  \bibinfo{author}{\bibfnamefont{S.}~\bibnamefont{Forte}},
  \bibinfo{journal}{Phys. Lett. B} \textbf{\bibinfo{volume}{405}},
  \bibinfo{pages}{317} (\bibinfo{year}{1997}), \eprint{hep-ph/9703417}.

\bibitem[{\citenamefont{Altarelli et~al.}(2002)\citenamefont{Altarelli, Ball,
  and Forte}}]{Altarelli:2001ji}
\bibinfo{author}{\bibfnamefont{G.}~\bibnamefont{Altarelli}},
  \bibinfo{author}{\bibfnamefont{R.~D.} \bibnamefont{Ball}}, \bibnamefont{and}
  \bibinfo{author}{\bibfnamefont{S.}~\bibnamefont{Forte}},
  \bibinfo{journal}{Nucl. Phys. B} \textbf{\bibinfo{volume}{621}},
  \bibinfo{pages}{359} (\bibinfo{year}{2002}), \eprint{hep-ph/0109178}.

\bibitem[{\citenamefont{Altarelli et~al.}(2003)\citenamefont{Altarelli, Ball,
  and Forte}}]{Altarelli:2003hk}
\bibinfo{author}{\bibfnamefont{G.}~\bibnamefont{Altarelli}},
  \bibinfo{author}{\bibfnamefont{R.~D.} \bibnamefont{Ball}}, \bibnamefont{and}
  \bibinfo{author}{\bibfnamefont{S.}~\bibnamefont{Forte}},
  \bibinfo{journal}{Nucl. Phys. B} \textbf{\bibinfo{volume}{674}},
  \bibinfo{pages}{459} (\bibinfo{year}{2003}), \eprint{hep-ph/0306156}.

\bibitem[{\citenamefont{Altarelli et~al.}(2006)\citenamefont{Altarelli, Ball,
  and Forte}}]{Altarelli:2005ni}
\bibinfo{author}{\bibfnamefont{G.}~\bibnamefont{Altarelli}},
  \bibinfo{author}{\bibfnamefont{R.~D.} \bibnamefont{Ball}}, \bibnamefont{and}
  \bibinfo{author}{\bibfnamefont{S.}~\bibnamefont{Forte}},
  \bibinfo{journal}{Nucl. Phys. B} \textbf{\bibinfo{volume}{742}},
  \bibinfo{pages}{1} (\bibinfo{year}{2006}), \eprint{hep-ph/0512237}.

\bibitem[{\citenamefont{Altarelli et~al.}(2008)\citenamefont{Altarelli, Ball,
  and Forte}}]{Altarelli:2008aj}
\bibinfo{author}{\bibfnamefont{G.}~\bibnamefont{Altarelli}},
  \bibinfo{author}{\bibfnamefont{R.~D.} \bibnamefont{Ball}}, \bibnamefont{and}
  \bibinfo{author}{\bibfnamefont{S.}~\bibnamefont{Forte}},
  \bibinfo{journal}{Nucl. Phys. B} \textbf{\bibinfo{volume}{799}},
  \bibinfo{pages}{199} (\bibinfo{year}{2008}), \eprint{0802.0032}.

\bibitem[{\citenamefont{White and Thorne}(2007)}]{White:2006yh}
\bibinfo{author}{\bibfnamefont{C.}~\bibnamefont{White}} \bibnamefont{and}
  \bibinfo{author}{\bibfnamefont{R.}~\bibnamefont{Thorne}},
  \bibinfo{journal}{Phys. Rev. D} \textbf{\bibinfo{volume}{75}},
  \bibinfo{pages}{034005} (\bibinfo{year}{2007}), \eprint{hep-ph/0611204}.

\bibitem[{\citenamefont{Marzani et~al.}(2008)\citenamefont{Marzani, Ball,
  Del~Duca, Forte, and Vicini}}]{Marzani:2008az}
\bibinfo{author}{\bibfnamefont{S.}~\bibnamefont{Marzani}},
  \bibinfo{author}{\bibfnamefont{R.~D.} \bibnamefont{Ball}},
  \bibinfo{author}{\bibfnamefont{V.}~\bibnamefont{Del~Duca}},
  \bibinfo{author}{\bibfnamefont{S.}~\bibnamefont{Forte}}, \bibnamefont{and}
  \bibinfo{author}{\bibfnamefont{A.}~\bibnamefont{Vicini}},
  \bibinfo{journal}{Nucl. Phys. B} \textbf{\bibinfo{volume}{800}},
  \bibinfo{pages}{127} (\bibinfo{year}{2008}), \eprint{0801.2544}.

\bibitem[{\citenamefont{Caola and Marzani}(2011)}]{Caola:2011wq}
\bibinfo{author}{\bibfnamefont{F.}~\bibnamefont{Caola}} \bibnamefont{and}
  \bibinfo{author}{\bibfnamefont{S.}~\bibnamefont{Marzani}},
  \bibinfo{journal}{Phys. Lett. B} \textbf{\bibinfo{volume}{698}},
  \bibinfo{pages}{275} (\bibinfo{year}{2011}), \eprint{1101.3975}.

\bibitem[{\citenamefont{Caola et~al.}(2011)\citenamefont{Caola, Forte, and
  Marzani}}]{Caola:2010kv}
\bibinfo{author}{\bibfnamefont{F.}~\bibnamefont{Caola}},
  \bibinfo{author}{\bibfnamefont{S.}~\bibnamefont{Forte}}, \bibnamefont{and}
  \bibinfo{author}{\bibfnamefont{S.}~\bibnamefont{Marzani}},
  \bibinfo{journal}{Nucl. Phys. B} \textbf{\bibinfo{volume}{846}},
  \bibinfo{pages}{167} (\bibinfo{year}{2011}), \eprint{1010.2743}.

\bibitem[{\citenamefont{Forte and Muselli}(2016)}]{Forte:2015gve}
\bibinfo{author}{\bibfnamefont{S.}~\bibnamefont{Forte}} \bibnamefont{and}
  \bibinfo{author}{\bibfnamefont{C.}~\bibnamefont{Muselli}},
  \bibinfo{journal}{JHEP} \textbf{\bibinfo{volume}{03}}, \bibinfo{pages}{122}
  (\bibinfo{year}{2016}), \eprint{1511.05561}.

\bibitem[{\citenamefont{Catani et~al.}(1990)\citenamefont{Catani, Ciafaloni,
  and Hautmann}}]{Catani:1990xk}
\bibinfo{author}{\bibfnamefont{S.}~\bibnamefont{Catani}},
  \bibinfo{author}{\bibfnamefont{M.}~\bibnamefont{Ciafaloni}},
  \bibnamefont{and} \bibinfo{author}{\bibfnamefont{F.}~\bibnamefont{Hautmann}},
  \bibinfo{journal}{Phys. Lett. B} \textbf{\bibinfo{volume}{242}},
  \bibinfo{pages}{97} (\bibinfo{year}{1990}).

\bibitem[{\citenamefont{Catani et~al.}(1991)\citenamefont{Catani, Ciafaloni,
  and Hautmann}}]{Catani:1990eg}
\bibinfo{author}{\bibfnamefont{S.}~\bibnamefont{Catani}},
  \bibinfo{author}{\bibfnamefont{M.}~\bibnamefont{Ciafaloni}},
  \bibnamefont{and} \bibinfo{author}{\bibfnamefont{F.}~\bibnamefont{Hautmann}},
  \bibinfo{journal}{Nucl. Phys. B} \textbf{\bibinfo{volume}{366}},
  \bibinfo{pages}{135} (\bibinfo{year}{1991}).

\bibitem[{\citenamefont{Collins and Ellis}(1991)}]{Collins:1991ty}
\bibinfo{author}{\bibfnamefont{J.~C.} \bibnamefont{Collins}} \bibnamefont{and}
  \bibinfo{author}{\bibfnamefont{R.}~\bibnamefont{Ellis}},
  \bibinfo{journal}{Nucl. Phys. B} \textbf{\bibinfo{volume}{360}},
  \bibinfo{pages}{3} (\bibinfo{year}{1991}).

\bibitem[{\citenamefont{Catani et~al.}(1993)\citenamefont{Catani, Ciafaloni,
  and Hautmann}}]{Catani:1993ww}
\bibinfo{author}{\bibfnamefont{S.}~\bibnamefont{Catani}},
  \bibinfo{author}{\bibfnamefont{M.}~\bibnamefont{Ciafaloni}},
  \bibnamefont{and} \bibinfo{author}{\bibfnamefont{F.}~\bibnamefont{Hautmann}},
  \bibinfo{journal}{Phys. Lett. B} \textbf{\bibinfo{volume}{307}},
  \bibinfo{pages}{147} (\bibinfo{year}{1993}).

\bibitem[{\citenamefont{Catani and Hautmann}(1993)}]{Catani:1993rn}
\bibinfo{author}{\bibfnamefont{S.}~\bibnamefont{Catani}} \bibnamefont{and}
  \bibinfo{author}{\bibfnamefont{F.}~\bibnamefont{Hautmann}},
  \bibinfo{journal}{Phys. Lett. B} \textbf{\bibinfo{volume}{315}},
  \bibinfo{pages}{157} (\bibinfo{year}{1993}).

\bibitem[{\citenamefont{Catani and Hautmann}(1994)}]{Catani:1994sq}
\bibinfo{author}{\bibfnamefont{S.}~\bibnamefont{Catani}} \bibnamefont{and}
  \bibinfo{author}{\bibfnamefont{F.}~\bibnamefont{Hautmann}},
  \bibinfo{journal}{Nucl. Phys. B} \textbf{\bibinfo{volume}{427}},
  \bibinfo{pages}{475} (\bibinfo{year}{1994}), \eprint{hep-ph/9405388}.

\bibitem[{\citenamefont{Ball}(2008)}]{Ball:2007ra}
\bibinfo{author}{\bibfnamefont{R.~D.} \bibnamefont{Ball}},
  \bibinfo{journal}{Nucl. Phys. B} \textbf{\bibinfo{volume}{796}},
  \bibinfo{pages}{137} (\bibinfo{year}{2008}), \eprint{0708.1277}.

\bibitem[{\citenamefont{Bonvini et~al.}(2016)\citenamefont{Bonvini, Marzani,
  and Peraro}}]{Bonvini:2016wki}
\bibinfo{author}{\bibfnamefont{M.}~\bibnamefont{Bonvini}},
  \bibinfo{author}{\bibfnamefont{S.}~\bibnamefont{Marzani}}, \bibnamefont{and}
  \bibinfo{author}{\bibfnamefont{T.}~\bibnamefont{Peraro}},
  \bibinfo{journal}{Eur. Phys. J. C} \textbf{\bibinfo{volume}{76}},
  \bibinfo{pages}{597} (\bibinfo{year}{2016}), \eprint{1607.02153}.

\bibitem[{\citenamefont{Bonvini et~al.}(2017)\citenamefont{Bonvini, Marzani,
  and Muselli}}]{Bonvini:2017ogt}
\bibinfo{author}{\bibfnamefont{M.}~\bibnamefont{Bonvini}},
  \bibinfo{author}{\bibfnamefont{S.}~\bibnamefont{Marzani}}, \bibnamefont{and}
  \bibinfo{author}{\bibfnamefont{C.}~\bibnamefont{Muselli}},
  \bibinfo{journal}{JHEP} \textbf{\bibinfo{volume}{12}}, \bibinfo{pages}{117}
  (\bibinfo{year}{2017}), \eprint{1708.07510}.

\bibitem[{\citenamefont{Bonvini}(2018)}]{Bonvini:2018iwt}
\bibinfo{author}{\bibfnamefont{M.}~\bibnamefont{Bonvini}},
  \bibinfo{journal}{Eur. Phys. J. C} \textbf{\bibinfo{volume}{78}},
  \bibinfo{pages}{834} (\bibinfo{year}{2018}), \eprint{1805.08785}.

\bibitem[{\citenamefont{Bonciani et~al.}(2003)\citenamefont{Bonciani, Catani,
  Mangano, and Nason}}]{Bonciani:2003nt}
\bibinfo{author}{\bibfnamefont{R.}~\bibnamefont{Bonciani}},
  \bibinfo{author}{\bibfnamefont{S.}~\bibnamefont{Catani}},
  \bibinfo{author}{\bibfnamefont{M.~L.} \bibnamefont{Mangano}},
  \bibnamefont{and} \bibinfo{author}{\bibfnamefont{P.}~\bibnamefont{Nason}},
  \bibinfo{journal}{Phys. Lett. B} \textbf{\bibinfo{volume}{575}},
  \bibinfo{pages}{268} (\bibinfo{year}{2003}), \eprint{hep-ph/0307035}.

\bibitem[{\citenamefont{de~Florian et~al.}(2006)\citenamefont{de~Florian,
  Kulesza, and Vogelsang}}]{deFlorian:2005fzc}
\bibinfo{author}{\bibfnamefont{D.}~\bibnamefont{de~Florian}},
  \bibinfo{author}{\bibfnamefont{A.}~\bibnamefont{Kulesza}}, \bibnamefont{and}
  \bibinfo{author}{\bibfnamefont{W.}~\bibnamefont{Vogelsang}},
  \bibinfo{journal}{JHEP} \textbf{\bibinfo{volume}{02}}, \bibinfo{pages}{047}
  (\bibinfo{year}{2006}), \eprint{hep-ph/0511205}.

\bibitem[{\citenamefont{Muselli et~al.}(2017)\citenamefont{Muselli, Forte, and
  Ridolfi}}]{Muselli:2017bad}
\bibinfo{author}{\bibfnamefont{C.}~\bibnamefont{Muselli}},
  \bibinfo{author}{\bibfnamefont{S.}~\bibnamefont{Forte}}, \bibnamefont{and}
  \bibinfo{author}{\bibfnamefont{G.}~\bibnamefont{Ridolfi}},
  \bibinfo{journal}{JHEP} \textbf{\bibinfo{volume}{03}}, \bibinfo{pages}{106}
  (\bibinfo{year}{2017}), \eprint{1701.01464}.

\bibitem[{\citenamefont{Forte et~al.}(2021)\citenamefont{Forte, Ridolfi, and
  Rota}}]{Forte:2021wxe}
\bibinfo{author}{\bibfnamefont{S.}~\bibnamefont{Forte}},
  \bibinfo{author}{\bibfnamefont{G.}~\bibnamefont{Ridolfi}}, \bibnamefont{and}
  \bibinfo{author}{\bibfnamefont{S.}~\bibnamefont{Rota}},
  \bibinfo{journal}{JHEP} \textbf{\bibinfo{volume}{08}}, \bibinfo{pages}{110}
  (\bibinfo{year}{2021}), \eprint{2106.11321}.

\bibitem[{\citenamefont{Bonvini and Marzani}(2018)}]{Bonvini:2018ixe}
\bibinfo{author}{\bibfnamefont{M.}~\bibnamefont{Bonvini}} \bibnamefont{and}
  \bibinfo{author}{\bibfnamefont{S.}~\bibnamefont{Marzani}},
  \bibinfo{journal}{Phys. Rev. Lett.} \textbf{\bibinfo{volume}{120}},
  \bibinfo{pages}{202003} (\bibinfo{year}{2018}), \eprint{1802.07758}.

\bibitem[{\citenamefont{Ball et~al.}(2013)\citenamefont{Ball, Bonvini, Forte,
  Marzani, and Ridolfi}}]{Ball:2013bra}
\bibinfo{author}{\bibfnamefont{R.~D.} \bibnamefont{Ball}},
  \bibinfo{author}{\bibfnamefont{M.}~\bibnamefont{Bonvini}},
  \bibinfo{author}{\bibfnamefont{S.}~\bibnamefont{Forte}},
  \bibinfo{author}{\bibfnamefont{S.}~\bibnamefont{Marzani}}, \bibnamefont{and}
  \bibinfo{author}{\bibfnamefont{G.}~\bibnamefont{Ridolfi}},
  \bibinfo{journal}{Nucl. Phys. B} \textbf{\bibinfo{volume}{874}},
  \bibinfo{pages}{746} (\bibinfo{year}{2013}), \eprint{1303.3590}.

\bibitem[{\citenamefont{Hautmann}(2002)}]{Hautmann:2002tu}
\bibinfo{author}{\bibfnamefont{F.}~\bibnamefont{Hautmann}},
  \bibinfo{journal}{Phys. Lett. B} \textbf{\bibinfo{volume}{535}},
  \bibinfo{pages}{159} (\bibinfo{year}{2002}), \eprint{hep-ph/0203140}.

\bibitem[{\citenamefont{Cipriano et~al.}(2013)\citenamefont{Cipriano, Dooling,
  Grebenyuk, Gunnellini, Hautmann, Jung, and Katsas}}]{Cipriano:2013ooa}
\bibinfo{author}{\bibfnamefont{P.}~\bibnamefont{Cipriano}},
  \bibinfo{author}{\bibfnamefont{S.}~\bibnamefont{Dooling}},
  \bibinfo{author}{\bibfnamefont{A.}~\bibnamefont{Grebenyuk}},
  \bibinfo{author}{\bibfnamefont{P.}~\bibnamefont{Gunnellini}},
  \bibinfo{author}{\bibfnamefont{F.}~\bibnamefont{Hautmann}},
  \bibinfo{author}{\bibfnamefont{H.}~\bibnamefont{Jung}}, \bibnamefont{and}
  \bibinfo{author}{\bibfnamefont{P.}~\bibnamefont{Katsas}},
  \bibinfo{journal}{Phys.\ Rev.\ D} \textbf{\bibinfo{volume}{88}},
  \bibinfo{pages}{097501} (\bibinfo{year}{2013}), \eprint{1308.1655}.

\bibitem[{\citenamefont{Anchordoqui et~al.}(2022)}]{Anchordoqui:2021ghd}
\bibinfo{author}{\bibfnamefont{L.~A.} \bibnamefont{Anchordoqui}}
  \bibnamefont{et~al.}, \bibinfo{journal}{Phys. Rept.}
  \textbf{\bibinfo{volume}{968}}, \bibinfo{pages}{1} (\bibinfo{year}{2022}),
  \eprint{2109.10905}.

\bibitem[{\citenamefont{Feng et~al.}(2022)}]{Feng:2022inv}
\bibinfo{author}{\bibfnamefont{J.~L.} \bibnamefont{Feng}} \bibnamefont{et~al.}
  (\bibinfo{year}{2022}), \eprint{2203.05090}.

\bibitem[{\citenamefont{Brodsky et~al.}(2006)\citenamefont{Brodsky,
  Kopeliovich, Schmidt, and Soffer}}]{Brodsky:2006wb}
\bibinfo{author}{\bibfnamefont{S.~J.} \bibnamefont{Brodsky}},
  \bibinfo{author}{\bibfnamefont{B.}~\bibnamefont{Kopeliovich}},
  \bibinfo{author}{\bibfnamefont{I.}~\bibnamefont{Schmidt}}, \bibnamefont{and}
  \bibinfo{author}{\bibfnamefont{J.}~\bibnamefont{Soffer}},
  \bibinfo{journal}{Phys. Rev. D} \textbf{\bibinfo{volume}{73}},
  \bibinfo{pages}{113005} (\bibinfo{year}{2006}), \eprint{hep-ph/0603238}.

\bibitem[{\citenamefont{Brodsky et~al.}(2009)\citenamefont{Brodsky, Goldhaber,
  Kopeliovich, and Schmidt}}]{Brodsky:2007yz}
\bibinfo{author}{\bibfnamefont{S.~J.} \bibnamefont{Brodsky}},
  \bibinfo{author}{\bibfnamefont{A.~S.} \bibnamefont{Goldhaber}},
  \bibinfo{author}{\bibfnamefont{B.~Z.} \bibnamefont{Kopeliovich}},
  \bibnamefont{and} \bibinfo{author}{\bibfnamefont{I.}~\bibnamefont{Schmidt}},
  \bibinfo{journal}{Nucl. Phys. B} \textbf{\bibinfo{volume}{807}},
  \bibinfo{pages}{334} (\bibinfo{year}{2009}), \eprint{0707.4658}.

\bibitem[{\citenamefont{Brodsky et~al.}(1980)\citenamefont{Brodsky, Hoyer,
  Peterson, and Sakai}}]{Brodsky:1980pb}
\bibinfo{author}{\bibfnamefont{S.~J.} \bibnamefont{Brodsky}},
  \bibinfo{author}{\bibfnamefont{P.}~\bibnamefont{Hoyer}},
  \bibinfo{author}{\bibfnamefont{C.}~\bibnamefont{Peterson}}, \bibnamefont{and}
  \bibinfo{author}{\bibfnamefont{N.}~\bibnamefont{Sakai}},
  \bibinfo{journal}{Phys. Lett. B} \textbf{\bibinfo{volume}{93}},
  \bibinfo{pages}{451} (\bibinfo{year}{1980}).

\bibitem[{\citenamefont{Brodsky et~al.}(2015)\citenamefont{Brodsky, Kusina,
  Lyonnet, Schienbein, Spiesberger, and Vogt}}]{Brodsky:2015fna}
\bibinfo{author}{\bibfnamefont{S.~J.} \bibnamefont{Brodsky}},
  \bibinfo{author}{\bibfnamefont{A.}~\bibnamefont{Kusina}},
  \bibinfo{author}{\bibfnamefont{F.}~\bibnamefont{Lyonnet}},
  \bibinfo{author}{\bibfnamefont{I.}~\bibnamefont{Schienbein}},
  \bibinfo{author}{\bibfnamefont{H.}~\bibnamefont{Spiesberger}},
  \bibnamefont{and} \bibinfo{author}{\bibfnamefont{R.}~\bibnamefont{Vogt}},
  \bibinfo{journal}{Adv. High Energy Phys.} \textbf{\bibinfo{volume}{2015}},
  \bibinfo{pages}{231547} (\bibinfo{year}{2015}), \eprint{1504.06287}.

\bibitem[{\citenamefont{Kniehl et~al.}(2009)\citenamefont{Kniehl, Kramer,
  Schienbein, and Spiesberger}}]{Kniehl:2009ar}
\bibinfo{author}{\bibfnamefont{B.~A.} \bibnamefont{Kniehl}},
  \bibinfo{author}{\bibfnamefont{G.}~\bibnamefont{Kramer}},
  \bibinfo{author}{\bibfnamefont{I.}~\bibnamefont{Schienbein}},
  \bibnamefont{and}
  \bibinfo{author}{\bibfnamefont{H.}~\bibnamefont{Spiesberger}},
  \bibinfo{journal}{Phys. Rev. D} \textbf{\bibinfo{volume}{79}},
  \bibinfo{pages}{094009} (\bibinfo{year}{2009}), \eprint{0901.4130}.

\bibitem[{\citenamefont{Bednyakov et~al.}(2014)\citenamefont{Bednyakov,
  Demichev, Lykasov, Stavreva, and Stockton}}]{Bednyakov:2013zta}
\bibinfo{author}{\bibfnamefont{V.~A.} \bibnamefont{Bednyakov}},
  \bibinfo{author}{\bibfnamefont{M.~A.} \bibnamefont{Demichev}},
  \bibinfo{author}{\bibfnamefont{G.~I.} \bibnamefont{Lykasov}},
  \bibinfo{author}{\bibfnamefont{T.}~\bibnamefont{Stavreva}}, \bibnamefont{and}
  \bibinfo{author}{\bibfnamefont{M.}~\bibnamefont{Stockton}},
  \bibinfo{journal}{Phys. Lett. B} \textbf{\bibinfo{volume}{728}},
  \bibinfo{pages}{602} (\bibinfo{year}{2014}), \eprint{1305.3548}.

\bibitem[{\citenamefont{Ball et~al.}(2015)\citenamefont{Ball, Bonvini, and
  Rottoli}}]{Ball:2015dpa}
\bibinfo{author}{\bibfnamefont{R.~D.} \bibnamefont{Ball}},
  \bibinfo{author}{\bibfnamefont{M.}~\bibnamefont{Bonvini}}, \bibnamefont{and}
  \bibinfo{author}{\bibfnamefont{L.}~\bibnamefont{Rottoli}},
  \bibinfo{journal}{JHEP} \textbf{\bibinfo{volume}{11}}, \bibinfo{pages}{122}
  (\bibinfo{year}{2015}), \eprint{1510.02491}.

\bibitem[{\citenamefont{Dulat et~al.}(2014)\citenamefont{Dulat, Hou, Gao,
  Huston, Pumplin, Schmidt, Stump, and Yuan}}]{Dulat:2013hea}
\bibinfo{author}{\bibfnamefont{S.}~\bibnamefont{Dulat}},
  \bibinfo{author}{\bibfnamefont{T.-J.} \bibnamefont{Hou}},
  \bibinfo{author}{\bibfnamefont{J.}~\bibnamefont{Gao}},
  \bibinfo{author}{\bibfnamefont{J.}~\bibnamefont{Huston}},
  \bibinfo{author}{\bibfnamefont{J.}~\bibnamefont{Pumplin}},
  \bibinfo{author}{\bibfnamefont{C.}~\bibnamefont{Schmidt}},
  \bibinfo{author}{\bibfnamefont{D.}~\bibnamefont{Stump}}, \bibnamefont{and}
  \bibinfo{author}{\bibfnamefont{C.~P.} \bibnamefont{Yuan}},
  \bibinfo{journal}{Phys. Rev. D} \textbf{\bibinfo{volume}{89}},
  \bibinfo{pages}{073004} (\bibinfo{year}{2014}), \eprint{1309.0025}.

\bibitem[{\citenamefont{Ball et~al.}(2016{\natexlab{a}})\citenamefont{Ball,
  Bertone, Bonvini, Forte, Groth~Merrild, Rojo, and Rottoli}}]{Ball:2015tna}
\bibinfo{author}{\bibfnamefont{R.~D.} \bibnamefont{Ball}},
  \bibinfo{author}{\bibfnamefont{V.}~\bibnamefont{Bertone}},
  \bibinfo{author}{\bibfnamefont{M.}~\bibnamefont{Bonvini}},
  \bibinfo{author}{\bibfnamefont{S.}~\bibnamefont{Forte}},
  \bibinfo{author}{\bibfnamefont{P.}~\bibnamefont{Groth~Merrild}},
  \bibinfo{author}{\bibfnamefont{J.}~\bibnamefont{Rojo}}, \bibnamefont{and}
  \bibinfo{author}{\bibfnamefont{L.}~\bibnamefont{Rottoli}},
  \bibinfo{journal}{Phys. Lett. B} \textbf{\bibinfo{volume}{754}},
  \bibinfo{pages}{49} (\bibinfo{year}{2016}{\natexlab{a}}),
  \eprint{1510.00009}.

\bibitem[{\citenamefont{Ball et~al.}(2016{\natexlab{b}})\citenamefont{Ball,
  Bertone, Bonvini, Carrazza, Forte, Guffanti, Hartland, Rojo, and
  Rottoli}}]{Ball:2016neh}
\bibinfo{author}{\bibfnamefont{R.~D.} \bibnamefont{Ball}},
  \bibinfo{author}{\bibfnamefont{V.}~\bibnamefont{Bertone}},
  \bibinfo{author}{\bibfnamefont{M.}~\bibnamefont{Bonvini}},
  \bibinfo{author}{\bibfnamefont{S.}~\bibnamefont{Carrazza}},
  \bibinfo{author}{\bibfnamefont{S.}~\bibnamefont{Forte}},
  \bibinfo{author}{\bibfnamefont{A.}~\bibnamefont{Guffanti}},
  \bibinfo{author}{\bibfnamefont{N.~P.} \bibnamefont{Hartland}},
  \bibinfo{author}{\bibfnamefont{J.}~\bibnamefont{Rojo}}, \bibnamefont{and}
  \bibinfo{author}{\bibfnamefont{L.}~\bibnamefont{Rottoli}}
  (\bibinfo{collaboration}{NNPDF}), \bibinfo{journal}{Eur. Phys. J. C}
  \textbf{\bibinfo{volume}{76}}, \bibinfo{pages}{647}
  (\bibinfo{year}{2016}{\natexlab{b}}), \eprint{1605.06515}.

\bibitem[{\citenamefont{Bodwin et~al.}(2013)\citenamefont{Bodwin, Petriello,
  Stoynev, and Velasco}}]{Bodwin:2013gca}
\bibinfo{author}{\bibfnamefont{G.~T.} \bibnamefont{Bodwin}},
  \bibinfo{author}{\bibfnamefont{F.}~\bibnamefont{Petriello}},
  \bibinfo{author}{\bibfnamefont{S.}~\bibnamefont{Stoynev}}, \bibnamefont{and}
  \bibinfo{author}{\bibfnamefont{M.}~\bibnamefont{Velasco}},
  \bibinfo{journal}{Phys. Rev. D} \textbf{\bibinfo{volume}{88}},
  \bibinfo{pages}{053003} (\bibinfo{year}{2013}), \eprint{1306.5770}.

\bibitem[{\citenamefont{Bodwin et~al.}(2014)\citenamefont{Bodwin, Chung, Ee,
  Lee, and Petriello}}]{Bodwin:2014bpa}
\bibinfo{author}{\bibfnamefont{G.~T.} \bibnamefont{Bodwin}},
  \bibinfo{author}{\bibfnamefont{H.~S.} \bibnamefont{Chung}},
  \bibinfo{author}{\bibfnamefont{J.-H.} \bibnamefont{Ee}},
  \bibinfo{author}{\bibfnamefont{J.}~\bibnamefont{Lee}}, \bibnamefont{and}
  \bibinfo{author}{\bibfnamefont{F.}~\bibnamefont{Petriello}},
  \bibinfo{journal}{Phys. Rev. D} \textbf{\bibinfo{volume}{90}},
  \bibinfo{pages}{113010} (\bibinfo{year}{2014}), \eprint{1407.6695}.

\bibitem[{\citenamefont{Kagan et~al.}(2015)\citenamefont{Kagan, Perez,
  Petriello, Soreq, Stoynev, and Zupan}}]{Kagan:2014ila}
\bibinfo{author}{\bibfnamefont{A.~L.} \bibnamefont{Kagan}},
  \bibinfo{author}{\bibfnamefont{G.}~\bibnamefont{Perez}},
  \bibinfo{author}{\bibfnamefont{F.}~\bibnamefont{Petriello}},
  \bibinfo{author}{\bibfnamefont{Y.}~\bibnamefont{Soreq}},
  \bibinfo{author}{\bibfnamefont{S.}~\bibnamefont{Stoynev}}, \bibnamefont{and}
  \bibinfo{author}{\bibfnamefont{J.}~\bibnamefont{Zupan}},
  \bibinfo{journal}{Phys. Rev. Lett.} \textbf{\bibinfo{volume}{114}},
  \bibinfo{pages}{101802} (\bibinfo{year}{2015}), \eprint{1406.1722}.

\bibitem[{\citenamefont{K\"onig and Neubert}(2015)}]{Konig:2015qat}
\bibinfo{author}{\bibfnamefont{M.}~\bibnamefont{K\"onig}} \bibnamefont{and}
  \bibinfo{author}{\bibfnamefont{M.}~\bibnamefont{Neubert}},
  \bibinfo{journal}{JHEP} \textbf{\bibinfo{volume}{08}}, \bibinfo{pages}{012}
  (\bibinfo{year}{2015}), \eprint{1505.03870}.

\bibitem[{\citenamefont{Brivio et~al.}(2015)\citenamefont{Brivio, Goertz, and
  Isidori}}]{Brivio:2015fxa}
\bibinfo{author}{\bibfnamefont{I.}~\bibnamefont{Brivio}},
  \bibinfo{author}{\bibfnamefont{F.}~\bibnamefont{Goertz}}, \bibnamefont{and}
  \bibinfo{author}{\bibfnamefont{G.}~\bibnamefont{Isidori}},
  \bibinfo{journal}{Phys. Rev. Lett.} \textbf{\bibinfo{volume}{115}},
  \bibinfo{pages}{211801} (\bibinfo{year}{2015}), \eprint{1507.02916}.

\bibitem[{\citenamefont{Han et~al.}(2022)\citenamefont{Han, Leibovich, Ma, and
  Tan}}]{Han:2022rwq}
\bibinfo{author}{\bibfnamefont{T.}~\bibnamefont{Han}},
  \bibinfo{author}{\bibfnamefont{A.~K.} \bibnamefont{Leibovich}},
  \bibinfo{author}{\bibfnamefont{Y.}~\bibnamefont{Ma}}, \bibnamefont{and}
  \bibinfo{author}{\bibfnamefont{X.-Z.} \bibnamefont{Tan}}
  (\bibinfo{year}{2022}), \eprint{2202.08273}.

\bibitem[{\citenamefont{Jiang and Qiao}(2016)}]{Jiang:2015pah}
\bibinfo{author}{\bibfnamefont{J.}~\bibnamefont{Jiang}} \bibnamefont{and}
  \bibinfo{author}{\bibfnamefont{C.-F.} \bibnamefont{Qiao}},
  \bibinfo{journal}{Phys. Rev. D} \textbf{\bibinfo{volume}{93}},
  \bibinfo{pages}{054031} (\bibinfo{year}{2016}), \eprint{1512.01327}.

\bibitem[{\citenamefont{Karyasov et~al.}(2016)\citenamefont{Karyasov,
  Martynenko, and Martynenko}}]{Karyasov:2016hfm}
\bibinfo{author}{\bibfnamefont{A.~A.} \bibnamefont{Karyasov}},
  \bibinfo{author}{\bibfnamefont{A.~P.} \bibnamefont{Martynenko}},
  \bibnamefont{and} \bibinfo{author}{\bibfnamefont{F.~A.}
  \bibnamefont{Martynenko}}, \bibinfo{journal}{Nucl. Phys. B}
  \textbf{\bibinfo{volume}{911}}, \bibinfo{pages}{36} (\bibinfo{year}{2016}),
  \eprint{1604.07633}.

\bibitem[{\citenamefont{Liao et~al.}(2018)\citenamefont{Liao, Deng, Yu, Wang,
  and Xie}}]{Liao:2018nab}
\bibinfo{author}{\bibfnamefont{Q.-L.} \bibnamefont{Liao}},
  \bibinfo{author}{\bibfnamefont{Y.}~\bibnamefont{Deng}},
  \bibinfo{author}{\bibfnamefont{Y.}~\bibnamefont{Yu}},
  \bibinfo{author}{\bibfnamefont{G.-C.} \bibnamefont{Wang}}, \bibnamefont{and}
  \bibinfo{author}{\bibfnamefont{G.-Y.} \bibnamefont{Xie}},
  \bibinfo{journal}{Phys. Rev. D} \textbf{\bibinfo{volume}{98}},
  \bibinfo{pages}{036014} (\bibinfo{year}{2018}), \eprint{1807.11918}.

\bibitem[{\citenamefont{Celiberto
  et~al.}(2022{\natexlab{c}})\citenamefont{Celiberto, Fucilla, Ivanov,
  Mohammed, and Papa}}]{Celiberto:2022fgx}
\bibinfo{author}{\bibfnamefont{F.~G.} \bibnamefont{Celiberto}},
  \bibinfo{author}{\bibfnamefont{M.}~\bibnamefont{Fucilla}},
  \bibinfo{author}{\bibfnamefont{D.~{\relax Yu}.} \bibnamefont{Ivanov}},
  \bibinfo{author}{\bibfnamefont{M.~M.~A.} \bibnamefont{Mohammed}},
  \bibnamefont{and} \bibinfo{author}{\bibfnamefont{A.}~\bibnamefont{Papa}},
  \bibinfo{journal}{JHEP, in press}  (\bibinfo{year}{2022}{\natexlab{c}}),
  \eprint{2205.02681}.

\bibitem[{\citenamefont{Hentschinski et~al.}(2021)\citenamefont{Hentschinski,
  Kutak, and van Hameren}}]{Hentschinski:2020tbi}
\bibinfo{author}{\bibfnamefont{M.}~\bibnamefont{Hentschinski}},
  \bibinfo{author}{\bibfnamefont{K.}~\bibnamefont{Kutak}}, \bibnamefont{and}
  \bibinfo{author}{\bibfnamefont{A.}~\bibnamefont{van Hameren}},
  \bibinfo{journal}{Eur. Phys. J. C} \textbf{\bibinfo{volume}{81}},
  \bibinfo{pages}{112} (\bibinfo{year}{2021}), \bibinfo{note}{[Erratum: Eur.
  Phys. J. C 81, 262 (2021)]}, \eprint{2011.03193}.

\bibitem[{\citenamefont{Nefedov}(2019)}]{Nefedov:2019mrg}
\bibinfo{author}{\bibfnamefont{M.~A.} \bibnamefont{Nefedov}},
  \bibinfo{journal}{Nucl. Phys. B} \textbf{\bibinfo{volume}{946}},
  \bibinfo{pages}{114715} (\bibinfo{year}{2019}), \eprint{1902.11030}.

\bibitem[{\citenamefont{Ivanov and Papa}(2012)}]{Ivanov:2012iv}
\bibinfo{author}{\bibfnamefont{D.~{\relax Yu}.} \bibnamefont{Ivanov}}
  \bibnamefont{and} \bibinfo{author}{\bibfnamefont{A.}~\bibnamefont{Papa}},
  \bibinfo{journal}{JHEP} \textbf{\bibinfo{volume}{07}}, \bibinfo{pages}{045}
  (\bibinfo{year}{2012}), \eprint{1205.6068}.

\bibitem[{\citenamefont{Harland-Lang et~al.}(2015)\citenamefont{Harland-Lang,
  Martin, Motylinski, and Thorne}}]{Harland-Lang:2014zoa}
\bibinfo{author}{\bibfnamefont{L.}~\bibnamefont{Harland-Lang}},
  \bibinfo{author}{\bibfnamefont{A.}~\bibnamefont{Martin}},
  \bibinfo{author}{\bibfnamefont{P.}~\bibnamefont{Motylinski}},
  \bibnamefont{and} \bibinfo{author}{\bibfnamefont{R.}~\bibnamefont{Thorne}},
  \bibinfo{journal}{Eur. Phys. J. C} \textbf{\bibinfo{volume}{75}},
  \bibinfo{pages}{204} (\bibinfo{year}{2015}), \eprint{1412.3989}.

\bibitem[{\citenamefont{Buckley et~al.}(2015)\citenamefont{Buckley, Ferrando,
  Lloyd, Nordstr\"om, Page, R\"ufenacht, Sch\"onherr, and
  Watt}}]{Buckley:2014ana}
\bibinfo{author}{\bibfnamefont{A.}~\bibnamefont{Buckley}},
  \bibinfo{author}{\bibfnamefont{J.}~\bibnamefont{Ferrando}},
  \bibinfo{author}{\bibfnamefont{S.}~\bibnamefont{Lloyd}},
  \bibinfo{author}{\bibfnamefont{K.}~\bibnamefont{Nordstr\"om}},
  \bibinfo{author}{\bibfnamefont{B.}~\bibnamefont{Page}},
  \bibinfo{author}{\bibfnamefont{M.}~\bibnamefont{R\"ufenacht}},
  \bibinfo{author}{\bibfnamefont{M.}~\bibnamefont{Sch\"onherr}},
  \bibnamefont{and} \bibinfo{author}{\bibfnamefont{G.}~\bibnamefont{Watt}},
  \bibinfo{journal}{Eur. Phys. J. C} \textbf{\bibinfo{volume}{75}},
  \bibinfo{pages}{132} (\bibinfo{year}{2015}), \eprint{1412.7420}.

\bibitem[{\citenamefont{Kneesch et~al.}(2008)\citenamefont{Kneesch, Kniehl,
  Kramer, and Schienbein}}]{Kneesch:2007ey}
\bibinfo{author}{\bibfnamefont{T.}~\bibnamefont{Kneesch}},
  \bibinfo{author}{\bibfnamefont{B.~A.} \bibnamefont{Kniehl}},
  \bibinfo{author}{\bibfnamefont{G.}~\bibnamefont{Kramer}}, \bibnamefont{and}
  \bibinfo{author}{\bibfnamefont{I.}~\bibnamefont{Schienbein}},
  \bibinfo{journal}{Nucl. Phys. B} \textbf{\bibinfo{volume}{799}},
  \bibinfo{pages}{34} (\bibinfo{year}{2008}), \eprint{0712.0481}.

\bibitem[{\citenamefont{Bertone et~al.}(2017)\citenamefont{Bertone, Carrazza,
  Hartland, Nocera, and Rojo}}]{Bertone:2017tyb}
\bibinfo{author}{\bibfnamefont{V.}~\bibnamefont{Bertone}},
  \bibinfo{author}{\bibfnamefont{S.}~\bibnamefont{Carrazza}},
  \bibinfo{author}{\bibfnamefont{N.~P.} \bibnamefont{Hartland}},
  \bibinfo{author}{\bibfnamefont{E.~R.} \bibnamefont{Nocera}},
  \bibnamefont{and} \bibinfo{author}{\bibfnamefont{J.}~\bibnamefont{Rojo}}
  (\bibinfo{collaboration}{NNPDF}), \bibinfo{journal}{Eur. Phys. J. C}
  \textbf{\bibinfo{volume}{77}}, \bibinfo{pages}{516} (\bibinfo{year}{2017}),
  \eprint{1706.07049}.

\bibitem[{\citenamefont{Bardeen et~al.}(1978)\citenamefont{Bardeen, Buras,
  Duke, and Muta}}]{PhysRevD.18.3998}
\bibinfo{author}{\bibfnamefont{W.~A.} \bibnamefont{Bardeen}},
  \bibinfo{author}{\bibfnamefont{A.~J.} \bibnamefont{Buras}},
  \bibinfo{author}{\bibfnamefont{D.~W.} \bibnamefont{Duke}}, \bibnamefont{and}
  \bibinfo{author}{\bibfnamefont{T.}~\bibnamefont{Muta}},
  \bibinfo{journal}{Phys. Rev. D} \textbf{\bibinfo{volume}{18}},
  \bibinfo{pages}{3998} (\bibinfo{year}{1978}).

\bibitem[{\citenamefont{Sabio~Vera}(2006)}]{Vera:2006un}
\bibinfo{author}{\bibfnamefont{A.}~\bibnamefont{Sabio~Vera}},
  \bibinfo{journal}{Nucl. Phys. B} \textbf{\bibinfo{volume}{746}},
  \bibinfo{pages}{1} (\bibinfo{year}{2006}), \eprint{hep-ph/0602250}.

\bibitem[{\citenamefont{Sabio~Vera and Schwennsen}(2007)}]{Vera:2007kn}
\bibinfo{author}{\bibfnamefont{A.}~\bibnamefont{Sabio~Vera}} \bibnamefont{and}
  \bibinfo{author}{\bibfnamefont{F.}~\bibnamefont{Schwennsen}},
  \bibinfo{journal}{Nucl. Phys. B} \textbf{\bibinfo{volume}{776}},
  \bibinfo{pages}{170} (\bibinfo{year}{2007}), \eprint{hep-ph/0702158}.

\bibitem[{\citenamefont{Marquet and Royon}(2009)}]{Marquet:2007xx}
\bibinfo{author}{\bibfnamefont{C.}~\bibnamefont{Marquet}} \bibnamefont{and}
  \bibinfo{author}{\bibfnamefont{C.}~\bibnamefont{Royon}},
  \bibinfo{journal}{Phys. Rev. D} \textbf{\bibinfo{volume}{79}},
  \bibinfo{pages}{034028} (\bibinfo{year}{2009}), \eprint{0704.3409}.

\end{thebibliography}

\end{document}